\documentclass{elsart}

\usepackage{amsfonts}

\DeclareFontFamily{U}{msa}{}
\DeclareFontShape{U}{msa}{m}{n}
    { <-> msam10}{}
\DeclareSymbolFont{AMSa}{U}{msa}{m}{n}

\DeclareFontFamily{U}{msb}{}
\DeclareFontShape{U}{msb}{m}{n}
     { <-> msbm10}{}
\DeclareSymbolFont{AMSb}{U}{msb}{m}{n}

\DeclareFontFamily{U}{euf}{}
\DeclareFontShape{U}{euf}{m}{n}
    { <-> eufm10}{}
\DeclareFontShape{U}{euf}{b}{n}
    { <-> eufb10}{}

\DeclareFontFamily{U}{eus}{}
\DeclareFontShape{U}{eus}{m}{n}
    { <-> eusm10}{}
\DeclareFontShape{U}{eus}{b}{n}
    { <-> eusb10}{}

\DeclareFontFamily{U}{eur}{}
\DeclareFontShape{U}{eur}{m}{n}
    { <-> eurm10}{}
\DeclareFontShape{U}{eur}{b}{n}
    { <-> eurb10}{}

\DeclareMathAlphabet{\matheurm}{U}{eur}{m}{n}
\DeclareMathAlphabet{\matheuf}{U}{euf}{m}{n}
\DeclareMathAlphabet{\matheurmbf}{U}{eur}{b}{n}
\DeclareMathAlphabet{\matheuscr}{U}{eus}{m}{n}
\DeclareMathAlphabet{\mathsfsl}{OT1}{cmss}{m}{sl}
\DeclareMathAlphabet{\mathsf}{OT1}{cmss}{m}{n}
\DeclareFontShape{OT1}{cmr}{bx}{n}{ <-> cmbx10 }{}

\DeclareMathSymbol{\smallfrown}{\mathrel}{AMSa}{"61}
\DeclareMathSymbol{\subsetneq}{\mathrel}{AMSb}{"24}
\DeclareMathSymbol{\therefore}{\mathrel}{AMSa}{"29}
\DeclareMathSymbol\compl{\mathord}{AMSb}{"73}

\newcommand{\disptitle}[1]{{\sc #1}}

\newcommand{\elsaxiom}[1]{{\sf #1}}
\newcommand{\cnum}[1]{\matheurm #1}
\newcommand{\tiffdef}{$\iffdef$}
\newcommand{\concat}{{}^\smallfrown}

\newcommand{\famind}[1]{_{(#1)}}
\newcommand{\mathand}{\mathrm{\ \&\ }}
\newcommand{\mathor}{\mathrm{\normalfont\ or\ }}
\newcommand{\opneg}{{\boldsymbol{\neg}}}
\newcommand{\opand}{{\boldsymbol{\wedge}}}
\newcommand{\opor}{{\boldsymbol{\vee}}}
\newcommand{\opimplies}{{\boldsymbol{\rightarrow}}}

\newcommand{\opiff}{{\boldsymbol{\leftrightarrow}}}
\newcommand{\opin}{{\boldsymbol{\in}}}

\newcommand{\opeq}{{\boldsymbol{=}}}
\newcommand{\opforall}{{\boldsymbol{\forall}}}
\newcommand{\opexists}{{\boldsymbol{\exists}}}
\newcommand{\visavis}{\emph{vis-\`a-vis\ }}
\newcommand{\lcneg}{{\neg}}

\newcommand{\lcimplies}{{\rightarrow}}

\newcommand{\lciff}{{\leftrightarrow}}

\newcommand{\bv}[1]{{[\![#1]\!]}}
\newcommand{\preset}[2]{{}^{#1}{}#2}
\newcommand{\tzf}{{\theory{ZF}}}
\newcommand{\tzfa}{{\theory{ZFA}}}
\newcommand{\tzfc}{{\theory{ZFC}}}
\newcommand{\tzfca}{{\theory{ZFCA}}}
\newcommand{\tr}{\operatorname{tr}}
\newcommand{\vecsp}[1]{{\mathsf #1}}
\newcommand{\eqdef}{{\,\stackrel{\rm{def}}=\,}}
\newcommand{\iffdef}{\stackrel{\mathrm{def}}{\iff}}
\newcommand{\image}{{}^{\scriptscriptstyle\rightarrow}}
\newcommand{\invimage}{{}^{\scriptscriptstyle\leftarrow}}

\newcommand{\elsalg}[1]{\mathfrak #1}
\newcommand{\ideal}[1]{\elsalg #1}
\newcommand{\structure}[1]{\mathfrak{#1}}
\newcommand{\theory}[1]{\mathsf {#1}}
\newcommand{\con}{{\operatorname{Con}}}
\newcommand{\Con}{{\operatorname{Con}}}
\newcommand{\lang}[1]{{\mathcal {#1}}}
\newcommand{\system}[1]{\mathfrak{#1}}
\newcommand{\proves}{\vdash}
\newcommand{\forces}{\Vdash}

\newcommand{\powerset}{{\operatorname{\mathcal P}}}

\newcommand{\im}{\operatorname{im}}
\newcommand{\dom}{\operatorname{dom}}

\newcommand{\direction}[1]{\cnum{#1}}

\newtheorem{premise}{Premise}
\newtheorem{premprime}{Premise}

\newtheorem{inference}{Inference}
\newtheorem{definition}{Definition}

\newtheorem{statement}{}
\newcommand{\Element}{\operatorname{Element}}
\newcommand{\Set}{\operatorname{Set}}
\newcommand{\Class}{\operatorname{Class}}
\renewcommand{\opneg}{{\boldsymbol{\compl}}}
\renewcommand{\disptitle}[1]{{\sc #1}}
\newcommand{\entails}{{\Vvdash}}
\newcommand{\nentails}{\not{\hspace{-1ex}\Vvdash}}

\newcommand{\subexp}[1]{{\hat{#1}}}
\newcommand{\parper}{\square}
\newcommand{\semifil}[1]{{[#1]}}
\newcommand{\regalg}{\operatorname{Reg}}
\newcommand{\complet}[1]{{\overline{#1}}}

\newcommand{\completsf}[1]{{#1^{\mathrm{sf}}}}
\newcommand{\completcsf}[1]{{#1^{\mathrm{csf}}}}
\newcommand{\setalg}{{\elsalg S}}

\newcommand{\Exp}{\operatorname{Exp}}
\newcommand{\Gcheck}{{\boldsymbol{\mathsf G}}}
\newcommand{\Mtilde}{{\tilde M}}

\newcommand{\Com}{\operatorname{Com}}

\newenvironment{elscases}{\left\{\begin{array}{ll}}{\end{array}\right.}
\newcommand{\boldsymbol}[1]{\mbox{\boldmath $#1$}}
\newcommand{\implies}{\Longrightarrow}

\newlength{\pmbwidth}
\newsavebox{\pmbbox}
\newsavebox{\pmbchar}
\newsavebox{\pmbcompl}
\newsavebox{\pmbvee}
\newsavebox{\pmbwedge}

\newsavebox{\pmbtbigvee}
\newsavebox{\pmbdbigvee}

\newsavebox{\pmbtbigwedge}
\newsavebox{\pmbdbigwedge}

\newsavebox{\pmblcimplies}
\newsavebox{\pmblciff}
\newsavebox{\pmbforall}
\newsavebox{\pmbexists}
\newsavebox{\pmble}

\savebox{\pmbcompl}{%
\savebox{\pmbbox}{$\compl$}%
\settowidth{\pmbwidth}{\usebox{\pmbbox}}%
\raisebox{-.03ex}{\usebox{\pmbbox}}%
\hspace{-\pmbwidth}%
\hspace{.03ex}\usebox{\pmbbox}%
\hspace{-\pmbwidth}%
\hspace{.03ex}\raisebox{.03ex}{\usebox{\pmbbox}}%
}%

\savebox{\pmbvee}{%
\savebox{\pmbbox}{$\vee$}%
\settowidth{\pmbwidth}{\usebox{\pmbbox}}%
\raisebox{.07ex}{\usebox{\pmbbox}}%
\hspace{-\pmbwidth}%
\hspace{.03ex}\usebox{\pmbbox}%
\hspace{-\pmbwidth}%
\hspace{.03ex}\raisebox{.07ex}{\usebox{\pmbbox}}%
}%

\savebox{\pmbwedge}{%
\savebox{\pmbbox}{$\wedge$}%
\settowidth{\pmbwidth}{\usebox{\pmbbox}}%
\usebox{\pmbbox}%
\hspace{-\pmbwidth}%
\hspace{-.03ex}\raisebox{.07ex}{\usebox{\pmbbox}}%
\hspace{-\pmbwidth}%
\hspace{.06ex}\raisebox{.07ex}{\usebox{\pmbbox}}%
\hspace{-\pmbwidth}%
\hspace{.03ex}{\usebox{\pmbbox}}%
}%

\savebox{\pmbtbigvee}{%
\savebox{\pmbbox}{$\textstyle\bigvee$}%
\settowidth{\pmbwidth}{\usebox{\pmbbox}}%
\raisebox{.06ex}{\usebox{\pmbbox}}%
\hspace{-\pmbwidth}%
\hspace{.03ex}\raisebox{-.06ex}{\usebox{\pmbbox}}%
\hspace{-\pmbwidth}%
\hspace{.03ex}\raisebox{.06ex}{\usebox{\pmbbox}}%
}%

\savebox{\pmbdbigvee}{%
\savebox{\pmbbox}{$\displaystyle\bigvee$}%
\settowidth{\pmbwidth}{\usebox{\pmbbox}}%
\raisebox{.06ex}{\usebox{\pmbbox}}%
\hspace{-\pmbwidth}%
\hspace{.03ex}\raisebox{-.06ex}{\usebox{\pmbbox}}%
\hspace{-\pmbwidth}%
\hspace{.03ex}\raisebox{.06ex}{\usebox{\pmbbox}}%
}%

\savebox{\pmbtbigwedge}{%
\savebox{\pmbbox}{$\textstyle \bigwedge$}%
\settowidth{\pmbwidth}{\usebox{\pmbbox}}%
\raisebox{-.06ex}{\usebox{\pmbbox}}%
\hspace{-\pmbwidth}%
\hspace{.03ex}\raisebox{.06ex}{\usebox{\pmbbox}}%
\hspace{-\pmbwidth}%
\hspace{.03ex}\raisebox{-.06ex}{\usebox{\pmbbox}}%
}%

\savebox{\pmbdbigwedge}{%
\savebox{\pmbbox}{$\displaystyle \bigwedge$}%
\settowidth{\pmbwidth}{\usebox{\pmbbox}}%
\raisebox{-.06ex}{\usebox{\pmbbox}}%
\hspace{-\pmbwidth}%
\hspace{.03ex}\raisebox{.06ex}{\usebox{\pmbbox}}%
\hspace{-\pmbwidth}%
\hspace{.03ex}\raisebox{-.06ex}{\usebox{\pmbbox}}%
}%

\savebox{\pmblcimplies}{%
\savebox{\pmbbox}{$\lcimplies$}%
\settowidth{\pmbwidth}{\usebox{\pmbbox}}%
\raisebox{-.025ex}{\usebox{\pmbbox}}%
\hspace{-\pmbwidth}%
\raisebox{.025ex}{\usebox{\pmbbox}}%
\hspace{-\pmbwidth}%
\hspace{.05ex}{\usebox{\pmbbox}}%
}%

\savebox{\pmblciff}{%
\savebox{\pmbbox}{$\lciff$}%
\settowidth{\pmbwidth}{\usebox{\pmbbox}}%
\usebox{\pmbbox}%
\hspace{-\pmbwidth}%
\hspace{.05ex}%
\raisebox{-.025ex}{\usebox{\pmbbox}}%
\hspace{-\pmbwidth}%
\raisebox{.025ex}{\usebox{\pmbbox}}%
\hspace{-\pmbwidth}%
\hspace{.05ex}\usebox{\pmbbox}%
}%

\savebox{\pmbforall}{%
\savebox{\pmbbox}{$\forall$}%
\settowidth{\pmbwidth}{\usebox{\pmbbox}}%
\raisebox{.06ex}{\usebox{\pmbbox}}%
\hspace{-\pmbwidth}%
\hspace{.03ex}\raisebox{-.06ex}{\usebox{\pmbbox}}%
\hspace{-\pmbwidth}%
\hspace{.03ex}\raisebox{.06ex}{\usebox{\pmbbox}}%
}%

\savebox{\pmbexists}{%
\savebox{\pmbbox}{$\exists$}%
\settowidth{\pmbwidth}{\usebox{\pmbbox}}%
\raisebox{.025ex}{\usebox{\pmbbox}}%
\hspace{-\pmbwidth}%
\raisebox{-.025ex}{\usebox{\pmbbox}}%
\hspace{-\pmbwidth}%
\hspace{.05ex}%
\raisebox{.025ex}{\usebox{\pmbbox}}%
\hspace{-\pmbwidth}%
\raisebox{-.025ex}{\usebox{\pmbbox}}%
}%

\savebox{\pmble}{%
\savebox{\pmbbox}{$\le$}%
\settowidth{\pmbwidth}{\usebox{\pmbbox}}%
\usebox{\pmbbox}%
\hspace{-\pmbwidth}%
\hspace{.05ex}%
\raisebox{-.025ex}{\usebox{\pmbbox}}%
\hspace{-\pmbwidth}%
\raisebox{.025ex}{\usebox{\pmbbox}}%
}%

\newcommand{\bigbv}[1]{{%
\settowidth{\pmbwidth}{$\big[$}%
\big[\hspace{-\pmbwidth}
\hspace{.25ex}\big[#1\big]%
\settowidth{\pmbwidth}{$\big]$}%
\hspace{-\pmbwidth}%
\hspace{.25ex}%
\big]%
}}

\newcommand{\bcompl}{\usebox{\pmbcompl}}
\newcommand{\bvee}{\usebox{\pmbvee}}
\newcommand{\bwedge}{\usebox{\pmbwedge}}
\newcommand{\btbigvee}{\usebox{\pmbtbigvee}}
\newcommand{\bdbigvee}{\usebox{\pmbdbigvee}}
\newcommand{\btbigwedge}{\usebox{\pmbtbigwedge}}
\newcommand{\bdbigwedge}{\usebox{\pmbdbigwedge}}
\newcommand{\bimplies}{\usebox{\pmblcimplies}}
\newcommand{\biff}{\usebox{\pmblciff}}
\newcommand{\bforall}{\usebox{\pmbforall}}
\newcommand{\bexists}{\usebox{\pmbexists}}
\newcommand{\ble}{\usebox{\pmble}}

\newsymbol\Vdash 130D
\newsymbol\Vvdash 130E
\newsymbol\nsupseteq 232B
\newsymbol\shortparallel 2371

\newcommand{\ie}{i.e.}
\newcommand{\viz}{viz.}

\newcommand{\tfrac}[2]{{\textstyle\frac{#1}{#2}}}

\begin{document}

\section{Introduction}

In the companion paper\cite{RVWmw:2005} to this we discuss two approaches to the quantum theory of observation:  the Copenhagen interpretation and the so-called many-worlds interpretation.  The position of that paper may be paraphrased as follows:  The Copenhagen interpretation correctly describes the experience of observation of quantum systems but is untenable as a description of reality; whereas the many-worlds ``interpretation'' is not really an interpretation of quantum mechanics but is simply the application of standard quantum theory to the compound system consisting of the observed and the observer, both considered as physical systems.  Its tenability as a description of reality requires that it yield the experience of observation described by the Copenhagen interpretation, which is that of a stochastic process with probabilities given by the squared norm.  The principal result of \cite{RVWmw:2005} is that it does.

The present paper addresses a third approach to modeling quantum reality:  the \emph{hidden-variables} program, which proposes to account for the apparent nondeterministic nature of quantum measurement by supposing that physical states have properties that are not ``observable'' in the ordinary sense.  These non-observable attributes are the \emph{hidden variables} for which the program is named, and it is supposed that they uniquely determine the outcome of any observation.

The principal objects of interest in \cite{RVWmw:2005} are propositional algebras, which are boolean algebras of commuting quantum propositions, and the present work begins with an examination of the hidden-variables program for such algebras.  In Section~\ref{sec genericity} it is shown that hidden-variables models in this setting are \emph{generic} in the set-theoretic sense, and---as with the so-called many-worlds interpretation---they give the appearance of randomness, with the usual role of the squared norm as probability, while being themselves entirely deterministic.  Section~\ref{sec gen meth force} presents the theory of genericity \emph{per se}, providing a more comprehensive setting for the results of the preceding section and a more flexible tool for the work of Section~\ref{sec prop sys}, which extends the hidden-variables construction to suitable systems of possibly noncommuting propositions.  It follows from familiar \emph{no-hidden-variables} theorems that this extension cannot be universal, but a natural sufficient condition is given for its existence.  A full appreciation of the significance of genericity is only possible with some understanding of mathematical logic and models of set theory, and a brief sketch of this theory is given in Section~\ref{sec mod set theory} for the interested reader.  As in \cite{RVWmw:2005}, it will be convenient to employ certain elementary conventions of modern set theory throughout this discussion; some of these are introduced in \cite{RVWmw:2005}, and the rest are summarized in the appendix to the present paper.  Familiarity with \cite{RVWmw:2005} is assumed.  As in \cite{RVWmw:2005}, theorems of a general nature are labeled `{\bf Proposition}'; proofs of these are straightforward or readily available in the existing literature, but some are nevertheless presented here for their pedagogic value.  Theorems specific to the argument of this paper are labeled `{\bf Theorem}'.

\section{Limitations on the hidden-variables program}

At its most ambitious the hidden-variables program proposes that the true state of a system is such it has a definite value---1 or 0---for any quantum proposition.  It is well known that if the dimension of the statevector space for the system is greater than 2 then this is impossible, so it is necessary from the outset to restrict our attention to specific sets of propositions, for which purpose we introduce the notion of a \emph{propositional system}.  Recall (Section II of \cite{RVWmw:2005}) that the meet and join of propositions are defined iff they commute with one another, in which case meet is intersection and join is orthogonal complement of intersection of orthogonal complements.
\begin{definition}
\label{edd}
A \emph{propositional system} (PS) on a Hilbert space $\vecsp V$ is a nonempty set $\elsalg R$ of propositions that is closed under complementation and binary meet (equivalently, finitary meet or join).

A \emph{boolean subsystem} of $\elsalg R$ is a subset of $\elsalg R$ that is a propositional algebra (PA)---in other words, a subsystem of commuting elements.

$\elsalg R$ is \emph{complete} iff it is also closed under general meet.
\end{definition}
The set of all propositions for a Hilbert space $\vecsp V$ is of course a complete PS, and the intersection of any collection of complete PSs is a complete PS.  Given any PS $\elsalg R$ we define $\overline{\elsalg R}$, the \emph{completion of $\elsalg R$}, to be the smallest complete PS that includes $\elsalg R$, which is the intersection of all complete PSs that include $\elsalg R$.

We will call the hypothetical states called for by the hidden-variables program \emph{pseudoclassical}.  Specifically, a pseudoclassical state $\Psi$ for a PS $\elsalg R$ has a definite value---either 1 or 0---for each proposition $P \in \elsalg R$.  We use `true' and `yes' interchangeably with 1, and `false' and `no' interchangeably with 0.  For a given $\Psi$ and $P,Q \in \elsalg R$,
\begin{enumerate}
\item if $P$ is true of $\Psi$ and $P \le Q$ then $Q$ is true of $\Psi$;
\item if $P$ and $Q$ commute and $P$ and $Q$ are true of $\Psi$ then $P\wedge Q$ is true of $\Psi$.
\end{enumerate}
A proof of the inconsistency of these conditions for a given PS or family of PSs is called a \emph{no-hidden-variables} (NHV) theorem, and of these a number exist, the first having been given by von Neumann\cite{vonNeumann:1955}.\footnote{Note that while Conditions~1 and 2 are sufficient for the typical NHV theorem, they do not necessarily fully express the hidden-variables premise.}  All rely on the presence of noncommuting propositions in $\elsalg R$, for the simple reason that Conditions~1 and 2 can always be satisfied if $\elsalg R$ is boolean.

In the hidden-variables literature, dependence of the value of an observable on the particular interaction by which it is measured, as represented by the set of all (necessarily commuting) observables measured by that interaction, is called \emph{contextuality}.  To assert that $\Psi$ is pseudoclassical for a nonboolean PS, therefore, is to assert some degree of noncontextuality.  Some of the interest in NHV theorems has to do with the sort of noncontextuality they assume.  Bell's first NHV theorem\cite{Bell:1964}, for example, assumes only that degree of noncontextuality that follows from the principle of \emph{locality}, which states that the result of an observation localized in space does not depend on what other observations are considered along with it as long as the latter are localized to regions of space sufficiently remote from the first (see \cite{Mermin:1993} for a nice presentation of related NHV results).

Positive results have received relatively little attention in the hidden-variables literature, and it is this territory that we explore in this paper.  In the NHV literature the PAs $\elsalg P$ that occur (as boolean subsystems of the nonboolean PSs of interest) are typically finite, and for these the positive result is trivial:  the simultaneous eigenstates of all $P \in \elsalg P$ are the pseudoclassical states for $\elsalg P$.  The same holds for any \emph{atomic algebra}.\footnote{$\elsalg P$ is atomic iff every $P \in \elsalg P$ is the join of the atoms below it, an atom in $\elsalg P$ being a minimal nonzero element of $\elsalg P$.  The 1-eigenstates of the atoms are the pseudoclassical states.}  If $\elsalg P$ is not atomic then at least some pseudoclassical states are not ordinary quantum states, and if $\elsalg P$ is atomless---as it is in cases of interest for the purposes of this paper---then pseudoclassical states cannot be ordinary quantum states.

To take a familiar example, suppose $A$ is a continuous quantum observable, say the $x$-coordinate of a particle at a given time.  Let $\elsalg B$ be the boolean algebra of Borel subsets of $\mathbb R$, which are obtained from intervals by repeated applications of the operations of complementation and countable union and intersection.  For $X \in \elsalg B$, let $\chi^X$ be the characteristic function of $X$, so that
\[
\chi^X(x) =
 \begin{elscases}
  1 & \mbox{if $x \in X$}\\
  0 & \mbox{if $x \notin X$}.
 \end{elscases}
\]
Then $\chi^X(A)$ is a well defined selfadjoint operator, which represents the quantum proposition that the $x$-coordinate of the particle at the given time is in $X$.  Let $\elsalg A$ be the algebra of propositions $\chi^X(A)$ for $X \in \elsalg B$.  These are quantum propositions in the ordinary sense:  true of some states, false of others, and indeterminate for the rest.  Concerning a pseudoclassical state, however, each of these is true or false.  It will follow from the analysis presented below that a pseudoclassical state for $\elsalg A$ has a precise numerical value for $A$, but that this value is \emph{random} in the sense of \cite{Solovay:1970}, so that, for example, it is not in any definable set of real numbers with ordinary Lebesgue measure 0; it is not, for example, rational.

\section{Genericity}
\label{sec genericity}

\subsection{The necessity of genericity}
\label{sec nec gen}

We begin by considering a boolean algebra ${\elsalg P}$ of propositions, i.e., a propositional algebra, in a Hilbert space $\vecsp V$ of statevectors of a physical system $\system S$, and we suppose that every proposition in ${\elsalg P}$ corresponds to a physical observable.

Suppose $\Psi$ is a pseudoclassical state.  Let $ F^{\Psi} = F^{\Psi, {\elsalg P}}$ be the set of propositions in ${\elsalg P}$ true of $\Psi$.  By virtue of Conditions~1 and 2, $F^\Psi$ is a filter on $\elsalg P$.  Since $\Psi$ has a definite value for each $P \in \elsalg P$, $F^\Psi$ is an ultrafilter.  Let $\mathcal H^{\elsalg P}$ be the set of all $ F^\Psi$ where $\Psi$ is pseudoclassical.  We will call these the \emph{pseudoclassical filters}.  The characterization of ${\mathcal H^{\elsalg P}}$ is central to the hidden-variables program.

The basic hidden-variables premise is that the ordinary quantum propositions are all we can know of pseudoclassical states, or, equivalently, regarded as questions are all we can pose of pseudoclassical states.  To put it the other way round, any question that may be posed of a pseudoclassical state $\Psi$ \visavis ${\elsalg P}$ is equivalent to `is ${ P}$ true of $\Psi$?' for some proposition ${ P} \in {\elsalg P}$.  This naturally entails some restriction on the class of posable questions under consideration.

The mathematical equivalent of `is ${ P}$ true of $\Psi$?' is
\begin{statement}
\normalfont `${ P} \in  F^\Psi$?'.
\end{statement}
Since a pseudoclassical state $\Psi$ determines an ultrafilter $ F^\Psi$, any set $C$ of ultrafilters on ${\elsalg P}$ could correspond \emph{a priori} to a question concerning a pseudoclassical state, viz., the question
\begin{statement}
\label{F in C?}
\normalfont `$ F^\Psi \in C$?'.
\end{statement}
A hidden-variables model will in general consider only some of these as actually posable.

We will show that pseudoclassical states must correspond to generic filters and that generic filters in turn have all the properties required of pseudoclassical states in a hidden-variables model, so that they constitute the unique realization of the hidden-variables program in this setting.  We do this by stating several premises that a hidden-variables model should fulfill, from which we draw several inferences, concluding with the genericity property.  We assume familiarity with the concepts covered in the appendix.

Let $ {\mathcal C}$ be the set of sets of ultrafilters on ${\elsalg P}$ that correspond to posable questions of the form {\bf \ref{F in C?}} in a given hidden-variables model.
\begin{premise}
\label{prem basic}
The basic hidden-variables premise restricts $ {\mathcal C}$ and
${\mathcal H} = \mathcal H^{\elsalg P}$ relative to ${\elsalg P}$ by the requirement that
\begin{equation}
\label{dvp}
(\forall C \in  {\mathcal C})(\exists { P} \in {\elsalg P})(\forall F \in
{\mathcal H}) ({ P} \in F \iff F \in C).
\end{equation}
\end{premise}
\begin{definition}
\label{def topol}
In general, if $\elsalg P$ is a BA, let $\mathcal U^{\elsalg P}$ be the set of all ultrafilters on $\elsalg P$.  For any $S \subseteq |\elsalg P|$, let $\mathcal U_S = \mathcal U^{\elsalg P}_S$ be the set of ultrafilters on $\elsalg P$ containing a member of $S$.  For $P \in \elsalg P$, let $\mathcal U_P = \mathcal U_{\{P\}}$.

The sets $\mathcal U^{\elsalg P}_{P}$ form a base for a topology on $\mathcal U^{\elsalg P}$.  The open sets are arbitrary unions of these, i.e., sets of the form $\mathcal U^{\elsalg P}_{\mathcal S}$.
\end{definition}
We can certainly pose the negations, conjunctions, and disjunctions of posable questions, so
\begin{premise}
\label{prem boolean ops}
$ {\mathcal C}$ is closed under finitary boolean operations.
\end{premise}
We have assumed that every ${ P} \in {\elsalg P}$ is a posable question, so
\begin{premise}
\label{prem F sub P}
for every $P \in \elsalg P$, $\mathcal U_{ P} \in  {\mathcal C}$.
\end{premise}
The fundamental existence premise for the hidden-variables program is that every physical proposition that can be true for an ordinary quantum state is true for some pseudoclassical state; hence,
\begin{premise}
\label{dvv}
for every ${ P} \in {\elsalg P}^-$, for some $F \in \mathcal H$, $P \in F$.
\end{premise}
The following inferences may be drawn from Premises 1--4.
\begin{inference}
\label{dvo}
${ P}$ in (\ref{dvp}) is uniquely determined by $C$.
\end{inference}
\begin{pf} Suppose ${ P}$ and ${ P}'$ both work for a given $C$, and ${ P} \neq { P}'$.  Then either ${ P}-{ P}'$ or ${ P}'-{ P}$ is nonzero.  Without loss of generality we suppose ${ P}-{ P}' \neq 0$.  Premise~\ref{prem F sub P} implies that for some $F \in \mathcal H$, $({ P}-{ P}') \in F$, so ${ P} \in F$ and ${ P}' \notin F$, which contradicts the hypothesis that both ${ P}$ and ${ P}'$ satisfy (\ref{dvp}) for $C$.\qed\end{pf}
\begin{definition}
We define $\bv C = \bv C^{{\elsalg P}}$ to be that ${ P} \in {\elsalg P}$ that satisfies (\ref{dvp}) for $C$.  We call $\bv C$ the \emph{boolean value} of $C$ or the \emph{truth
value} of $C$, thinking of $C$ as a unary predicate applied to members of $\mathcal H$.
\end{definition}
\begin{inference}
\label{inf bvs}
If $C$ and $C'$ are in ${\mathcal C}$ then
\begin{eqnarray}
\label{eq inf bvs a}\compl C, C \cup C', C\cap C' &\in&  {\mathcal C},\\
\label{eq inf bvs b}\bv{\compl C}
&=&
\compl \bv C,\\
\label{eq inf bvs c}\bv{C \cup C'}
&=&
\bv C \vee \bv {C'},\\
\label{eq inf bvs d}\bv{C \cap C'}
&=&
\bv C \wedge \bv {C'},\\
\label{eq inf bvs e}\mbox{\normalfont and }C \subseteq C'
&\implies&
\bv C \le \bv{C'}.
\end{eqnarray}
\end{inference}
\begin{pf} (\ref{eq inf bvs a}) is just Premise~\ref{prem boolean ops}.  To prove (\ref{eq inf bvs b}) we observe that for all $F
\in \mathcal H$,
\[
\bv{\compl C} \in F \iff F \in \compl C \iff F \notin C
 \iff \bv C \notin F.
\]
It follows from Premise~\ref{dvv} that $\bv{\compl C} \vee \bv{C}
= 1$ and $\bv{\compl C} \wedge \bv{C} = 0$, i.e., that $\bv{\compl
C} = \compl \bv C$, as claimed.

To prove (\ref{eq inf bvs c}) we observe that for all $F \in \mathcal H$
\begin{eqnarray*}
\bv{C \cup C'} \in F
&\iff& F \in (C \cup C')\\
&\iff& (F \in C \mathor F \in C')\\
&\iff& (\bv C \in F \mathor \bv{C'} \in F).
\end{eqnarray*}
As before, Premise~\ref{dvv} yields the desired conclusion.

(\ref{eq inf bvs d}) follows from (\ref{eq inf bvs b}) and (\ref{eq inf bvs c}), and
(\ref{eq inf bvs e}) follows from the fact that for any ${ P}, { P}' \in \elsalg
P$, ${ P} \le { P}' \iff { P} \wedge { P}' = { P}$ by definition.\qed\end{pf}

(\ref{eq inf bvs c}) is only a special case of the
following inference, which deals with possibly infinite unions (and
by implication intersections) of members of $ {\mathcal C}$ and is proved similarly.
\begin{inference}
\label{inf dwq}
Suppose $\mathcal C' \subseteq  {\mathcal C}$ and $\bigcup
\mathcal C' \in  {\mathcal C}$.  Then $\textstyle \bv {\bigcup \mathcal C'} = \bigvee \{ \bv C \mid C \in \mathcal C' \}$.
\end{inference}
This is important for the hidden-variables program if it is to explain the statistics of observation in quantum mechanics.  In particular, the set
\begin{equation}
\label{eay}
C_q = \bigcap_{M > 0}
 \bigcup_{N \in \omega}
 \bigcap_{n > N}
 \bigcup_{\stackrel{\scriptstyle \sigma \in \preset n2}
  {\scriptstyle \big|q - \frac1n \sum\limits_{m=0}^{n-1} \sigma(m)\big| < \frac1M}}
 \bigcap_{m < n} \mathcal U_{(\compl)^{1-\sigma(m)} P_m}
\end{equation}
plays a critical role in this regard, where $\langle P_n \mid n \in \omega \rangle$ is a sequence of projections, just as the corresponding set $L_q \subseteq \preset\omega2$ and projection $P(L_q)$ do in \cite{RVWmw:2005}.

Given (\ref{dvp}), Inference~\ref{inf dwq} implies that if $\bigcup \mathcal C' \in {\mathcal C}$ and $S = \{ \bv C \mid C \in \mathcal C' \}$ then for all $F \in {\mathcal H}$, $\bigvee S \in F \iff (\exists  P \in S)  P \in F$. Note that the $\Longleftarrow$-direction follows from the fact that $F$, being a filter, is closed upward.  The
$\Longrightarrow$-direction imposes an additional requirement on $F$.  When in Section~\ref{sec suff gen} we reverse the current chain of inference it will be convenient to have a statement of this requirement that does not explicitly presume that $\bigvee S$ exists in ${\elsalg P}$.  Although we are currently interested only in boolean algebras, we will introduce the relevant notion (that of genericity) in the more general context of partial orders, as this is the most natural setting for the concept, and we will use it in this way later on.\footnote{As we will see, any partial order corresponds to a boolean algebra in a way that is quite natural in the context of genericity, and the theory of generic filters in POs is not essentially more general than the theory restricted to BAs.  (This is a different correspondence than that of a BA to the partial order defined by $P \le Q \iff P \wedge Q = P$.)  When we come to deal with propositional systems, however, the relevant PO will be easier to work with than the associated BA, which is one step more abstract than the partial order, which itself is a step more abstract than the initial propositional system.  It should be noted that the BAs that arise in this way are only tenuously related to the propositional algebras with which we have dealt so far, and dealing with the PO is a way of avoiding confusion that might otherwise result from the simultaneous use of two very different algebras.}  For technical reasons it will be convenient for us to work with prefilters---i.e., directed sets---rather than exclusively with filters.
\begin{definition}
\label{dyb}
Suppose $\mathbb P$ is a PO and $\mathcal S$ is a class.  A prefilter $D$ on $\mathbb P$ is \emph{$\mathcal
S$-generic} iff for all $S \in \mathcal S$, if $S \subseteq |\mathbb P|$ then there exists $p \in
D$ such that $p \in S$ or $p$ is incompatible with all elements of $S$.  We are often particularly concerned with maximal prefilters (which are necessarily filters), and we let $\mathcal G^{\mathcal S, \mathbb P} \eqdef$ the set of maximal $\mathcal S$-generic filters on $\mathbb P$.
\end{definition}
If $\elsalg P$ is a BA, the above definitions apply with $\elsalg P^-$ for $\mathbb P$.  For terminologic convenience later on we have not required that all members of $\mathcal S$ be subsets of $|\mathbb P|$, although we are only concerned with those that are.  Note that a prefilter $D$ is $\mathcal S$-generic iff the filter $\lfloor D \rfloor$ it generates is $\mathcal S$-generic.

It should be noted that `generic' is usually defined with reference
to \emph{dense} sets, as follows.
\begin{definition}
\label{dvk}
Suppose $\mathbb P$ is a PO.  A subset $S$ of $|\mathbb P|$
is \emph{dense} iff any condition in $|\mathbb P|$ has an
extension in $S$.  A prefilter $D$ is $\mathcal S$-generic according to this definition iff $D$
meets every dense set in $\mathcal S$.
\end{definition}
For technical reasons Def.~\ref{dyb} is preferable for this discussion.

Note that for any $S \subseteq |\mathbb P|$, the set
\begin{equation}
\label{dyc}
S' = \big\{ p \in |\mathbb P|\ \big|\ (\exists q \in S)\, p \le q \mathor
(\forall q \in S)\, p|q\big\}
\end{equation}
is dense, so for any class $\mathcal S$ and any \emph{filter} $D$, $D$ is $\mathcal S$-generic in the sense of
Def.~\ref{dyb} iff $D$ is $\mathcal S'$-generic in the usual
sense, where $\mathcal S' = \{ S' \mid S \in \mathcal S \}$.  (Remember that filters are by definition closed upward.)
\begin{inference}
\label{dwj}
Let $\mathcal S$ consist of all sets of the form
\[
\{ \bv C \mid C \in \mathcal C' \},
\]
where $\mathcal C' \subseteq {\mathcal C}$ and $\bigcup \mathcal C'
\in {\mathcal C}$.  Then every $F \in {\mathcal H}$ is $\mathcal S$-generic.
\end{inference}
\begin{pf} Suppose $\mathcal C' \subseteq {\mathcal C}$ and $\bigcup \mathcal C'
\in {\mathcal C}$, and let $S = \{ \bv C \mid C \in \mathcal C' \}$.  If $F \in \bigcup \mathcal C'$ then $F \in C$ for some $C \in \mathcal C'$, whence $F$ contains $\bv C \in S$.  If $F \in \compl \bigcup \mathcal C'$ then $\bv{\compl \bigcup \mathcal C'} \in F$.  But $\bv{\compl \bigcup \mathcal C'} = \compl \bv{\bigcup \mathcal C'} = \compl \bigvee S$, which is incompatible with all members of $S$.\qed\end{pf}
  
Note that if $\mathcal C' = \{ \mathcal U_{ P} \}$ for some $ P \in
{\elsalg P}$, then $\bigcup \mathcal C' = \mathcal U_{ P} \in {\mathcal C}$.  Since $\bv{\mathcal U_{ P}} =  P$, every $F \in {\mathcal H}$ is $\{  P \}$-generic, so either $ P \in F$ or $\compl { P} \in
F$.  Thus, even if we had not stipulated at the outset that $\mathcal H$ consists of ultrafilters, this would follow from Premises~1--4.

\subsection{The sufficiency of genericity for propositional algebras}
\label{sec suff gen}

\subsubsection{The existence of generic filters}

Inference~\ref{dwj} tells us that the filters that arise
naturally in any hidden-variables model are generic to the extent
that the class $\mathcal C$ of ``posable questions'' is closed
under unions (equivalently, intersections).  In this section we will show conversely that generic filters have all the properties required of pseudoclassical states, including the familiar statistical properties of quantum observation.  Hence generic filters on PAs constitute the unique realization of the hidden-variables program for propositional algebras.

The following observations give some information about the
existence of $\mathcal S$-generic filters.
\begin{prop}
\label{prop count filter}
Suppose $\mathbb P$ is a PO and $\mathcal S$ is a
countable class.  Then for any condition $p$
there is an $\mathcal S$-generic filter $F$ on $\mathbb P$ that contains
$p$.  By Prop.~\ref{prop max filter}, $F$ may be taken to be maximal.
\end{prop}
\begin{pf} Let $\mathcal S = \{ S_n \mid n \in \omega \}$.  Construct a descending sequence $\langle p_n
\mid n \in \omega \rangle$ as follows.  Let $p_0 = p$ and for each $n \in \omega$, if $p_n$ is compatible with some $q \in S_n$ let $p_{n+1}$ be some common extension of $p_n$ and some such $q$, otherwise let $p_{n+1} = p_n$.  Let $F = \big\{ p \in |\mathbb P|\ \big|\ (\exists n \in \omega) p_n \le p \big\}$.  Clearly $F$ is as desired.\qed\end{pf}

The existence of $\mathcal S$-generic filters for uncountable
$\mathcal S$ is a matter of great interest in set theory, and we
will see why---or rather how---this is in due course.  For now we simply point out that if every member of $\mathbb P$ has incompatible extensions and $F$ is a filter on $\mathbb P$ then $F$ is not $\{|\mathbb P| \setminus F\}$-generic, so there are no $\mathcal S$-generic filters if $\mathcal S$ is the set of all subsets of $|\mathbb P|$.

In general, we may anticipate some difficulty in attempting to
construct $\mathcal S$-generic filters for uncountable classes
$\mathcal S$, although Proposition~\ref{prop max filter} shows that this
can be done for some interesting uncountable classes \big(viz., $\big
\{ \{ p \}\ \big|\ p \in |\mathbb P| \big\}$ for uncountable POs
$\mathbb P$\big).  Let us defer this issue for the moment while we show that generic filters do indeed have the properties of pseudoclassical states as promised above.

\subsubsection{Boolean expressions}
\label{sec bool exp}

It is useful to separate the notion of boolean value from the issue of existence of generic filters, which we do by working with \emph{boolean expressions}.  Boolean expressions (BEs) are constructed from primitive expressions using complementation, join, and meet.  (Technically, one defines \emph{boolean expression of rank $\alpha$} for $\alpha \in \Omega$, the class of ordinals, by $\in$-recursion, as described in Sec.~\ref{sec axiom set theory}.  Def.~\ref{def rank BE} is subsumed in this definition.)  It is convenient to suppose that the primitive expressions in any instance are indexed by a set in some standard way.  If $I$ is the index set, we use `$\epsilon_i$' to indicate the $i^{\mbox{th}}$ primitive expression for $i \in I$.  We use `$\bcompl$', `$\epsilon_\iota = \{ f \in \preset I2 \mid f(i) = 1 \}$', and `$\btbigwedge$' to indicate the operations that form the complement, join, and meet, respectively.  These are unary operations, with $\bcompl$ applying to expressions, and $\btbigvee$ and $\btbigwedge$ applying to sets of expressions.  The only requirement we impose on $\epsilon_\cdot$, $\bcompl \cdot$, $\btbigvee \cdot$ and $\btbigwedge \cdot$ is that they be one-one with disjoint images, so that every expression $E$ is uniquely of one of the forms $\epsilon_i$, $\bcompl E'$, $\btbigvee \mathcal E$, or $\btbigwedge \mathcal E$ for some $i \in I$, some expression $E'$, or some set $\mathcal E$ of expressions.  A BE whose primitive expressions are indexed by a set $I$ is said to be $I$-ary.  For example, $\bcompl \epsilon_0$, $\btbigvee\{ \bcompl \epsilon_0, \epsilon_1 \}$, and $\btbigwedge_{m \in \omega} \btbigvee_{n \in \omega} \epsilon_{2^m 3^n}$ are examples of $\omega$-ary BEs.
\begin{definition}
\label{def rank BE}
The \emph{rank} $\rho(E)$ and the set $\subexp E$ of \emph{subexpressions} of a boolean expression $E$ are defined recursively.
\begin{enumerate}
\item If $E$ is primitive then $\rho(E) = 0$ and $\subexp E = \{ E \}$.
\item If $E = \bcompl E'$ then $\rho(E) = \rho(E') + 1$ and $\subexp E = \{ E \} \cup \subexp{E'}$.
\item If $E = \btbigvee \mathcal E$ or $\btbigwedge \mathcal E$ then $\rho(E) = \sup_{E' \in \mathcal E} (\rho(E')+1)$, and $\subexp E = \{ E \} \cup \bigcup_{E' \in \mathcal E} \subexp {E'}$.
\end{enumerate}
\end{definition}
What we have said so far about $\bcompl$, $\btbigvee$, and $\btbigwedge$ has simply defined an abstract structure.  The following definition confers meaning on these operations.
\begin{definition}
Suppose $I$ is a class, $\elsalg A$ is a complete boolean algebra, and $\iota : I \to |\elsalg A|$.  The following conditions uniquely determine an $|\elsalg A|$-valued function $\bv{\cdot}^\iota$ on the class $\mathcal{B}^I$ of $I$-ary boolean expressions.
\begin{eqnarray*}
\bv{\epsilon_i}^\iota &=& \iota (i)\\
\bv{\bcompl E}^\iota &=& \compl \bv E^\iota\\
\textstyle\bigbv{\btbigvee \mathcal E}^\iota &=& \textstyle\bigvee \{ \bv E^\iota \mid E \in \mathcal E \}\\
\textstyle\bigbv{\btbigwedge \mathcal E}^\iota &=& \textstyle\bigwedge \{ \bv E^\iota \mid E \in \mathcal E \}.
\end{eqnarray*}
An \emph{$\elsalg A$-valuation} or \emph{$\elsalg A$-interpretation of $\mathcal{B}^I$} is any function $\bv\cdot^\iota$ obtained in this way.  We will also refer to $\iota$ itself as an interpretation.
\end{definition}
If $\elsalg A$ is a propositional algebra we refer to $\elsalg A$-interpretations as \emph{propositional interpretations}.  Note that since boolean operations have a natural interpretation for (commuting) propositions, any proposition-valued function $\iota$ with domain $I$, all of whose values commute with one another, is extendible to a propositional interpretation of all $I$-ary BEs; it is not necessary to specify the algebra $\elsalg A$, which may be taken to be any PA that includes the image of $\iota$ and is sufficiently complete that the meets and joins occurring in the BEs under consideration may be formed within it.

In addition to the propositional interpretations, several algebras $\elsalg A$ are of particular interest.  If $\elsalg A = \boldsymbol 2$, the 2-element algebra, then we may identify 1 with \emph{true} and 0 with \emph{false}.  A $\boldsymbol 2$-interpretation is an interpretation in the usual sense of assigning a truth value to expressions, where $\bcompl$, $\btbigvee$, and $\btbigwedge$ are logical negation, disjunction, and conjunction, respectively.  These expressions may be considered to belong to a language that permits infinite disjunction and conjunction.  Of course, letting 0 and 1 be respectively the zero and unit propositions, $\boldsymbol 2$-interpretations may be viewed as propositional interpretations.

We may also take $\elsalg A$ to be a boolean set-algebra, for which $\bcompl$, $\btbigvee$, and $\btbigwedge$ correspond to set-theoretic complementation (relative to some fixed set), union, and intersection.  One important such case is that of the algebra of all sets of subsets of the index set $I$, with the interpretation $\bv{\epsilon_i} = \{ X \subseteq I \mid i \in X \}$.  Equivalently via the correspondence of a subset $X$ of $I$ with its characteristic function $\chi^X \in \preset I2$ (the set of functions from $I$ to the set $2 = \{0,1\}$), given by
\[
\chi^X(i) =
 \begin{elscases}
  1 & \mbox{if $i \in X$}\\
  0 & \mbox{if $i \notin X$},
 \end{elscases}
\]
we take $\elsalg A$ to be the algebra of all subsets of $\preset I2$, with the interpretation $\epsilon_\iota = \{ f \in \preset I2 \mid f(i) = 1 \}$.

Several variations may be played on this theme by forming subalgebras and quotient algebras.  One such is defined in terms of a partial order $\mathbb P$.  Let $\mathcal D^{\mathbb P}$ be the set of prefilters on $\mathbb P$.  For $X \subseteq |\mathbb P|$, $\mathcal D_X \eqdef \{ D \in \mathcal D^{\mathbb P} \mid X \subseteq D \}$; and for $p \in |\mathbb P|$, $\mathcal D_p \eqdef \mathcal D_{\{ p \}}$.  If we let $\elsalg A = \setalg(\mathcal D^{\mathbb P})$, the set-algebra consisting of all subsets of $\mathcal D^{\mathbb P}$, we may interpret $|\mathbb P|$-ary expressions according to the prescription $\bv{\epsilon_p} = \mathcal D_p$.

If $I$ is countable and $\ideal n$ is a countably complete ideal in the Borel set-algebra $\elsalg B$ for $\preset I2$, the quotient algebra $\elsalg B / \ideal n$, obtained by identifying elements $P$ and $Q$ of $\elsalg B$ whose difference $P-Q$ is in $\ideal n$, is a complete BA, which is a natural valuation algebra for $I$-ary BEs.  Of particular interest for the present work is the case that $\ideal n = \{ X \in \elsalg B \mid \mu(X) = 0 \}$ for a Borel measure $\mu$.

The property of genericity may be used to relate $\elsalg A$-interpretations and $\setalg(\mathcal D^{\elsalg A})$-interpretations.  We defer a general discussion of this connection for the moment, and focus on the case of propositions.
\begin{definition}
Recall that we have defined a directed set, or prefilter, in a partial order to be a set $D$ such that any two elements of $D$ have a common extension in $D$.  We say that a set $D$ of propositions is \emph{directed} iff all members of $D$ commute with one another and $D$ is directed in the usual partial ordering of propositions (i.e., with $\le = \subseteq$).  A \emph{propositional prefilter} is a directed set of propositions.  Let $\mathcal D$ be the set of all propositional prefilters, and let $\elsalg D$ be the boolean set-algebra of subsets of $\mathcal D$.
\end{definition}
Suppose $\mathcal P$ is a set of commuting projections.  We define two interpretations for $\mathcal P$-expressions, a proposition-valued interpretation $\bv\cdot^\pi$ and a $\elsalg D$-valued interpretation $\bv\cdot^\delta$, which are uniquely determined by the conditions
\begin{eqnarray*}
\label{eqn bv pi}
\bv{\epsilon_P}^\pi &= &P\\
\label{eqn bv delta}
\bv{\epsilon_P}^\delta &= &\{ D \in \mathcal D \mid P \in D \}
\end{eqnarray*}
for all $P \in \mathcal P$.

Suppose $E$ is a $\mathcal P$-ary BE.  Let $\mathcal S$ consists of the sets
\begin{enumerate}
\item $\{ \bv{E'}^\pi \}$, for all $E' \in \hat E$;
\item $\{ \bv{E'}^\pi \mid E' \in \mathcal E \}$, for all $\mathcal E$ such that $\btbigvee \mathcal E \in \hat E$; and
\item $\{ \compl \bv{E'}^\pi \mid E' \in \mathcal E \}$, for all $\mathcal E$ such that $\btbigwedge \mathcal E \in \hat E$.
\end{enumerate}
The following theorem is converse to Inference~\ref{dwj} and is the key to implementing the hidden-variables program.
\begin{thm}
\label{thm P E S}
Suppose $\mathcal P$, $E$, and $S$ are as above, and suppose $D$ is an $\mathcal S$-generic propositional prefilter all of whose members commute with all the members of $\mathcal P$.  Then
\begin{equation}
\label{eq pi eq delta}
D \in \bv E^\delta \iff \bv E^\pi \in D.
\end{equation}
\end{thm}
\begin{pf} By induction on the complexity (i.e., rank) of $E$.  (\ref{eq pi eq delta}) is trivially true if $E = \epsilon_P$ for some $P \in \mathcal P$.

Suppose $E = \bcompl E'$ and $E'$ satisfies (\ref{eq pi eq delta}). Let $P = \bv{E'}^\pi$; then $\bv E^\pi = \compl P$.  Since by definition $E$ and $E'$ are in $\hat E$, $\{ \compl P \}$ and $\{ P \}$ are in $\mathcal S$.  As $D$ is $\mathcal S$-generic, $D$ contains $\compl P$ or an element $Q$ such that $Q | \compl P$, and $D$ contains $P$ or an element $Q$ such that $Q | P$.  It is not possible that $D$ contain elements $Q$ and $Q'$ such that $Q | \compl P$ and $Q' | P$, for in that case $Q | Q'$; hence, either $\compl P$ or $P$ is in $D$, and
\[
D \in \bv E^\delta
\iff
D \notin \bv{E'}^\delta
 \iff
P \notin D
\iff
\compl P \in D,
\]
so $E$ satisfies (\ref{eq pi eq delta}).

Now suppose $E = \btbigvee \mathcal E$ and each $E' \in \mathcal E$ satisfies (\ref{eq pi eq delta}).  By hypothesis, either $D$ contains $\bv{E'}^\pi$ for some $E' \in \mathcal E$ or for some $Q \in D$, $Q$ is incompatible with $\bv{E'}^\pi$ for all $E' \in \mathcal E$, in which case $Q$ is incompatible with $\bigvee_{E' \in \mathcal E} \bv{E'}^\pi = \bv E^\pi$, so $D$ cannot contain $\bv E^\pi$; in other words,
\[
\bv E^\pi \in D \implies (\exists E' \in \mathcal E)\, \bv{E'}^\pi \in D.
\]
Inversely, as we have just seen, if $\bv E^\pi$ is not in $D$ then $\compl \bv E^\pi \in D$, i.e., $\bigwedge_{E' \in \mathcal E} \compl \bv{E'}^\pi \in D$, whence it follows that $\bv{E'}^\pi \notin D$ for all $E' \in \mathcal E$.  Thus, 
\begin{eqnarray*}
\textstyle
D \in \bigbv E^\delta
&\iff&
(\exists E' \in \mathcal E)\, D \in \bv{E'}^\delta
\iff
(\exists E' \in \mathcal E)\, \bv{E'}^\pi \in D\\
&\iff&
\bv E^\pi \in D,
\end{eqnarray*}
so $E$ satisfies (\ref{eq pi eq delta}).

If $E = \btbigwedge \mathcal E$ the desired identity follows from the identity $\btbigwedge \mathcal E = \bcompl \bigvee_{E' \in \mathcal E} \bcompl E'$.\qed\end{pf}

\begin{definition}
A boolean expression $E$ is \emph{Borel} iff for every subexpression of $E$ of the form $\btbigvee \mathcal E$ or $\btbigwedge \mathcal E$, $\mathcal E$ is countable.
\end{definition}
\begin{prop}
\label{thm Borel exp}
An expression $E$ is Borel iff $\hat E$ is countable.
\end{prop}
\begin{pf} The if-direction is immediate.  We prove the only if-direction by induction on the rank of $E$.  Let $I$ be the index set (which need not be countable).  For each $i \in I$, $\widehat{\epsilon_i} = \{ \epsilon_i \}$ has cardinality 1, so it is countable.

Suppose $E = \bcompl E'$ and $\hat E'$ is countable.  Then $\hat E = \{ E \} \cup \hat E'$ is countable.

Suppose $E$ is of the form $\btbigvee \mathcal E$ or $\btbigwedge \mathcal E$, where for each $E' \in \mathcal E$, $\hat E'$ is countable.  Then $\hat E = \{ E \} \cup \bigcup_{E' \in \mathcal E} \hat E'$ is countable.\qed\end{pf}

\subsubsection{Statistics of generic states}

As an application of Thm.~\ref{thm P E S} and Prop.~\ref{thm Borel exp} suppose $\elsalg P$ is an algebra of propositions that correspond to physical observables for a system $\system S$.  (Note that this cannot consist of all binary observables, since the propositions in $\elsalg P$ commute by definition.)  If we suppose that by `physical observable' we mean a property of $\system S$ for which there is a describable measurement, then $\elsalg P$ is countable, because there are only countably many descriptions in any effective system of descriptions.  Let $\mathcal E$ be the set of Borel $\elsalg P$-ary expressions that are definable in the same sense.  Then $\mathcal E$ is likewise countable.  Since any definable boolean expression applied to a describable measurement yields a description of a measurement, we may suppose that $\elsalg P$ is closed under the action of the expressions in $\mathcal E$---i.e., letting $\bv{\epsilon_P}^\pi = P$ for all $P \in \elsalg P$ as in (\ref{eqn bv pi}), $\bv E^\pi \in \elsalg P$ for all $E \in \mathcal E$.  Let $\mathcal S$ be the union of the sets $\mathcal S$ defined for Theorem~\ref{thm P E S} for each $E \in \mathcal E$.  The members of $\mathcal S$ are subsets of $\elsalg P$.

Let $\mathcal H_0$ be the set of $\mathcal S$-generic ultrafilters on $\elsalg P$.  Letting $\bv {\epsilon_P}^\delta = \{ F \in \mathcal U^{\elsalg P} \mid P \in F \}$ as in (\ref{eqn bv delta}) (with the technical modification that we deal exclusively with ultrafilters), the set $\mathcal C$ of posable questions is the set of sets of the form $\bv E^\delta$, for $E \in \mathcal E$.  Premises~1--3 are easily shown to be satisfied.  Premise~4 follows from Prop.~\ref{prop count filter}, so $\mathcal H_0$ qualifies as $\mathcal H^{\elsalg P}$.  By Inference~\ref{dwj}, $\mathcal H_0$ is the largest set that does.

We now examine the statistics of models $\mathcal H \subseteq \mathcal H_0$ with $\mathcal H$ satisfying Prem\-ises~1--4.  As in \cite{RVWmw:2005} suppose $\psi$ is $q$-homogeneous for a sequence $\langle P_n \mid
n \in \omega \rangle$ of commuting propositions in $\elsalg P$.  Let $\elsalg P$ be the PA generated by $\{ P_n \mid n \in \omega \}$, and let $\hat{\elsalg P}$ be the PA generated by $\mathcal E$ acting on $\{ P_n \mid n \in \omega \}$.  Note that $\elsalg P \subseteq \hat{\elsalg P}$, and these algebras have the same completion as PAs.  Let $\tilde{\vecsp V} = [\psi]^{\elsalg P}$, the smallest closed subspace of $\vecsp V$ that contains $\psi$ and is closed under all projections in $\elsalg P$.  Note that $\tilde{\vecsp V} = [\psi]^{\complet{\elsalg P}} = [\psi]^{\hat{\elsalg P}}$.  We may regard $\hat{\elsalg P}$ as a PA in $\tilde{\vecsp V}$.

Let $\mu_q$ be the $q$-homogeneous measure on the algebra $\elsalg B$ of Borel subsets of $\preset\omega2$, and let $P(\cdot)$ be the canonical homomorphism of $\elsalg B$ to $\overline{\elsalg P}$ as in \cite{RVWmw:2005}.  Then $\mu_q(X) = 1 \iff P(X) = 1$.  Let $\hat{\elsalg B}$ be the subalgebra $\elsalg B$ generated from the basic open sets by the BEs in $\mathcal E$.  Then $P(\cdot)$ maps $\hat{\elsalg B}$ into $\hat{\elsalg P}$.  Given an ultrafilter $F$ on $\elsalg P$, define $f^F \in \preset\omega2$ by: $f^F(n) = 1 \iff P_n \in F$.  Let $C_q$ be as in (\ref{eay}) and let $L_q \subseteq \preset\omega2$ be the corresponding Borel set as in \cite{RVWmw:2005}.\footnote{Note that $M$, $N$, and $n$ in (\ref{eay}) take on integer values, so the unions and intersections are countable.}  Let $E_q$ be the corresponding boolean expression.  In the terminology of Theorem~\ref{thm P E S}, $C_q = \bv{E_q}^\delta \cap \mathcal U^{\elsalg P}$ and $P(L_q) = \bv{E_q}^\pi$.  $L_q$ itself is the value of $E_q$ in the natural interpretation of $E_q$ in the set-algebra $\setalg(\preset\omega2)$.  We suppose $E_q \in \mathcal E$, i.e., we suppose that $E_q$ represents a posable question; it is, after all, the principal question we propose to pose.  By virtue of Theorem~\ref{thm P E S}, since $\mathcal H \subseteq \mathcal H_0$ by necessity, the boolean value $\bv{C_q}$ (in the terminology of Section~\ref{sec genericity}) is $\bv{E_q}^\pi$, i.e., $P(L_q)$.  As before, by the law of large numbers, $\mu_q(L_q) = 1$, so $\bv{C_q} = \bv{E_q}^\pi = P(L_q) = 1$.  Thus every pseudoclassical filter is in $C_q$, \ie, every pseudoclassical state exhibits the correct limiting behavior of the frequency statistic.
\begin{thm}
\label{thm first rand var}
The same argument applies to all definable tests of randomness as discussed in \cite{RVWmw:2005}, so observations performed on a pseudoclassical state will give every appearance of a stochastic process with the usual interpretation of the squared norm as probability even though their outcomes are predetermined.
\end{thm}

\section{The general method of forcing}
\label{sec gen meth force}

\subsection{The regular algebra}
\label{sec reg alg}

The power of the genericity property---as expressed for propositional algebras in Theorem~\ref{thm P E S}---lies in the fact that it renders a great variety of assertions regarding a prefilter $D$, as represented by the left side of (\ref{eq pi eq delta}), equivalent to assertions of the very limited type represented by the right side of (\ref{eq pi eq delta}).  If we let $\epsilon_P$ be the sentence `$P \in D$' for each proposition $P$, and we interpret $\bcompl$, $\btbigvee$, and $\btbigwedge$ as logical negation, disjunction, and conjunction, as discussed previously, we obtain a linguistic expression for the left side of (\ref{eq pi eq delta}).  In general, of course this expression involves infinite boolean operations and is therefore somewhat nonstandard, and we will eventually drop these in favor of conventional linguistic constructs.  For the moment, however, let us view a BE $E$ as a predicate with one free variable, which a prefilter will make true or false according to the left side--equivalently the right side--of (\ref{eq pi eq delta}).  As is customary in discussions of this topic, without any significant loss of generality and with some gain in convenience, we restrict our attention to generic ultrafilters, for which we use the variable `$G$'.

The notion of genericity as we have developed it for propositional algebras applies essentially verbatim to an arbitrary complete BA, but to explore this concept fully we must broaden our outlook to include not just boolean algebras, but partial orders (POs) in general.  We must do this also as a practical matter, so that we may extend our results about algebras of commuting propositions to certain systems of propositions that do not necessarily commute, and achieve a more comprehensive realization of the hidden-variables program.

As the following series of definitions and the subsequent discussion make clear, boolean algebras arise naturally in the theory of generic filters on POs.
\begin{definition}
\label{dxq}
Suppose $\mathbb P$ is a PO and $X \subseteq |\mathbb P|$.  Let $X^\perp \eqdef \big\{ p \in |\mathbb P|\ \big|\ (\forall q \in X) p|q \big\}$, and $\overline X \eqdef X^{\perp\perp}$.  Note that $X \subseteq \overline X$.  $X$ is \emph{regular} iff $X = \overline X$.
\end{definition}
\begin{definition}
\label{def separative}
Suppose $\mathbb P$ is a PO and $p \in |\mathbb P|$.  Then
\begin{eqnarray*}
\lceil p \rceil &\eqdef \big\{ q \in |\mathbb P|\ \big|\ q \le p \big\}\\
\lfloor p \rfloor &\eqdef \big\{ q \in |\mathbb P|\ \big|\ q \ge p \big\}.
\end{eqnarray*}
$\mathbb P$ is \emph{separative} iff for all $p \in |\mathbb P|$, $\lceil p \rceil$ is regular.
\end{definition}
The separativity property gets its name from the equivalent formulation: $p \not\le q \implies (\exists r \le p)\, r|q$, i.e., $r$ ``separates'' $p$ from $q$.

Let $\regalg{\mathbb P}$ be the set of regular subsets of $|\mathbb P|$.
$\regalg{\mathbb P}$ has a natural structure as a BA for which $\le = \subseteq$.  In this algebra
\begin{eqnarray*}
\textstyle\compl X
&=&
X^\perp\\
\textstyle\bigwedge \mathcal X
&=&
\textstyle\bigcap \mathcal X\\
\textstyle\bigvee \mathcal X
&=&
\compl \left( \textstyle\bigwedge_{X \in \mathcal X} \compl X \right)\\
&=& \left( \textstyle\bigcap_{X \in \mathcal X} X^\perp\right)^\perp
 = \left( \textstyle\bigcup \mathcal X \right)^{\perp\perp}\\
&=& \overline { \textstyle\bigcup \mathcal X }.
\end{eqnarray*}
Note that meets and joins are defined for all subsets $\mathcal X$ of $\regalg{\mathbb P}$, so
\begin{prop}
$\regalg{\mathbb P}$ is a complete boolean algebra.
\end{prop}
The map $p \mapsto \overline{\lceil p \rceil}$ is a homomorphism of $\mathbb P$ onto a dense subset of $\regalg{\mathbb P}^-$.  If $\mathbb P$ is separative this is the map $p \mapsto \lceil p \rceil$ and it is an isomorphism.  $\regalg{\mathbb P}$ is---up to isomorphism---the unique extension of $\mathbb P$ to a complete BA in which $|\mathbb P|$ is dense.  That is, if $\iota : \mathbb P \to \elsalg A$ is an isomorphism of $\mathbb P$ with a dense subset of $\elsalg A^-$ (treated as a partial order), then $\iota$ may be extended (uniquely, in fact) to an isomorphism $\iota' : \regalg{\mathbb P} \to \elsalg A$.  We call $\regalg{\mathbb P}$ the \emph{regular algebra for $\mathbb P$}.

For any BA $\elsalg A$ we define $\regalg{\elsalg A}$ to be $\regalg{(\elsalg A^-)}$, and---identifying $P \in \elsalg A$ with $\lceil P \rceil \in \regalg{\elsalg A}$---we call this the \emph{regular completion of $\elsalg A$}.  Note that $\elsalg A^-$ is a separative PO for any BA $\elsalg A$, so $\elsalg A^-$ is dense in $\regalg{\elsalg A}$.\footnote{We note in that if $\elsalg A$ is a propositional algebra then there is a natural map $\iota: \regalg{\elsalg A} \to \complet{\elsalg A}$, where $\complet{\elsalg A}$ is the completion of $\elsalg A$ as a PA, \viz, $\iota (X) = \bigvee X$ for all regular $X \subseteq \elsalg A^-$.

\hspace{1em}$\im \iota$ need not be dense in $\complet{\elsalg A}$.  For example, suppose $\elsalg A$ is the PA generated by propositions $\langle P_n \mid n \in \omega \rangle$ for which there exists a $q$-homogeneous vector with $0 < q < 1$.  (This is just a convenient choice---many other PAs would serve.)  Following the conventions of \cite{RVWmw:2005}, we let $I_\sigma = \{ f \in \preset\omega2 \mid \sigma \subseteq f \}$, which we call a \emph{basic interval}, and we let $P(\cdot)$ be the standard map of Borel sets to propositions defined from $\langle P_n \mid n \in \omega \rangle$.  The propositions $P(I_\sigma)$ form a dense subset of $\elsalg A$.  Let $\langle K_n \mid n \in \omega \rangle$ be an enumeration of the basic intervals, and for each $n \in \omega$, let $K'_n \subseteq K_n$ be a basic interval such that $\mu_q(K'_n) < 2^{-n}$.  Let $X = \{ P(K'_n) \mid n \in \omega \}$.  By construction, $X$ is dense in $\elsalg A$, so $X^\perp = 0$, \ie, $\emptyset$, and $\complet X = X^{\perp\perp} = 1$, \ie, $|\elsalg A|^-$.  $\complet X$ is the smallest element of $\regalg \elsalg A$ that includes $X$.

\hspace{1em}By construction, however, $\mu_q \left( \bigvee^P_{n \in \omega} K'_n \right) < 1$, so $\bigvee X < 1$.  Let $Q = 1 - \bigvee X$.  We claim that there is no nonzero element of $\iota\image \regalg \elsalg A$ below $Q$.  Suppose to the contrary that $X'$ is a nonempty regular subset of $\elsalg A^-$ such that $\bigvee X' = \iota(X') \le Q$.  Then for all $P \in X$ and all $P' \in X'$, $P \perp P'$, which is not possible, as $X$ is dense in $\elsalg A$.

This corresponds to the fact that the open sets in $\preset\omega2$ are not dense in the measure algebra for $\mu_q$, which may be defined as the algebra of Borel subsets of $\preset\omega2$ divided by the ideal of sets of measure 0.  The closed sets, it may be noted, do form a dense set in this algebra.}

\subsection{The forcing language}
\label{sec forcing language}

Suppose $\mathbb P$ is a partial order.  By definition, $\mathbb P$ is of the form $(|\mathbb P|, \le )$, where $|\mathbb P|$ is a class and $\le$ is a binary relation on $|\mathbb P|$ that satisfies Def.~\ref{def PO}.  A language appropriate to this structure would have a predicate symbol for $\le$ along with all the other paraphernalia of a formal language: variables; symbols for negation, conjunction, and other logical connectives; quantifier symbols `$\forall$' and `$\exists$'; a symbol `$=$' for identity; and grouping symbols, say `$($' and `$)$'.  We define $\lang L^{\mathbb P}$ to be such a language with the following modifications.  First, we add a unary predicate symbol $\Gcheck$.  By this we mean that an interpretation of $\lang L^{\mathbb P}$ assigns to $\Gcheck$ a subclass of $|\mathbb P|$.  Second, for each $p \in |\mathbb P|$ we add a constant symbol $\check p$.  We will restrict the interpretation of $\lang L^{\mathbb P}$ so that $\check p$ always denotes $p$.  Finally, we allow arbitrary disjunction and conjunction, i.e., for any set $\mathcal E$ of $\lang L^{\mathbb P}$-expressions we have expressions $\btbigvee \mathcal E$ and $\btbigwedge \mathcal E$, which have the usual interpretation \visavis the interpretations of the expressions in $\mathcal E$.\footnote{Note that we have used the same symbols for the disjunction- and conjunction-forming operations of $\lang L$ as we have been using for the corresponding abstract BE-forming operations, which are the boldface form of the corresponding symbols for disjunction and conjunction in the metalanguage, which is the language in which this paper is written.  In this way we maintain in our typography the distinction between use (lightface) and mention (boldface) of linguistic entities.  We will do this for several other expression-building operations but will otherwise tolerate some ambiguity on this score.  We will also use quotation marks, as in the paragraph containing this footnote, to effect this distinction.}  As long as we restrict interpretations of $\lang L$ to structures $(|\mathbb P|, \le, \Gcheck, \dots, \check p, \dots)$ with $\check p$ denoting $p$ for all $p \in |\mathbb P|$, we may replace any existential quantifier construction $\bexists p\, \phi(p)$ by the disjunction $\btbigvee_{p \in |\mathbb P|} \phi(\check p)$ and any universal quantifier construction $\bforall p\, \phi(p)$ by the conjunction $\btbigwedge_{p \in |\mathbb P|} \phi(\check p)$.  In this way we may replace any expression of $\lang L$ by a quantifier-free expression, i.e., a boolean combination of expressions of the form $\check p \le \check q$ and $\Gcheck(\check p)$.  We may eliminate any expression of the former type in favor of its truth value because we know that $\check p \ble \check q$ is true or false according as $p \le q$ or $p \not\le q$.  This leaves a boolean combination of expressions $\Gcheck(\check p)$.  Letting $\epsilon_p$ be the sentence $\Gcheck(\check p)$ for each $p \in |\mathbb P|$, every sentence of $\lang L$ is equivalent to a $|\mathbb P|$-ary boolean expression.

We are interested in the structure of boolean interpretations of $\lang L^{\mathbb P}$.  One interpretation suggests itself immediately.
\begin{definition}
The \emph{canonical interpretation} $\iota^{\mathbb P}$ of $\lang L^{\mathbb P}$ is the $\regalg{\mathbb P}$-interpretation defined by $\bv {\Gcheck(\check p)}^{\iota^{\mathbb P}} = \bv{\epsilon_p}^{\iota^{\mathbb P}} = \lceil p \rceil$, where $\regalg{\mathbb P}$ is the complete BA of regular subsets of $|\mathbb P|$ and $\lceil p \rceil = \{ q \in |\mathbb P| \mid q \le p \}$.
\end{definition}
\begin{definition}
In general, if $\iota$ is an interpretation of $\lang L^{\mathbb P}$, an \emph{$\iota$-validity} is a sentence $\sigma$ of $\lang L^{\mathbb P}$ such that $\bv\sigma^\iota = 1$.  \emph{Validity} unqualified means \emph{$\iota^{\mathbb P}$-validity}.
\end{definition}
Note that we have used the same symbols for the respective logical operations of disjunction and conjunction in $\lang L^{\mathbb P}$ as for the corresponding algebraic operations of join and meet.  We extend this convention to negation/complementation, denoted by `$\compl$', implication, denoted by `$\lcimplies$', and logical equivalence or bi-implication, denoted by `$\lciff$'.  That is, we let
\begin{eqnarray}
P \lcimplies Q &\eqdef& (\compl P) \vee Q\label{eq ba implies}\\
P \lciff Q &\eqdef& (P \lcimplies Q) \wedge (Q \lcimplies P) = (P \wedge Q) \vee (\compl P \wedge \compl Q),\label{eq ba iff}
\end{eqnarray}
for elements $P, Q$ of any BA.  We indicate the corresponding expression-building operations in $\lang L$ by the boldface versions of these.

We leave it as an instructive exercise for the reader to show that each of the following sentences is an $\iota^{\mathbb P}$-validity:
\begin{eqnarray}
\textstyle \btbigwedge_{p}\btbigwedge_{q \ge p}\big( \Gcheck(\check p) \bimplies \Gcheck(\check q) \big),\label{eq bv filter a}\\
\textstyle \btbigwedge_{p,q}\left( \big(\Gcheck(\check p) \bwedge \Gcheck(\check q)\big)
 \bimplies \btbigvee_{r \ble p,q}\Gcheck(\check r) \right),\label{eq bv filter b}\\
\textstyle \btbigwedge_{X \subseteq |\mathbb P|} \btbigvee_{q \in X \cup X^\perp} \Gcheck(\check q),\label{eq bv filter c}
\end{eqnarray}
where $p$, $q$, and $r$ range over $|\mathbb P|$.  Together these sentences say that $\Gcheck$ is a generic filter,\footnote{We are treating `$\Gcheck$' as a unary predicate symbol, but a unary predicate is identifiable with the class of elements that satisfy it, so we may also regard `$\Gcheck$' as denoting a class, in this case a subclass of $|\mathbb P|$.  In other words, `$\Gcheck(x)$' corresponds to `$x \in \Gcheck$'.  It is in the latter sense that we say `$\Gcheck$ is a generic filter'.} so
\begin{equation}
\label{eq gen fil}
\bv{\mbox{$\Gcheck$ is a generic filter}}^{\iota^{\mathbb P}} = 1,
\end{equation}
where by `generic' we mean generic with respect to \emph{all} subsets of $|\mathbb P|$.  It must be emphasized that the fact that in general a filter on $\mathbb P$ cannot be generic with respect to all subsets of $|\mathbb P|$ (specifically, in an atomless algebra no filter is generic with respect to its complement) is not at odds with (\ref{eq gen fil}), which makes no claim as to the existence of generic filters.  The logical position of this sort of statement will become clear in Section~\ref{sec mod set theory}.
\begin{thm}
\label{thm universal}
$\iota^{\mathbb P}$ is \emph{universal} among boolean interpretations of $\mathcal B^{|\mathbb P|}$ for which (\ref{eq bv filter a}--\ref{eq bv filter c}) are validities, i.e., for any BA $\elsalg A$ and any $\iota : |\mathbb P| \to |\elsalg A|$, if
\[
\bv{\mbox{$\Gcheck$ is a generic filter}}^{\iota} = 1
\]
then there exists a complete homomorphism $h : \regalg {\mathbb P} \to \elsalg A$ such that for any sentence $\sigma$ of $\lang L^{\mathbb P}$, $\bv\sigma^\iota = h \left(\bv\sigma^{\iota^{\mathbb P}}\right)$.
\end{thm}
\begin{pf} Suppose $X \subseteq |\mathbb P|$.  Then
\[
\Big(\bigvee_{p \in X} \bv{\Gcheck(\check p)}^\iota\Big)
 \wedge \Big(\bigvee_{p \in X^\perp} \bv{\Gcheck(\check p)}^\iota\Big)
=  \bigvee_{\stackrel{\scriptstyle p \in X}{\scriptstyle q \in X^\perp}}
 \big( \bv{\Gcheck(\check p)}^\iota \wedge \bv{\Gcheck(\check q)}^\iota \big) = 0,
\]
as follows from the $\iota$-validity of (\ref{eq bv filter b}).  (If $p \in X$ and $q \in X^\perp$ then there is no $r \le p,q$, and the join of the empty set is 0.)  Moreover,
\[
\Big(\bigvee_{p \in X} \bv{\Gcheck(\check p)}^\iota\Big)
 \vee \Big(\bigvee_{p \in X^\perp} \bv{\Gcheck(\check p)}^\iota\Big)
 = 1,
\]
as follows from the $\iota$-validity of (\ref{eq bv filter c}).  Hence,
\begin{equation}
\label{eq bigvee compl}
\textstyle \left(\bigvee\nolimits_{p \in X} \bv{\Gcheck(\check p)}^\iota\right)
 =
\compl \left(\bigvee\nolimits_{p \in X^\perp} \bv{\Gcheck(\check p)}^\iota\right).
\end{equation}
Let $h$ be defined by
\[
\textstyle h(X) = \bigvee_{p \in X} \iota(p)
\]
for every $X \in \regalg{\mathbb P}$, i.e., for every regular $X \subseteq |\mathbb P|$.  Since by definition $\compl X = X^\perp$, (\ref{eq bigvee compl}) implies that $h(\compl X) = \compl h(X)$.  

We now show that $h\left(\bigvee \mathcal X\right) = \bigvee_{X \in \mathcal X} h(X)$, i.e.,
\[
\bigvee_{p \in \bigvee \mathcal X} \bv{\Gcheck(\check p)}^\iota
 =
\bigvee_{X \in \mathcal X} \bigvee_{p \in X} \bv{\Gcheck(\check p)}^\iota,
\]
for all sets $\mathcal X$ of regular subsets of $|\mathbb P|$.  By the definition of join in $\regalg{\mathbb P}$, this is equivalent to
\[
\textstyle \bigvee_{p \in \complet{\bigcup \mathcal X}} \bv{\Gcheck(\check p)}^\iota
 =
\bigvee_{p \in \bigcup \mathcal X} \bv{\Gcheck(\check p)}^\iota,
\]
which follows from (\ref{eq bigvee compl}) applied first with $\bigcup \mathcal X$ for $X$ and then with $\left(\bigcup \mathcal X\right)^\perp$ for $X$, keeping in mind that $\complet Y$ is by definition $Y^{\perp\perp}$.

Since $\bv\sigma^\iota = h (\bv\sigma^{\iota^{\mathbb P}})$ for $\sigma = \Gcheck(\check p)$ for all $p \in |\mathbb P|$, it follows by induction on the rank of $\sigma$ that $\bv\sigma^\iota = h (\bv\sigma^{\iota^{\mathbb P}})$ for $\sigma \in \lang L^{\mathbb P}$.\qed\end{pf}

Note that (\ref{eq bv filter b}) and (\ref{eq bv filter c}) together imply (\ref{eq bv filter a}), which explains why we did not have to use (\ref{eq bv filter a}) explicitly in the proof of Theorem~\ref{thm universal}.
\begin{definition}
\label{def forcing}
Suppose $\mathbb P$ is a partial order and $\sigma$ is a sentence of $\lang L^{\mathbb P}$.  We say that an element $p \in |\mathbb P|$ \emph{forces $\sigma$}, $p \forces \sigma$, iff $p \in \bv{\sigma}^{\iota^{\mathbb P}}$.
\end{definition}
Recall that the starting point for our realization of the hidden-variables program for a propositional algebra $\elsalg P$ was the premise that any question that can be formulated from the propositions of $\elsalg P$ is equivalent for pseudoclassical states to a single proposition, which is in the completion $\complet{\elsalg P}$ of $\elsalg P$ as a PA.  We were careful to keep track of the size of the algebras involved and the number of conditions in the set $\mathcal S$ with respect to which genericity was defined, so that we could make use of the fact that $\mathcal S$-generic filters exist if $\mathcal S$ is countable.

We may streamline the argument by supposing at the outset that $\elsalg P$ is complete and dealing with $\elsalg P$-valued interpretations of $\lang L^{\elsalg P}$, in effect using $\elsalg P$-valued logic to manipulate $\lang L^{\elsalg P}$-expressions.  It must be emphasized that this characterization of the algebra of $\elsalg P$-valued expressions should not be taken to suggest that truth is ``really'' $\elsalg P$-valued.

At any time we may restate our results in terms of $\mathcal S$-generic filters for suitable $\mathcal S$.  Of course, the existence of such filters then becomes an issue, and we will gain more insight into this in Section~\ref{sec mod set theory}.

Let us formulate the discussion in Section~\ref{sec suff gen} leading up to Theorem~\ref{thm first rand var}
in these terms.  Let $\phi_q$ be the sentence
\[
\bdbigwedge_{M<0}
 \bdbigvee_{N \in \omega}
 \bdbigwedge_{n > N}
 \bdbigvee_{\stackrel
  {\scriptstyle \sigma \in \preset n2\settowidth{\pmbwidth}{$\scriptstyle \sigma \in \preset n2$}\hspace{-\pmbwidth}\phantom{\Big|q - \frac1n \sum\limits_{m=0}^{n-1} \sigma(m)\Big| < \frac1M}}
  {\scriptstyle \Big|q - \frac1n \sum\limits_{m=0}^{n-1} \sigma(m)\Big| < \frac1M}}
\bdbigwedge_{m < n} \Gcheck\left((\compl)^{1-\sigma(m)} \check P_m \right).
\]
Then by the same reasoning as before $\bv{\phi_q} = 1$.  Hence, if $F$ is a sufficiently generic filter on $\elsalg P$, $\phi$ holds with $F$ for $\Gcheck$. 

The set $\theory T^{\mathbb P}$ of validities of $\lang L^{\mathbb P}$ is easily shown to be closed under logical inference.\footnote{In logic one may define a theory in a language $\lang L$ to be any set of formulas of $\lang L$ and define $\complet{\theory T}$, the \emph{deductive closure of $\theory T$}, to be the closure of $\theory T$ under logical inference, or deduction.  For most purposes, $\theory T$ is equivalent to $\complet{\theory T}$, and when we speak of \emph{the} theory of something, we refer to a deductively closed theory.}  In the special case that $\mathbb P$ is $\elsalg P^-$ for an algebra $\elsalg P$ of 
propositions for a quantum system $\system S$, $\theory T^{\mathbb P}$ may be regarded as the pseudoclassical theory of $\system S$ (as regards $\elsalg P$):  it consists of all those and only those sentences that are true of all pseudoclassical states.  In this terminology Theorem~\ref{thm first rand var} becomes
\[
\bv{\mbox{it looks like the Copenhagen interpretation is right}} = 1.
\]

\subsection{Contextual hidden-variables models}
\label{sec contex hid var mod}

Recall that $F^{\Psi, \elsalg P}$ is the set of propositions $P \in \elsalg P$ true of a pseudoclassical state $\Psi$, where $\elsalg P$ is a PA.  Up to now the discussion has taken place in the context of a single PA, but we can easily carry out the key forcing construction for any number of PAs simultaneously if we allow for full contextuality, i.e., independence of $F^{\Psi, \elsalg P}$ from $F^{\Psi, \elsalg P'}$ for $\elsalg P \neq \elsalg P'$, so that a given proposition $P$ might be in $F^{\Psi, \elsalg P}$ but not in $F^{\Psi, \elsalg P'}$ for some $\elsalg P$ and $\elsalg P'$ that both contain $P$ and therefore serve as different \emph{contexts} for the measurement of $P$.  We do this as follows.

Suppose $\mathcal P$ is a set of PAs, which could be the set of all PAs for a Hilbert space $\vecsp V$.  Let $\mathbb P = (|\mathbb P|, \le)$, where $|\mathbb P|$ is the set of functions $p$ with domain $\mathcal P$ such that for all $\elsalg P \in \mathcal P$, $p(\elsalg P) \in \elsalg P^-$, and $p \le q$ iff for all $\elsalg P \in \mathcal P$, $p(\elsalg P) \le q(\elsalg P)$.  For any $\elsalg P \in \mathcal P$, and any dense $D \subseteq \elsalg P$, the set
\[
\{ p \in |\mathbb P| \mid p(\elsalg P) \in D \}\]
is dense in $\mathbb P$.  Thus, if $G$ is a generic filter on $\mathbb P$ then $\{ p(\elsalg P) \mid p \in G \}$ is a generic filter on $\elsalg P$ for each $\elsalg P \in \mathcal P$.

Note that no condition is imposed relating $p(\elsalg P)$ and $p(\elsalg P')$ for distinct PAs $\elsalg P$ and $\elsalg P'$.  If we identify pseudoclassical states with generic filters on $\mathbb P$ we obtain therefore a hidden-variables model that is maximally contextual.

Certain other partial orders $\mathbb P' \subseteq \mathbb P$ may be used in place of $\mathbb P$.  It is sufficient that $\{ p \in |\mathbb P'| \mid p(\elsalg P) \in D \}$ be dense in $\mathbb P'$ for any $\elsalg P \in \mathcal P$ and any dense $D \subseteq \elsalg P$.  For example, we may let $|\mathbb P'|$ be the set of $p \in \mathbb P$ such that $p(\elsalg P) = 1$ for all but finitely many $\elsalg P \in \mathcal P$.

In the next section we will see how a degree of noncontextuality may be incorporated into hidden-variables models.

\section{Propositional systems}
\label{sec prop sys}

\subsection{Pseudoclassical states and generic filters for propositional systems}

The creation of hidden-variables models by the method generic filters can be generalized to certain propositional systems (PSs) that are not boolean.
\begin{definition}
\label{def semifilter}
Given a PS $\elsalg R$, let $\elsalg R^- = \elsalg R \setminus \{0\}$.  A \emph{semifilter on $\elsalg R$} is a nonempty subset $F$ of $\elsalg R^-$ such that
\begin{enumerate}
\item $(\forall P \in F)(\forall Q \in \elsalg R) (P \le Q \implies Q \in F)$,
\item $(\forall P, Q \in F) (P\parper Q \implies P \wedge Q \in F)$.
\end{enumerate}
Recall that `$P \parper Q$' means $P$ and $Q$ commute.  A \emph{complete semifilter on $\elsalg R$} is a semifilter on $\complet{\elsalg R}$ that is closed under general meet.  Given any $S \subseteq \elsalg R$ there is a smallest semifilter $\completsf S$ and a smallest complete semifilter $\completcsf S$ that include $S$.

$F$ is \emph{full} iff for all $P \in \elsalg R$, either $P$ or $\compl P$ is in $F$.  We also call a full semifilter an \emph{ultrasemifilter}.
\end{definition}
If $\Psi$ is a pseudoclassical state for $\elsalg R$ then the set $F^{\Psi, \elsalg R}$ of propositions in $\elsalg R$ true of $\Psi$ is an ultrasemifilter.  As in the special case of PAs, $F^{\Psi, \elsalg R}$ contains all relevant information about $\Psi$, and any question that we may pose of pseudoclassical states may be interpreted as a question about semifilters.  We may identify such a question with the set of semifilters it defines, as we did in Section~\ref{sec genericity} for the case of filters on a PA.  This may be characterized as the \emph{extensional} point of view.  As we saw there, this approach necessitates consideration of the issue of existence of generic filters, which we prefer to defer; so for the present discussion we will take the \emph{intensional} approach developed in Section~\ref{sec bool exp}, i.e., we will deal directly with boolean values of expressions built from primitive expressions of the form $\Gcheck(\check P)$, i.e., $\check P \in \Gcheck$.

As in Section~\ref{sec bool exp} these expressions have a natural interpretation as statements in a language $\lang L = \lang L^{\elsalg R}$ appropriate to structures
\begin{equation}
\label{eq structure S R}
\structure S = ( \elsalg R, \le, \parper, \Gcheck, \dots, \check P, \dots),
\end{equation}
where $\elsalg R$ is the propositional system of interest, $\le$ has the usual meaning, $\parper$ is a binary relation symbol denoting commutativity of propositions, $\Gcheck$ is a unary predicate symbol, and for each $P \in \elsalg R$, $\check P$ is a constant that denotes $P$.  As before, $\lang L$ is an ordinary language except that we allow infinitary disjunction and conjunction, and we may replace any existential or universal quantifier construction by the corresponding disjunction or conjunction over $P \in \elsalg R$, so every expression of $\lang L$ is equivalent to a boolean combination of expressions of the form $\check P \le \check Q$, $\check P \parper \check Q$, and $\Gcheck(\check P)$.  Expressions of the first two types may be eliminated in favor of their truth values, which are constant, leaving a boolean combination of expressions $\Gcheck(\check P)$.  Letting $\epsilon_P$ be the sentence $\Gcheck(\check P)$ for each $P \in \elsalg R$, every sentence of $\lang L$ is given canonically by an $\elsalg R$-ary boolean expression.  We will represent the expression-building operations of $\lang L$, other than quantification, by the same symbols as we use for the abstract BE-forming operations, which are the boldface forms of the corresponding symbols of the metalanguage, which is the language in which this paper is written.  In this way we can maintain the distinction between making an assertion and mentioning it.

We wish to define an $\lang L$-theory $\theory T^{\elsalg R}$ that expresses `$\Gcheck$ is a pseudoclassical semifilter', and an interpretation $\hat\iota$ of $\lang L$ in a BA $\elsalg A^{\elsalg R}$ such that $\bv{\theory T^{\elsalg R}}^{\hat\iota} = 1$ and $\hat\iota$ is universal in the sense that for any interpretation $\iota$ of the class $\mathcal B^{\elsalg R}$ of $\elsalg R$-ary boolean expressions in a BA $\elsalg A$, if $\bv{\theory T^{\elsalg R}}^\iota = 1$ then there exists a complete homomorphism $h : \elsalg A^{\elsalg R} \to \elsalg A$ such that for any sentence $\sigma$ of $\lang L^{\elsalg R}$, $\bv{\sigma}^\iota = h\left( \bv{\sigma}^{\hat\iota}\right)$.
\begin{definition}
Let $\theory T^{\elsalg R}$ consist of the sentences
\begin{eqnarray}
\bdbigwedge_P \bdbigwedge_{Q \ge P} \big(\Gcheck(\check P) \bimplies \Gcheck(\check Q)\big),\label{eq bv semifilter a}\\
\bdbigwedge_{S \in \Com^{\elsalg R}} \bigg( \Big(\bdbigwedge_{P \in S} \Gcheck(\check P)\Big)
 \bimplies \Gcheck\Big(\check {\bigwedge S}\Big) \bigg),\label{eq bv semifilter b}\\
\bdbigwedge_P \big(\Gcheck(\check P) \bvee \Gcheck(\check{\compl P})\big),\label{eq bv semifilter c}
\end{eqnarray}
where $P$ and $Q$ range over $\elsalg R$ and $\Com^{\elsalg R}$ is the set of all subsets of $\elsalg R$ all of whose members commute with one another.  $\check {\bigwedge S}$ and $\check{\compl P}$ are the constant symbols of $\lang L$ that denote the elements $\bigwedge S$ and $\compl P$ of $\elsalg R$.
\end{definition}
\begin{definition}
If $p$ is a subset of $\elsalg R$ we let $E_{p} \eqdef \btbigwedge_{P \in p} \epsilon_P = \btbigwedge_{P \in p} \Gcheck(\check P)$.
\end{definition}
Suppose $\iota$ is an interpretation of $\mathcal B^{\elsalg R}$ in a BA $\elsalg A$ such that $\bv{\mbox{\normalfont(\ref{eq bv semifilter a}) \bvee (\ref{eq bv semifilter b})}}^\iota = 1$.  Then for any $p \subseteq \elsalg R$, $\bv{E_p}^\iota = \bv{E_{\completcsf p}}^\iota$, where $\completcsf p$ is the completion of $p$ to complete semifilter.  That is, the sentence
\[
\bdbigwedge_{P \in p} \Gcheck(\check P) \biff \bdbigwedge_{P \in \completcsf{p}} \Gcheck(\check P)
\]
is an $\iota$-validity.  To prove this one represents the process of completing $p$ to $\completcsf p$ as a well-ordered sequence $p = p_0 \subseteq p_1 \subseteq \cdots p_\alpha \subseteq \cdots p_\eta = \completcsf p$.  This process consists of steps of two sorts:  in one sort we close $p_\alpha$ upward, while in the other sort we add all meets of commuting subsets of $p_\alpha$.  To accommodate both sorts of step we extend the usual partition of natural numbers into even and odd to all ordinals by declaring that for any limit ordinal $\eta$ and any $n \in \omega$, $\eta+n$ is even or odd according as $n$ is even or odd.  We define $p_\alpha$ recursively by
\[
p_\alpha =
 \left\{
\begin{array}{ll}
\bigcup \{ p_\beta \mid \beta < \alpha \}&
\mbox{if $\alpha$ is a limit ordinal}\\
\{ Q \in \elsalg R \mid (\exists P \in p_{\alpha-1}) P \le Q \}&
\mbox{if $\alpha$ is an odd ordinal}\\
\{ \bigwedge S \mid S \in \Com^{p_\alpha} \}&
\mbox{if $\alpha$ is an even successor ordinal},
\end{array}\right.
\]
where $\Com^{p_\alpha}$ is the set of all commuting subsets of $p_\alpha$.  We show by induction on $\alpha$ that $\bigwedge_{P \in p_\alpha} \bv{\Gcheck(\check P)}^\iota = \bigwedge_{P \in p} \bv{\Gcheck(\check P)}^\iota$, using (\ref{eq bv semifilter a}) and (\ref{eq bv semifilter b}) in alternation to handle the successor steps.  The limit case of the induction is immediate.

Thus, as far as meets of primitive expressions are concerned we may restrict our attention to the expressions of the form $E_{p}$ for $p$ a complete semifilter.
\begin{definition}
Let $\mathbb P^{\elsalg R}$ be the set of complete semifilters on $\elsalg R$, which we make a partial order by reverse inclusion, i.e., $p \le q \iff p \supseteq q$.
\end{definition}
In terms of the forcing relation (Def.~\ref{def forcing}), Premise~1 states that anything that is true of a pseudoclassical state is forced to be true by some proposition that it satisfies.  In the general case of a propositional system $\elsalg R$ we cannot expect that single propositions will suffice to force everything about a pseudoclassical state.  For example, if $P$ and $Q$ are propositions that are not orthogonal, but $P \cap Q = \{ 0 \}$, then $P \bwedge Q$ (using `$\bwedge$' to indicate the formation of the conjunction of $P$ and $Q$ as predicates applicable to $\Psi$) may be true of a pseudoclassical state $\Psi$, but neither $P$ nor $Q$ nor any other proposition forces both $P$ and $Q$.  (Note that $P$ and $Q$ cannot commute in this circumstance.  If $P$ and $Q$ do commute then $P \bwedge Q$ is equivalent to the proposition $P \wedge Q$ since $F^\Psi$ is an ultrafilter.)  The premise that follows is the natural generalization of Premise~1 to propositional systems.

\begin{premprime}
Suppose $\Psi$ is a pseudoclassical state for a complete propositional system $\elsalg R$.  Then for any $\sigma \in \lang L^{\elsalg R}$, $\sigma$ is true of $\Psi$, i.e, $\sigma$ is true with $\Gcheck$ interpreted as $F^{\Psi, \elsalg R}$, iff for some $p \in \mathbb P^{\elsalg R}$, $p \subseteq F^{\Psi, \elsalg R}$ and $\sigma$ is true of all pseudoclassical $\Psi'$ such that $p \subseteq F^{\Psi', \elsalg R}$.
\end{premprime}
Recall that the language $\lang L^{\elsalg R}$ includes the class $\mathcal B^{\elsalg R}$ of boolean expressions built up from the primitive expressions $\epsilon_P$, i.e., $\Gcheck(\check P)$.  As discussed above, in the interest of efficiency we consider all of these to be questions ``posable'' of a pseudoclassical state in its incarnation as a semifilter $G$.  Thus Premises~\ref{prem boolean ops} and \ref{prem F sub P} are inherent in this formulation.  Premise~\ref{dvv}, which in our original formulation is the fundamental existence premise, corresponds to (\ref{eq bv semifilter c}), the only purely positive statement of membership in $\Gcheck$ among the sentences of $\theory T^{\elsalg R}$.

The reasoning of Section~\ref{sec nec gen} applies to show that pseudoclassical semifilters $F^{\Psi, \elsalg R}$ on $\elsalg R$ correspond to generic filters $F^{\Psi, \mathbb P^{\elsalg R}}$ on $\mathbb P^{\elsalg R}$ via the relations
\begin{eqnarray}
\label{eq alg sys}
F^{\Psi, \elsalg R} &=& \bigcup F^{\Psi, \mathbb P^{\elsalg R}}\\
F^{\Psi, \mathbb P^{\elsalg R}} &=& \mathcal P F^{\Psi, \elsalg R} \eqdef \{ p \in \mathbb P^{\elsalg R} \mid p \subseteq F^{\Psi, \elsalg R} \}.
\end{eqnarray}
As is customary in the context of forcing, we will refer to the members of $\mathbb P = \mathbb P^{\elsalg R}$ as conditions.

\subsection{The sufficiency of genericity for (some) propositional systems}

Let us now examine under what circumstances generic $\mathbb P^{\elsalg R}$-filters fulfill the requirements of the hidden-variables program. 

Given the importance of filters on $\mathbb P^{\elsalg R}$, it is convenient to expand our language $\lang L^{\elsalg R}$ to allow explicit reference to elements of $\mathbb P^{\elsalg R}$.  Thus we consider structures of the form $\structure S = (\elsalg R, \mathbb P^{\elsalg R}, \in, \le, \parper, \Gcheck, \dots, \check P, \dots, \dots, \check p, \dots)$, where $\Gcheck$ is a unary predicate on $\elsalg R$, and the language $\lang L^{\mathbb P^{\elsalg R}}$ appropriate to them.  $\lang L^{\mathbb P^{\elsalg R}}$ contains a name $\check p$ for each $p \in \mathbb P^{\elsalg R}$, as well as a name $\check P$ for each $P \in \elsalg R$, so we may eliminate quantifiers in favor of boolean operations, as we could for $\lang L^{\elsalg R}$.  We may also replace primitive expressions of the form $\check P \in \check p$, $\check P \le \check Q$, and $\check P \parper \check Q$ by their truth values, since---as in the case of $\lang L^{\elsalg R}$---the interpretation of these formulas is invariable.  It follows that every expression of $\lang L^{\mathbb P^{\elsalg R}}$ is equivalent to a boolean combination of the primitive expressions $\epsilon_P \eqdef \Gcheck(\check P)$, just as for $\lang L^{\elsalg R}$.

We note that for any $P \in \elsalg R$, the smallest $p \in \mathbb P^{\elsalg R}$ containing $P$ (smallest as a set, therefore largest as an element of $\mathbb P^{\elsalg R}$), i.e., the completion $\completcsf{ \{ P \} }$ of $\{ P \}$ to a complete semifilter, is
\[
\{ Q \in \elsalg R \mid Q \ge P \},
\]
and the smallest element of $\regalg{\mathbb P^{\elsalg R}}$ containing $\completcsf{ \{ P \} }$, i.e., the smallest regular subset of $\mathbb P^{\elsalg R}$ containing $\completcsf{ \{ P \} }$, is
\begin{eqnarray*}
\overline{\mbox{\Large$\lceil$} \completcsf{ \{ P \} } \mbox{\Large$\rceil$}}
&=&
\overline{\{ p \in \mathbb P^{\elsalg R} \mid (\forall Q \ge P)\, Q \in p \}}\\
&=&
\{ p \in \mathbb P^{\elsalg R} \mid (\forall q \supseteq p)(\exists r \supseteq q)\, P \in r \},
\end{eqnarray*}
where $q$ and $r$ range over $\mathbb P^{\elsalg R}$ and we have used `$\supseteq$' in place of `$\le$' to denote the order relation on $\mathbb P^{\elsalg R}$ to avoid confusion with the order relation on propositions.
\begin{definition}
\label{def semifil}
For $P \in \elsalg R$, let $\semifil P \eqdef \overline{\mbox{\Large$\lceil$} \completcsf{ \{ P \} } \mbox{\Large$\rceil$}}$.  The \emph{canonical interpretation} $\iota^{\mathbb P^{\elsalg R}}$ of $\lang L^{\mathbb P^{\elsalg R}}$ is the $\regalg \mathbb P^{\elsalg R}$-interpretation defined by
$
\bv {\Gcheck(\check P)}^{\iota^{\mathbb P^{\elsalg R}}}
 =
\semifil P.
$
\end{definition}
It is straightforward to show that
\begin{equation}
\label{eq bv semifilter a val}
\bv{\mbox{(\ref{eq bv semifilter a})}}^{\iota^{\mathbb P^{\elsalg R}}} = 1.
\end{equation}
The viability of the hidden-variables program depends on (\ref{eq bv semifilter b}) and (\ref{eq bv semifilter c}) being $\iota^{\mathbb P^{\elsalg R}}$-validities as well, but this is not true for all PSs $\elsalg R$.  For example, suppose $\elsalg R$ is finite.  Then every $\elsalg R$-semifilter is complete, and (\ref{eq bv semifilter c}) implies that for every $\elsalg R$-semifilter $p$, for every $P \in \elsalg R$, either $p \cup \{ P \}$ or $p \cup \{ \compl P \}$ may be extended to a $\elsalg R$-semifilter.  It follows that any $\elsalg R$-semifilter may be extended to an $\elsalg R$-ultrasemifilter.  A number of no-hidden-variables theorems provide examples of finite PSs for which ultrasemifilters do not exist.  In particular, $\elsalg R$ may be chosen to correspond to a system of physically meaningful operators, e.g., in Sec.~IV of \cite{Mermin:1993}, operators relating to measurements of the angular momentum of a spin-1 particle.   

A sufficient condition on $\elsalg R$ to ensure the validity of $\theory T^{\elsalg R}$ may be formulated in terms of the following notion of \emph{entailment}
\begin{definition}
A set $S \subseteq \elsalg R$ \emph{entails} a proposition $P$, $S \entails^{\elsalg R} P$, iff $P \in \completcsf S$.  $S$ is \emph{consistent} iff $S \nentails^{\elsalg R} 0$.
\end{definition}
Note that if every $P \in S$ holds for a pseudoclassical $\Psi$ and $S \entails^{\elsalg R} P$
then $P$ holds for $\Psi$, whence the name.

This notion of entailment differs from ordinary logical
entailment in that it does not in general incorporate the inference principle
of \emph{reductio ad absurdum}, which states that if $S$ is a set of
propositions and $P$ is a proposition then\footnote{The principle of \emph{reductio ad absurdum} is related to the \emph{law of the excluded
middle}, which asserts $\phi \bvee \bcompl \phi$ for all formulas $\phi$, and which is often advanced as a salient difference between
``quantum'' logic and ordinary or ``classical'' logic, inasmuch as
a quantum state may not satisfy either $P$ or $\compl P$ with
certainty, for $P$ a quantum proposition.  We prefer not to put it this way because we are
ultimately interested in pseudoclassical states, for which either
$\phi$ or $\bcompl \phi$ does hold with certainty for any formula $\phi$---\emph{a fortiori} for $\phi = P$, where $P$ is a quantum proposition; because $P \vee \compl P$ actually is 1; and because we quite specifically wish to avoid any suggestion that there is any necessary special quantum logic---the usefulness of boolean-valued logic in the context of hidden-variables models notwithstanding.}
\begin{equation}
\label{eq reduction}
\big(S \cup \{
P\} \entails 0 \big) \implies \big(S \entails \compl P\big).
\end{equation}
It is interesting that this classical logical principle, interpreted as a condition on a quantum propositional system, should be just what's needed to create the pseudoclassical states called for in the hidden-variables program.
\begin{definition}
\label{ede}
Suppose $\elsalg R$ is a propositional system.  We say that
$\elsalg R$ is \emph{reductive} iff for any $S \subseteq \elsalg R$ and $P \in \elsalg R$, $S \cup \{ \compl P \} \entails^{\elsalg R} 0 \implies S \entails^{\elsalg R} P$.
\end{definition}
\begin{thm}
\label{thm red sep}
Suppose $\elsalg R$ is a reductive PS.  Then $\mathbb P^{\elsalg R}$ is separative,
\begin{equation}
\label{eq red bv G}
\bv{\Gcheck(\check P)}^{\iota^{\mathbb P^{\elsalg R}}}
 =
\{ p \in \mathbb P^{\elsalg R} \mid P \in p \},
\end{equation}
and
\begin{equation}
\label{eq red bv compl G}
\bv{\bcompl \Gcheck(\check P)}^{\iota^{\mathbb P^{\elsalg R}}}
 =
\bv{ \Gcheck(\check{\compl P}) }.
\end{equation}
\end{thm}
\begin{pf} We first show that $\mathbb P = \mathbb P^{\elsalg R}$ is separative.  Suppose therefore that $p, q \in \mathbb P$ and $p \not\le q$, i.e., $p \nsupseteq q$.  We must show that for some $r \supseteq p$, $r | q$, i.e., $r \cup q$ is inconsistent.  To this end let $P$ be a proposition in $q \setminus p$.  Since $\mathbb P$ is reductive, $p \cup \{ \compl P \}$ is consistent.  Let $r = \completcsf{ (p \cup \{ \compl P \}) }$.  Then $r \supseteq p$ and $r \cup q $ contains $P$ and $\compl P$, so it is inconsistent as desired.

(\ref{eq red bv G}) now follows immediately from the definitions of $\iota^{\mathbb P^{\elsalg R}}$ and $[\,\cdot\,]$.

To prove (\ref{eq red bv compl G}) we observe that by definition
\begin{eqnarray*}
\bv{\bcompl \Gcheck(\check P)}
 &=&
\bv{\Gcheck(\check P)}^\perp\\
 &=&
\{ p \in \mathbb P \mid (\forall q \in \bv{ \Gcheck(\check P})\, p|q \},
\end{eqnarray*}
i.e.,  the conditions in $\bv{\bcompl \Gcheck(\check P)}$ are just those that are inconsistent with every condition that contains $P$.  Since $\elsalg R$ is reductive, these are just the conditions that contain $\compl P$.\qed\end{pf}

\begin{thm}
\label{edh}
Suppose $\elsalg R$ is a reductive PS.  Then $\bv{\theory T^{\elsalg R}}^{\iota^{\mathbb P^{\elsalg R}}} = 1$.
\end{thm}
\begin{pf} We have already noted that (\ref{eq bv semifilter a}) is valid for any PS $\elsalg R$.  The validity of (\ref{eq bv semifilter b}) follows quite directly from (\ref{eq red bv G}), and that of (\ref{eq bv semifilter c}) from (\ref{eq red bv compl G}).\qed\end{pf}

We should like to know that the hidden-variables theory just presented for reductive PSs extends the theory we have developed for PAs; in other words, if $\elsalg R$ is a reductive PS then for any boolean $\elsalg A \subseteq \elsalg R$, pseudoclassical $\elsalg R$-semifilters correspond to pseudoclassical $\elsalg A$-filters under the map $F \mapsto F \cap \elsalg A$.  Clearly, for any $\elsalg R$-semifilter, $F\cap \elsalg A$ is an $\elsalg A$-filter, so we only have to check that if $F$ is generic, then $F \cap \elsalg A$ is generic.  The following theorem states this in the language of boolean values.
\begin{thm}
Suppose $\elsalg R$ is a reductive PS and $\elsalg A$ is a boolean subsystem of $\elsalg R$.  Then for all $S \subseteq \elsalg A$
\begin{equation}
\label{eq bv semifilter T2}
\bigbv{\textstyle\big(\btbigvee_{P \in S} \Gcheck(\check P)\big) \bvee \big(\btbigvee_{P \in S^\perp} \Gcheck(\check P)\big)} = 1,
\end{equation}
where $S^\perp = \big\{ P \in \elsalg A\ \big|\ (\forall Q \in S)\, P | Q \big\}$.
\end{thm}
\begin{pf} This is a straightforward generalization of (\ref{eq bv semifilter c}), which is the special case of (\ref{eq bv semifilter T2}) obtained by letting $\elsalg A = \{ 0, P, \compl P, 1 \}$ and $S = \{ P \}$.  

(\ref{eq bv semifilter T2}) is equivalent to
\[
\textstyle\bigbv{\btbigwedge_{P \in S \cup S^\perp} \bcompl \Gcheck(\check P)} = 0.
\]
Using Theorem~\ref{thm red sep} and the fact that conditions in $\mathbb P^{\elsalg R}$ are by definition closed under general meet, we rewrite the left side as
\begin{eqnarray*}
\bigcap_{P \in S \cup S^\perp} \{ p \in \mathbb P \mid (\compl P) \in p \}
 &=&
\{ p \in \mathbb P \mid (\forall P \in S \cup S^\perp)\, (\compl P) \in p \}\\
&=&
\textstyle\big\{ p \in \mathbb P\ \big|\ \big(\bigwedge_{P \in S \cup S^\perp} \compl P\big) \in p \big\}\\
&=&
\textstyle\big\{ p \in \mathbb P\ \big|\ \big(\compl \bigvee_{P \in S \cup S^\perp} P\big) \in p \big\}.
\end{eqnarray*}
Since $\bigvee (S \cup S^\perp) = 1$ this proves the theorem.\qed\end{pf}

We close this section with some observations and questions.  It is easy to see that any boolean PS, i.e., any PA, is reductive.  As noted above, not all PSs are reductive; however, at least one physically interesting system is reductive.  Consider a pure spin system with spin $s = \tfrac12$, by which we mean a system with statevector space $\mathbb C^{2s + 1} = \mathbb C^2$ whose observables are the angular momentum operators $L^{\direction a}$, where $\direction a$ is a \emph{direction} in $\mathbb R^3$, and $L^{\direction a}$ measures the angular momentum in that direction.  The binary observables are the operators $P^{\direction a, S}$, where $S \subseteq \{ -\tfrac12, \tfrac12 \}$, and $P^{\direction a, S}$ is 1 for $L^{\direction a}$-eigenstates with eigenvalue in $S$ and 0 for $L^{\direction a}$-eigenstates with eigenvalue not in $S$.  If $\direction a$ and $\direction a'$ are distinct directions not opposite to one another, the only time a relation of the form $P^{\direction a, S} \le P^{\direction a', S'}$ can hold is if either $S = \emptyset$, i.e., $P^{\direction a, S} = 0$, or $S' = \{ -\tfrac12, \tfrac12 \}$, i.e., $P^{\direction a', S'} = 1$.  It follows immediately that $\elsalg R$ is reductive.  This is a rather trivial example in that the minimal elements of $\elsalg R$ in this case form a dense set, and every generic filter $G$ on $\mathbb P^{\elsalg R}$ is therefore principal, i.e., for some (necessarily minimal) $q \in \mathbb P^{\elsalg R}$, $G = \{ p \in \mathbb P^{\elsalg R} \mid p \ge q \}$.  We may therefore dispense with the whole machinery of generic filters in the 2-dimensional case:  a pseudoclassical state is any function that selects one member from each (unordered) pair $\{ P, \compl P \}$ of complementary propositions, as long as it selects 1 from $\{ 0, 1 \}$.

Are there more interesting examples of reductive PSs?  Is the system of position and momentum measurements on a quantum particle in one dimension reductive?  (This is known to be false for a particle in two or more dimensions\cite{Clifton:2000}.)  In general, how widespread is the reduction property?

In Sec.~\ref{sec contex hid var mod} we saw how to define a maximally contextual hidden-variables model for a PS.  Is there a natural notion of a minimally contextual hidden-variables model for a system $\elsalg R$, perhaps using reductive subsystems of $\elsalg R$?

\subsection{Comment on a result of Malley}

Malley\cite{Malley:2004} has recently shown that certain assumptions regarding a hidden-variables model for a Hilbert space of dimension at least 3 imply that all propositions in the relevant PS commute.  The setting for this result (slightly and inessentially modified to conform with the conventions of this paper) consists of a PS $\elsalg R$; a set $\Lambda$ of hidden-variables states, which by virtue of assumptions HV(a) and HV(b) of \cite{Malley:2004} may be identified with semifilters on $\elsalg R$; and a countably complete algebra $\mathcal B$ of subsets of $\Lambda$.  It is implicit in \cite{Malley:2004} that $\mathcal B$ contains the sets $\mathcal F_P$ for all $P \in \elsalg R$, where $\mathcal F_P$ is the set of semifilters on $\elsalg R$ that contain $P$.  If $\psi$ is a statevector then the \emph{expectation} $\Exp(P;\psi)$ of $P$ for $\psi$ is given by $\|P\psi\|^2/ \|\psi\|^2$ for $P \in \elsalg R$.  The map $P \mapsto E(P) = \Exp(P; \psi)$ is a \emph{probability measure on $\elsalg R$}, i.e.,
\begin{eqnarray}
E(1) &=& 1,\label{eq disjoint additivity a}
\\
\textstyle E\big(\bigvee_{n \in \omega} P_n\big) &=& \textstyle\sum_{n \in \omega} E(P)\label{eq disjoint additivity b},
\end{eqnarray}
for any family $\langle P_n \mid n \in \omega \rangle$ of pairwise orthogonal propositions.  Note that since $P$ can only have the values 0 and 1, $\Exp(P;\psi)$ is the probability that a measurement of $P$ on the state represented by $\psi$ has the value 1.  If $T$ is a statistical ensemble of such states the expectation $\Exp(P;T)$ of $P$ for the ensemble is the corresponding weighted average of $\Exp(P;\psi)$ over the ensemble.  Clearly, (\ref{eq disjoint additivity a}) and (\ref{eq disjoint additivity b}) are satisfied for $E(T) = \Exp(P; T)$.  By Gleason's theorem\cite{Gleason:1957}, if $\elsalg R$ consists of all propositions for a Hilbert space of dimension at least 3 then (\ref{eq disjoint additivity a}) and (\ref{eq disjoint additivity b}) imply that for some positive trace-class operator $D$ with $\tr D = 1$, for all $P \in \elsalg R$,
\[
E(P) = \tr(DP).
\]
$D$ is the \emph{density operator} for the ensemble.

A statistical ensemble of pseudoclassical states is a probability measure $\mu$ on a suitable countably complete algebra $\elsalg A$ of sets of pseudoclassical states.  The sets in $\elsalg A$ represent questions ``posable'' of the (states in the) ensemble.  Regarding pseudoclassical states as suitably generic filters on $\mathbb P^{\elsalg R}$, we have seen that posable questions may be identified with elements of the regular algebra $\regalg {\mathbb P^{\elsalg R}}$ of $\mathbb P^{\elsalg R}$.  Hence, a pseudoclassical ensemble may be identified with a unit measure $\mu$ on $\regalg{\mathbb P^{\elsalg R}}$.  The elements of $\regalg{\mathbb P^{\elsalg R}}$ are literally regular sets of complete semifilters on $\elsalg R$, but as we have seen, each may be identified with its union, which is a semifilter---typically not complete.  Extending Def.~\ref{def semifil}, we define $\semifil S$ to be $\overline{\mbox{\Large$\lceil$} \completcsf{ S } \mbox{\Large$\rceil$}}$, i.e., the set of complete semifilters that include the complete semifilter generated by $S$, which is just the set of complete semifilters that include $S$.  Note that for $P \in \elsalg R$, Def.~\ref{def semifil} defines $\semifil P$ to be $\semifil{\{P\}}$.  As we have seen, if $\elsalg R$ is reductive, $\mathbb P^{\elsalg R}$ is separative, so the sets $\semifil S$ are regular.

If a pseudoclassical ensemble $\mu$ is to mimic the ordinary-state ensemble with density operator $D$, then for each $P \in \elsalg R$
\[
\mu\big(\semifil P\big) = \tr(DP). 
\]
Does this give us any guidance as to the measure of sets $S \subseteq \elsalg R$ consisting of more than one proposition?  In particular, what can we say about $\mu\big(\semifil{\{P,Q\}}\big)$?

Suppose we have an ordinary-state ensemble with density operator $D$, and suppose $P$ is a proposition.  We may create a new ensemble by executing a measurement of $P$ on each system of the first ensemble (or on a representative sample of it) and only keeping the systems that give $P=1$---in other words, applying $P$ as a filter.  It is easily shown that the density of the new ensemble is $PDP/\tr(DP)$---the \emph{von Neumann conditional density}.\footnote{This is derived in Section~IV.3 of \cite{vonNeumann:1955} and essentially stated on  p.~341, albeit without the normalization factor.  Much of von Neumann's treatment of density operators is in terms of ``relative probability'', which does not require a total probability of 1.}  The expectation of $Q$ for this new ensemble is $\tr(PDPQ)/\tr(DP)$, or, as it may be written, $\tr(DPQP)/\tr(DP)$, using the identity $\tr(AB) = \tr(BA)$.  This may be regarded as the probability of $Q$ (i.e., $Q=1$) conditioned on $P$ (i.e., $P=1$).  Analogously, if we first filter with $Q$ and then measure $P$, the expectation is $\tr(DQPQ)/\tr(DQ)$, which is the probability of $P$ conditioned on $Q$.  If the filtering operation did not change the state of system being filtered, then we could conclude that the joint probability of $P$ and $Q$ is given by  $\tr(DPQP)$---which is the probability that $P$ is found to be true and then $Q$ is found to be true---as well as by $\tr(DQPQ)$, which is the probability that $Q$ is found to be true and then $P$ is found to be true.  It follows from this assumption that
\[
\tr(DPQP) = \tr(DQPQ).
\]
Malley\cite{Malley:2004} shows that this identity for all density operators $D$ implies that $P$ and $Q$ commute.  Since a pseudoclassical ensemble $\mu$ does have to assign a measure to $P \bwedge Q$ \big(it is $\mu\big(\{P,Q\}\big)$ in the present terminology\big), it would appear that such ensembles exist only for boolean propositional systems, i.e., for propositional algebras, which is the strongest possible no-hidden-variables conclusion.

The argument of course depends on the assumption that the joint probability $\mu\big(\{P,Q\}\big)$ is given in terms of the von Neumann conditional density.  The rationale for this choice seems to be that since any proposition has a definite value for any pseudoclassical state, we may filter an ensemble by a proposition without changing the (states of the) elements of the ensemble.  Without going into an analysis of the theory of measurement in a hidden-variables model\cite{Beltrametti:1981,Mittelstaedt:1998,Busch:1996}, it seems quite plausible that subjecting a pseudoclassical state $F$ to an interaction with an observer $\system O$ that may be regarded as a measurement of a proposition $P$, and then filtering on the basis of that measurement, which amounts to ``collapsing the wave packet'', will yield a result from which the answer to `$Q \in F$?', for $Q$ not commuting with $P$, cannot be inferred.  At any rate, the assumption of the applicability of the von Neumann formula seems to require some justification.

Given the importance of statistical ensembles for the theory of measurement in ordinary quantum theory and, by extension, in the theory of hidden-variables models, it would be desirable to have an informative characterization of measures on $\regalg\mathbb P^{\elsalg R}$ for a propositional system $\elsalg R$.  In particular, given the positive results of the present paper for the case of reductive systems, it is natural to ask what are the implications of the reduction property for this class of measures.

A concrete example may serve to highlight the issues involved here.  Consider the simple PS $\elsalg R = \{ 0, P, \compl P, Q, \compl Q, 1 \}$, where $P$ and $Q$ do not commute.  $\mathbb P^{\elsalg R}$ consists of the complete semifilters
\begin{eqnarray*}
&\{ 1 \},&\\
&\{ P, 1 \},\ \{ \compl P, 1 \},\ \{ Q, 1 \},\ \{ \compl Q, 1 \},&\\
&\{ P, Q, 1 \},\ \{ \compl P, Q, 1 \},\ \{ P, \compl Q, 1 \},\ \{ \compl P, \compl Q, 1 \}&.
\end{eqnarray*}
The semifilters in the last row are the maximal semifilters and represent the pseudoclassical states for this system.  The algebra $\regalg\mathbb P^{\elsalg R}$ of regular subsets of $\mathbb P^{\elsalg R}$ is atomic.  Its atoms (i.e., minimal nonzero elements) are the sets consisting of a single maximal semifilter, viz.,
\[
\big\{ \{ P, Q, 1 \} \big\}, \big\{ \{ \compl P, Q, 1 \} \big\}, \big\{ \{ P, \compl Q, 1 \} \big\}, \big\{ \{ \compl P, \compl Q, 1 \} \big\},
\]
and every element of $\regalg\mathbb P^{\elsalg R}$ is a join of these.  For a proposition, say $P$, the corresponding element of $\regalg\mathbb P^{\elsalg R}$ is the set of maximal filters containing $P$, viz.,
\[
\big\{ \{ P, 1 \}, \{ P, Q, 1 \}, \{ P, \compl Q, 1 \} \big\}.
\]
Any assignment of non-negative values $\mu_{P,Q}$, $\mu_{\compl P, Q}$, $\mu_{P, \compl Q}$, $\mu_{\compl P, \compl Q}$ to the atoms such that
\[
\mu_{P,Q} + \mu_{\compl P, Q} + \mu_{P, \compl Q} + \mu_{\compl P, \compl Q} = 1
\]
defines a measure on $\regalg \mathbb P^{\elsalg R}$.  With this assignment we have
\[
\begin{array}{ccccc}
              &   & \mu(Q)           &  & \mu(\compl Q)\\
              &   & \shortparallel                &  & \shortparallel\\
\mu(P)        & = & \mu_{P,Q}        &+ & \mu_{P, \compl Q}\\
              &   & +                &  & +\\
\mu(\compl P) & = & \mu_{\compl P,Q} &+ & \mu_{\compl P, \compl Q}
\end{array}
\]
There are no necessary relations among these measures beyond the above; and if $D$ is a density operator such that $\mu(P) = \tr DP$ and $\mu(Q) = \tr DQ$, there is no formula involving $D$, $P$, and $Q$ that necessarily gives $\mu_{P,Q}$, etc.

\section{Models of set theory and the method of forcing}
\label{sec mod set theory}

\subsection{Formal language and logic}

Prop.~\ref{prop count filter} has so far been our sole engine for
generating generic filters, and the hypothesis of the countability
of $\mathcal S$ in that theorem has accounted for the
restriction of our conclusions to sets $\mathcal C$ consisting of
countably many Borel sets.  This restriction is a bit
awkward, and it is also somewhat unsatisfying in that the meaning
of `pseudoclassical'---which resides in the set $\mathcal
S$---depends on the set $\mathcal C$ of ``posable questions''.  For
a proper physical theory it is preferable that `pseudoclassical'
have an absolute meaning.

We have seen how to develop the theory in terms of boolean values of expressions without reference to existence of generic filters, and we might leave it at this.  We would then regard a physical system $\system S$ with propositional system $\elsalg R = \elsalg R^{\system S}$ as having a single pseudoclassical state for each boolean subsystem $\elsalg P$ of $\elsalg R$, of which only boolean-valued (specifically $\regalg \mathbb P^{\elsalg P}$-valued) statements can be made.  The ``hiddenness'' of this state is embodied in the logic appropriate to it.  Pseudoclassical states in this sense are not states in the usual sense, according to which $\system S$ has many states that it can be \emph{in}, and it is only $\system S$-in-a-particular-state that has physical reality; $\system S$, like all physical systems, being an abstract entity in terms of which we understand relationships among physical states.  If pseudoclassical states are to be states in this sense then their corresponding filters must exist in an appropriate mathematical sense, a sense that we now explore.

The issue of the existence of generic filters is a matter of some delicacy that is most naturally addressed in the context of models of set theory.  The purpose of this section is to present the relevant mathematical concepts and to show their position within the broader context of mathematical logic and set theory.  Familiarity with the related topics in the appendix is assumed.  A strictly proper presentation of this part of mathematics is necessarily very formal, and we have applied a relatively relaxed standard of exposition here, consistent---one hopes---with the demands of clarity.  \cite{Jech:2003} provides a thorough treatment of the subject.  Section~\ref{sec ontol hid var} summarizes the ontology of pseudoclassical states in terms of generic filters.

A \emph{formal language} consists of \emph{expressions} formed from
relation and operation symbols, variables, logical connectives
(corresponding to `not', `and', etc.), and quantifiers
(corresponding to `for all' and `there exists').  The variables,
logical connectives, and quantifiers are common to all formal
languages.  All the languages we consider will also have a special
binary relation symbol `$=$', the intended interpretation of which is
identity.  The remaining relation and operation symbols are
particular to a given language.  The language of set theory, for
example, has one relation symbol in addition to `$=$', viz., the
binary symbol `$\in$'.

Operation symbols (which are interpreted as functions) can always
be eliminated in favor of relations, and we will largely limit our
remarks to purely relational languages, although we will find it
convenient occasionally to consider languages with constant
symbols, i.e., operation symbols of rank 0, which are interpreted as functions that take no arguments---in other words as constants.

Suppose $\lang L$ is a relational language with $n$
relation symbols $\rho_0, \dots, \rho_{n-1}$ (in addition to `$=$'). 
A \emph{class structure} $\structure S = ( U, R_0, \dots, R_{n-1} )$ for $\lang L$ is a class $U$ together with
relations $R_0, \dots, R_{n-1}$ on $U$ that correspond to $\rho_0,
\dots, \rho_{n-1}$ in terms of rank (i.e., unary, binary, etc.).  `$=$' is always interpreted as the identity relation $\{ \langle a,a \rangle \mid a \in U \}$.  $\structure S = ( U, R_0, \dots, R_{n-1} )$ a \emph{set structure} iff $U$ is a set (\ie, not a proper class), or, equivalently, $\structure S$ is a set.  The term \emph{structure} unqualified means set structure.  A class structure $\structure S$ provides an interpretation
of the expressions of $\lang L$ whereby a formula $\phi(x_0, \dots,
x_{n-1})$ with the free variables shown is either true or false
when $x_0, \dots, x_{n-1}$ are set equal to $a_0, \dots, a_{n-1}$,
respectively---$a_0, \dots, a_{n-1}$ being any members of $U$---and we say that $\phi$ is or is not \emph{satisfied at $\langle a_0, \dots, a_{n-1} \rangle$} accordingly.
\begin{definition}
\label{ebz}
We write
\begin{equation}
\label{def satisfaction}
\structure S \models \phi[a_0, \dots, a_{n-1}]
\end{equation}
for `$\phi$ is \emph{satisfied} at $\langle a_0, \dots, a_{n-1}
\rangle$ when its relation and operation symbols are interpreted according to $\structure S$'.  $\models$ is the \emph{satisfaction relation}.
\end{definition}
For a given formula $\phi$ there is an obvious way to express $\structure S \models \phi[a_0, \dots, a_{n-1}]$ as a formula with the free variables $\structure S$ and $\langle a_0, \dots, a_{n-1} \rangle$:  we just replace the relation and operation symbols in $\phi$ by their interpretations in $\structure S$, replace the free variables $v_0, \dots, v_{n-1}$ of $\phi$ by $a_0, \dots, a_{n-1}$ in order, and restrict the quantifiers in $\phi$ to $|\structure S|$.  Let $\hat \phi$ be this formula.  Its free variables are $\structure S, a_0, \dots, a_{n-1}$, and it defines the binary relation
\[
\cdot \models \phi [\cdot].
\]
Note that we do not obtain in this way a single formula for the ternary relation $\models$---there is no way to combine all the formulas $\hat \phi$ into a single formula $\theta$ with the additional free variable $\phi$, if for no other reason than that their quantifier depth---\ie, the maximum number of alternations of quantifier occurring along any branch of the tree of subformulas created when $\phi$ is parsed---can be arbitrarily large, while the quantifier depth of $\theta$ is some (finite) number.

Note also, however, that $\hat \phi$ is only one way to define $\cdot \models \phi [\cdot]$.  If we restrict $\structure S$ to (set) structures, there is a formula $\theta$ as above.  Very briefly one shows by induction that partial satisfaction relations exist for all finite sets of formulas and that all such relations agree on any formula in their common domain for every assignment of its variables.   $\structure S \models \phi[a_0, \dots, a_{n-1}]$ is then defined by either of the following formulas:
\begin{enumerate}
\item There exists a partial satisfaction relation for $\phi$ and all its subformulas that contains $\langle \phi, \langle a_0, \dots, a_{n-1} \rangle \rangle$.
\item Every partial satisfaction relation for $\phi$ and all its subformulas contains $\langle \phi, \langle a_0, \dots, a_{n-1} \rangle \rangle$.
\end{enumerate}
If we have the axiom of infinity we can show that there is a unique total satisfaction relation for any (set) $\structure S$, so we could then define $\structure S \models \cdot[\cdot]$ to be that relation.  Note that if $\system S$ is a proper class structure the partial satisfaction relations are also proper classes, so we cannot quantify over them to formulate either of formulas (1) and (2) (nor can we show that a total satisfaction relation exists with the axiom of infinity).

To summarize:  In the context of (set) structures, `$\models$' denotes a single formula, which may be any of those above.  In the context of class structures, there is no such formula, and for a given $\phi$, `$\structure S \models \phi[a_0, \dots, a_{n-1}]$' refers to the formula $\hat \phi = \hat \phi(\structure S, \langle a_0, \dots, a_{n-1} \rangle )$, whose free variables are $\structure S$ and $\langle a_0, \dots, a_{n-1} \rangle$ and whose quantifier depth increases with that of $\phi$.  We will largely confine our attention to set structures.

A \emph{sentence} is a formula with no free variables and is either satisfied or not satisfied by a  structure $\structure S$, \ie, it is either true or false for $\structure S$.  A \emph{theory} is a set of sentences.
\begin{definition}
If $\structure S$ is an $\lang L$-structure and $\theory T$ is an $\lang L$-theory, we say that $\structure S$ \emph{models
$\theory T$} or \emph{$\structure S$ is a model of $\theory T$}, written $\structure S \models \theory T$, iff $\structure S \models \sigma$ for all $\sigma \in \theory T$.
\end{definition}
\begin{definition}
We say that a theory $\theory T$ \emph{entails} a sentence $\sigma$ iff for any structure $\structure S$, if $\structure S \models \theory T$ then $\structure S \models \sigma$.  We say that $\theory T$ \emph{proves} $\sigma$, $\theory T \proves \sigma$, iff $\sigma$ follows from sentences in $\theory T$ by the usual rules of deduction.
\end{definition}
This definition of `entails' is the usual one in logic and differs from the \emph{ad hoc} definition given in Section~\ref{sec prop sys}.  Logical entailment has the reduction property, \ie, \emph{reductio ad absurdum} is a valid principle for entailment.  For suppose $\theory T$ is a theory and $\sigma$ is a sentence and $\theory T \cup \{ \lcneg \sigma \}$ entails $\sigma$, \ie, every model of $\theory T \cup \{ \lcneg \sigma \}$ also models $\sigma$.  Then there are no models of $\theory T \cup \{ \lcneg \sigma \}$, so every model of $\theory T$ satisfies $\sigma$, \ie, $\theory T$ entails $\sigma$.

Note that while the meaning of `entails' is absolute, the meaning of `proves' depends on the ``rules of deduction''.  It is nice if `entails' and `proves' are synonymous; in other words, if
\begin{equation}
\theory T \proves \sigma \iff \mbox{$\theory T$ entails $\sigma$}.
\end{equation}
The $\Longrightarrow$ direction is the \emph{correctness} property of the deductive system with respect to which `$\proves$' is defined.  The $\Longleftarrow$ direction is the \emph{completeness} property.  A system of deduction that is not correct is of little use.  On the other hand, systems of deduction that are not complete have generated some interest, particularly \emph{intuitionistic} systems that reject the \emph{law of the excluded middle}, i.e., the axiom schema $(\phi \mathor \lcneg \phi)$ for all formulas $\phi$, where `$\lcneg \phi$' means `not $\phi$'.  Nevertheless,
\begin{prop}
\disptitle{Completeness theorem}
the usual deductive systems are complete.  
\end{prop}
This is G\"odel's \emph{completeness theorem}.  The (essentially unique) correct complete deductive system is the familiar system of ordinary deductive discourse, called \emph{classical logic}, and this is the system to which we refer (and which we use!) in this paper.  

Suppose $\theory T$ is a theory and for some sentence $\sigma$, both $\sigma$ and
$\lcneg \sigma$ are theorems of $\theory
T$.  Then we can derive any sentence from $\theory T$.  As we have assumed a complete correct deductive system, this is equivalent to the corresponding statement for entailment:  if every model of $\theory T$ satisfies both $\sigma$ and $\lcneg \sigma$, then there are no models of $\theory T$, so every model of $\theory T$ satisfies everything, so $\theory T$ entails everything.  Of course we can prove any individual instance of this within any of the various descriptions of classical logic.  Letting $\phi$ be the conjunction of the premises in $\theory T$ used to prove the given $\sigma$ and $\lcneg \sigma$, this amounts to proving $\big((\phi \implies \sigma) \mathand (\phi \implies \lcneg \sigma)\big) \implies (\phi \implies \theta)$ for the given $\theta$.  Any reasonable formulation of classical logic will permit the construction of a short proof schema into which $\phi$, $\sigma$, and $\theta$ may be inserted to obtain a proof of the given instance.
\begin{definition}
A theory $\theory T$ is \emph{inconsistent} iff all sentences are theorems of $\theory T$, equivalently, if $\sigma$ and $\lcneg \sigma$ are both theorems of $\theory T$ for some $\sigma$.  Otherwise $\theory T$ is
\emph{consistent}.  
\end{definition}
The completeness theorem may be stated as follows.
\begin{prop}
Any consistent theory has a model.
\end{prop}
Note that `model', like `structure', means set model unless qualified with `class'.

Note that entailment is defined in terms of models, which are in
general infinitary objects, while inference is defined entirely in
terms of linguistic expressions and rules of inference, which are
finitary.  (The infinitary languages of Section~\ref{sec bool exp} are special constructs that do not reflect ordinary mathematical discourse.)  The great power of the completeness property is that it
permits the former to be expressed in terms of the latter.
\begin{definition}
\label{ebw}
Suppose $\theory T$ is an $\lang L$-theory.  \emph{$\theory T$ is deductively
closed} iff for
all $\lang L$-sentences $\sigma$, $T \proves \sigma \implies \sigma
\in T$.  $\overline T$, the (deductive) \emph{closure of $T$}, is the smallest closed
theory that includes $T$, which is defined explicitly as the set of
theorems of $T$.
\end{definition}
Of special interest is $\overline\emptyset$, the closure of the empty
theory, because it is included in any closed theory.  The members
of $\overline\emptyset$ are the theorems of pure logic, which are called
\emph{tautologies} because they are true by virtue of their form,
independent of the interpretation of the relation symbols of $\lang
L$.  For example, if $\lang L$ has two unary relation symbols $R_0$
and $R_1$, then
\[
\big(\forall x\, (R_0(x) \implies R_1(x)) \mathand \exists x\,
R_0(x) \big) \implies \exists x\, R_1(x)
\]
is a tautology of $\lang L$.

Proofs are finitary things, and a proof can utilize only
finitely many premises; hence, if $\theory T \proves \sigma$ then
for some finite $\Sigma \subseteq \theory T$, $\Sigma \proves
\sigma$.  Let $\langle \sigma_0, \dots, \sigma_{n-1} \rangle$ be an
enumeration of $\Sigma$.  Then
\begin{equation}
\label{ebx}
(\sigma_0 \mathand \sigma_1 \mathand \dots \mathand \sigma_{n-1})
 \implies \sigma
\end{equation}
is a tautology.

Note that one can easily program a computer, given unlimited memory
capacity, to generate exactly the tautologies simply
by systematically generating all proofs without
premises.\footnote{It is not to be thought that such a procedure
would necessarily facilitate the quest for understandable
mathematical theorems.  Most such procedures generate an
inconceivable number of ``junk theorems'', which are utterly beyond
comprehension---not because they are deep, but because they are
junk.  Of course, there is an effective theorem generator that is a veritable cornucopia of (humanly) understandable theorems:  the
human mathematics community.}  We call programmable procedures in
this sense \emph{effective} or \emph{recursive procedures}.  We
will consider two types of recursive procedure in this discussion. 
The first type accepts as input a sentence, presented as a string
of symbols in whatever (effective) representation suits our fancy. 
The procedure either never halts, or it halts with a specific
output value, which in this discussion is 0 or 1.  We call the
corresponding function \emph{recursive}.  We say that a recursive
function $f$ \emph{halts for $\sigma$} iff $f(\sigma)$ is
defined, i.e., $\sigma \in \dom f$.  A recursive function is
\emph{total} iff it halts for every input.

A set $S$ of sentences is \emph{recursive} iff there is a
total recursive function $f$ such that for all $\sigma$, $\sigma
\in S \iff f(\sigma) = 1$.  In other words, there is a recursive
procedure that always gives the right answer (1 for `yes' and 0 for
`no') to the question `$\sigma \in S$?'.

A set $S$ of sentences is \emph{recursively enumerable} iff
for some recursive $f$, $S = \dom f$.  This term originates
in the consideration of the second type of effective procedure.  A
procedure of this type does not accept an input: it simply runs,
occasionally putting out a value, which in this case is a sentence. 
We say that such a procedure \emph{enumerates} a set $S$ of
sentences iff it never outputs a sentence not in $S$ and for all
$\sigma \in S$ it eventually outputs $\sigma$.  It is not hard to
show that a set is recursively enumerable as defined above iff it can be enumerated in this way, i.e., recursively.  

Note that a recursive set is recursively enumerable, and if a set
and its complement are both recursively enumerable then they are both
recursive.

Since $\overline\emptyset$ is recursively enumerable, it is natural to
inquire whether it is recursive.  The answer is that it is iff all
the relation symbols of $\lang L$ other than `$=$' have rank 1 or 0 (a relation of
rank 0 is a primitive proposition that is either true or false in
a given interpretation).  All theories of intrinsic interest are in
languages with at least one relation symbol other than `$=$' of rank greater than 1,
and
\begin{prop}
\disptitle{Undecidability theorem of Church}
for these, $\overline\emptyset$ is not recursive.
\end{prop}
In other words there is no effective way to decide whether a sentence is a tautology.

A theory $\theory T$ is \emph{complete} iff for all sentences
$\sigma$, either $\sigma \in T$ or $(\lcneg \sigma) \in T$.\footnote{The notion of completeness of a theory is essentially unrelated to the notion of completeness of a system of deduction.}  A consistent complete theory need
not be recursively enumerable---indeed, if $\lang L$ has a relation
symbol of rank greater than 1, there are uncountably many complete
theories, so, as there are only countably many effective
procedures, most complete theories are not recursively enumerable. 

Nevertheless there are some interesting theories that are
complete, consistent, and recursively enumerable.  All these
theories have in common the feature that they are unable to
interpret the grammatical and logical structure of formal
languages---unable to talk about logic, as it were.  G\"odel's
celebrated incompleteness theorem states that
\begin{prop}
\disptitle{First incompleteness theorem}
any consistent
recursively enumerable theory that is ``able to talk about logic'' is
incomplete.
\end{prop}
If $\theory T$ is recursively enumerable, so is its deductive closure.  To prove the incompleteness theorem for a recursively enumerable theory $\theory T$, which we assume to be deductively closed, we suppose that $f$ recursively enumerates $\theory T$, and we form the sentence `this sentence is not
generated by $f$', which is to say,
\begin{equation}
\label{eby}
\mbox{this sentence is not a theorem of $\theory T$},
\end{equation}
and we show that this is not a theorem of $\theory T$.

We have to do a deal of rather meticulous work to construct a sentence that may be
regarded as saying (\ref{eby}) and to show that such a sentence is indeed not a theorem of $\theory T$ (assuming $\theory T$ is consistent), but the reward is well worth the effort.  By an ingenious turn of the screw, we can go on to show that
\begin{prop}
\disptitle{Second incompleteness theorem}
if $\theory T$ is a consistent recursively enumerable theory that is able to talk about logic then the sentence
\begin{equation}
\label{Con T}
\mbox{\rm$\theory T$ is consistent}
\end{equation}
is not a theorem of $\theory T$.
\end{prop}
Set theory is an example of a theory that can interpret logic, but
it is far richer in expression and deduction than is necessary (it
is, after all, capable of interpreting all of mathematics as we
know it)---indeed, a relatively meager fragment of the bare-bones theory of
the natural numbers known as \emph{Peano arithmetic} is sufficient for this purpose.

The undecidability theorem of Church and the incompleteness theorem of G\"odel have a common theme, which is also shared by the undecidability theorem of Tarski, which states that if a theory $\theory T$ in a language $\lang L$ can talk about logic and $\structure S$ is a model of $\theory T$, then there is no formula $\phi$ in $\lang L$ such that for all sentences $\sigma$ in $\lang L$,
\begin{equation}
\label{eq tarski}
\structure S \models \phi(\hat \sigma) \iff \structure S \models \sigma,
\end{equation}
where $\hat \sigma$ is the canonical term in $\lang L$ for $\sigma$.  Such a formula is called a \emph{truth predicate} for $\structure S$, true of exactly the sentences that are true (as well as their canonical terms) , and it can be used to construct a sentence $\sigma$ such that $\structure S \models \big(\sigma \lciff \lcneg \phi(\hat \sigma)\big)$, which contradicts (\ref{eq tarski}).  $\sigma$ is, in effect, the sentence uttered by Epimenides the Cretan, who said `all Cretans are liars', which has been interpreted in such a way that it is true iff it is false.  In our case $\sigma$ is a formulation of `this sentence is false'.

In a similar way we can use Richard's paradox, as given in the last footnote in Section 4 of \cite{RVWmw:2005}, to prove that if $\structure S$ is a model of Peano arithmetic ($\theory{PA}$), there does not exist a formula $\delta(\cdot,\cdot)$ in the language $\lang L$ of $\theory{PA}$ such that for all formulas $\phi(\cdot)$ of $\lang L$,
\begin{equation}
\label{eq richard}
\big(\structure S \models \delta[\hat \phi, n]\big) \iff\ \big(\mbox{$\phi$ is uniquely satisfied by $n$ in $\structure S$}\big),
\end{equation}
where $\hat \phi$ is the G\"odel number of $\phi$.  Note that the nonexistence of such a $\delta$ implies the nonexistence of a full Tarski truth predicate, but not obviously \emph{vice versa}.  If there were such a $\delta$ we could use Richard's schema to derive a contradiction.

Arithmetic is not able to talk about models in
general, because it can only represent finitary objects, and models are
generally infinite.  Set theory, on the other hand, is able to
discuss models.  For the sake of definiteness consider the Zermelo-Fraenkel theory, $\tzf$, which is the standard axiomatic theory of sets to be defined presently.  It can discuss models of theories and it can prove that if
$\tzf$ has a model then it is consistent, so by the incompleteness
theorem, assuming $\tzf$ (or at least the fragment of it that we
make use of) is consistent (and if it isn't our work is as dust in the wind),
then $\tzf$ cannot prove that a model of $\tzf$ exists.

It is convenient nevertheless to frame the discussion of generic
filters initially in terms of models of set theory.  Hence we must explicitly assume the
existence of a model of set theory.  When we thoroughly understand
generic extensions of such models we will be able to dispense with
this assumption.  For now we need an even stronger assumption,
viz., that there is a well-founded model of $\tzf$, i.e., a set
model $(M,E)$ for which the relation $E$ is well-founded (a notion to be defined presently).

\subsection{Axiomatic set theory}
\label{sec axiom set theory}

We first develop a useful fragment of set theory.  In doing this we
will present the axioms of set theory and a description of models
of set theory.  As discussed in the appendix, the most general version of set theory allows for objects that are not classes, which are referred to as
\emph{urelements} or \emph{atoms}.  From the standpoint of set
theory, atoms are regarded as structureless entities; all structure
is contained in the relations that hold among atoms and the sets
constructed from them.  In the interest of simplicity we will emphasize the theory of classes without atoms, but all our remarks apply \emph{mutatis mutandis} to the theory with atoms.  We will use `set theory' as a generic term for all such theories.

A model of set theory is best viewed hierarchically.  For
simplicity we will assume that the collection of all atoms (if we
permit atoms) is a set (as opposed to a proper class or not a class
at all).
\begin{definition}
\label{ebl}
Let $V_0$ be the set of atoms.  
\end{definition}
Let $V_1 = V_0 \cup \powerset V_0$, where for any set $X$,
$\powerset X \eqdef$ the set consisting of the subsets of $X$.  Let
$V_2 = V_1 \cup \powerset V_1$, and continue \emph{ad infinitum} to
define $V_n$ for any natural number $n$.  Recall that $\omega = \{ 0, 1, \dots
\}$.  Let $V_\omega \eqdef \bigcup_{n \in \omega} V_n$.  We continue
this construction into the transfinite, defining $V_{\omega + 1} = V_\omega \cup \powerset V_\omega$, and so on, where $\omega + 1 = \omega \cup \{ \omega \}$, $\omega + 2 = (\omega+1) \cup \{ \omega + 1 \}$, etc.  The sets $0, 1, 2, \dots, \omega, \omega + 1, \dots$ that occur here are the \emph{von Neumann ordinals}, which we will presently discuss at greater length.  In general,
\begin{equation}
\label{ebm}
V_\alpha = 
 \begin{elscases}
  \{ x \mid \mbox{$x$ is an atom} \} &\mbox{if $\alpha = 0$}\\
  V_\beta \cup \powerset V_\beta &\mbox{if $\alpha = \beta + 1$}\\
  \bigcup \{V_\beta \mid \beta \in \alpha\} &\mbox{if $\alpha$ is
a limit ordinal}.
 \end{elscases}
\end{equation}
To obtain the universe of pure sets we begin the above induction with $V_0 = \emptyset$, whether atoms are present or not.

The standard system of axioms for pure set theory is $\tzf$,
the \emph{Zermelo-Fraenkel} axioms.  For the purpose of this article we will use a version of $\tzf$ that allows for the possibility of
proper classes without positing any axioms that assert the
existence of a class that is not also asserted to be a set.  The corresponding system that
also allows atoms is $\tzfa$.  The corresponding systems with the
axiom of choice are $\tzfc$ and $\tzfca$.  The following discussion presents the axioms of $\tzfa$ in the above sense, with the axiom of choice brought in as
necessary or convenient.  As a
convenience we will introduce proper classes by
definition and use them in such a way that they may always be
eliminated in favor of their definitions (suitably incorporated
into expressions).  In particular $V$ is defined to be the class of
all sets.  We will use the predicates $\Element$, $\Set$, and $\Class$ for the several types of entities, and quantifiers will be qualified by the subscripts $E$, $S$, $C$, according as they govern variables that range over elements, sets, or classes, respectively.  An unsubscripted quantifier indicates a variable ranging over everything.

The first two axioms essentially define the three domain predicates:
\begin{eqnarray*}
(\forall x,y)\big(x \in y \implies (\Element(x) \mathand \Class(y) ) \big)\\
\forall x\, \big(\Set(x) \iff (\Element(x) \mathand \Class(x)) \big).
\end{eqnarray*}
The \emph{\bf existence} axiom
\begin{equation}
\label{eck}
\exists_S x\, x = x
\end{equation}
states that there exists a set.

The \emph{\bf extensionality} axiom
\begin{equation}
\label{ecl}
\forall_C X, Y\, \Big( X = Y \iff \big(\forall z\, ( z \in X \iff z
\in Y ) \big) \Big)
\end{equation}
states that the identity of a class $X$ is uniquely determined by
the collection $\{ x \mid x \in X \}$, i.e., if two classes have
the same members then they are identical, i.e., they are the same
class.\footnote{Note that (\ref{ecl}) only holds for classes.  Urelements have no members, so (\ref{ecl}) does not hold if $X$ and $Y$ are distinct urelements.  Note also that we do not have to specify a domain for `$z$'; if $z\in X$ then $z$ is an element.}  Note that only the reverse implication in
(\ref{ecl}) need be stated.  The forward direction is a tautology, \ie, a theorem of pure logic with identity.

The \emph{\bf comprehension} axiom schema states, for each formula
\[
\phi(
    \underbrace{\cdot, \cdot, \dots, \cdot}_
      {\stackrel{\mbox{\scriptsize$n+1$ free}}{\mbox{\scriptsize variables}}}
),
\]
that
\begin{eqnarray}
\label{ecm}
\forall z_0, \dots, z_{n-1}\, \forall_S x\, \exists_S y\,
\forall u\, \Big(u \in y \iff\\
 \big(u \in x \mathand \phi(u, z_0,
\dots, z_{n-1})\big)\Big),  
\end{eqnarray}
or, somewhat informally, `if $x$ is a set then $\{ u \in x \mid \phi(u, z_0, \dots, z_{n-1}) \}$ is a set for any formula $\phi$ and any $z_0, \dots, z_{n-1}$'.  With the comprehension schema we can turn Russell's paradox to our advantage and show that $V$ is a proper class.  For suppose to the contrary that it is a set and observe that $Y = \{ u \in V \mid u \notin u \}$ is then a set and therefore in
$V$.  It is evident that if $Y \in Y$ then $Y \notin Y$ and
\emph{vice versa}, which is a contradiction (which is, after all, what a paradox is).

It follows from the comprehension and existence axioms that there
is a set with no members, which by the axiom of extensionality is
unique.  We call the \emph{empty} set `$\emptyset$'.

The \emph{\bf pair} axiom
\begin{equation}
\label{ecn}
\forall_E x, y\, \exists_S z\, \forall w\, \big(w \in z
\iff (w = x \mathor w = y) \big)
\end{equation}
states that for any elements $x$ and $y$ there is a set $z$ such that $x$
and $y$ are members of $z$ and $z$ has no other members.  We denote
this set by `$\{ x, y \}$', the \emph{(unordered) pair of $x$ and
$y$}; $\{ x \} \eqdef \{ x, x \}$ is the \emph{singleton} of an element $x$.

The \emph{\bf powerset} axiom
\begin{equation}
\label{eco}
\forall_S x\, \exists_S y\, \forall w\, \big(w \in y \iff (\Set(w) \mathand w \subseteq x\big)
\end{equation}
states that for any set $x$ there is a set $y$ such that the
members of $y$ are exactly the subsets of $x$.  We denote this set
by `$\powerset x$', the \emph{powerset of $x$}.

The \emph{\bf union} axiom
\begin{equation}
\label{ecp}
\forall_S x\, \exists_S y\, \forall z\, ( z \in y \iff \exists
w\, z \in w \in x )
\end{equation}
states that for any set $x$ there exists a set $y$ that consists of
exactly the members of members of $x$.  We denote this set by
`$\bigcup x$', the \emph{union of $x$}, or---in terms perhaps more familiar---\emph{union of the members of $x$}.

The \emph{\bf replacement} axiom schema states, for each formula
\[
\phi(
    \underbrace{\cdot, \cdot, \dots, \cdot}_
      {\stackrel{\mbox{\scriptsize$n+2$ free}}{\mbox{\scriptsize variables}}}
)
\]
that\label{ecq}
\[
\begin{array}{l}
\forall z_0, \dots, z_{n-1}\, \forall_S x\, \exists_S y\, \forall_E v\\
\quad v \in y  \iff (\exists u \in x) \Big(\phi(u,v,z_0, \dots, z_{n-1})\\
 \phantom{\quad v \in y  \iff (\exists u \in x) \Big(}\mathand \forall_E v'\,
  \big(\phi(u, v', z_0, \dots, z_{n-1})
   \implies v' = v\big) \Big).
\end{array}
\]
In other words, given a formula $\phi$, if we fix $z_0, \dots, z_{n-1}$ and restrict our attention to those $u$ for which there is a unique $v$ such that $\phi(u,v,z_0, \dots, z_{n-1})$, i.e., the \emph{functional part} of $\phi(\cdot, \cdot, z_0, \dots, z_{n-1})$, which we will represent by `$F(\cdot)$', then if we restrict $u$ to range over a set $x$, $F(u)$ also ranges over a set.\footnote{Note
that the comprehension axiom for $\phi$, $z_0, \dots, z_{n-1}$, and
$x$ follows from the replacement axiom for $\phi'$, $z_0, \dots,
z_{n-1}$, and $x$, where
\[
\phi'(u,v,z_0, \dots, z_{n-1})
 \iffdef
\phi(u, z_0, \dots, z_{n-1}) \mathand u = v,
\]
so the comprehension schema is redundant once we have the
replacement schema; nevertheless, it is conceptually and
pedagogically useful to consider the comprehension schema
separately.}

Using these axioms we can derive the existence of a rich variety of
sets.  We have already derived the existence of a unique empty set
$\emptyset$.  We also denote this set by `0', a symbol that has
other meanings in various settings.  The intended denotation of any
instance of `0' may usually be inferred from its context.  In
(\ref{ebl}), which defines $V_0$ as the set of atoms, the
intended denotation is $\emptyset$.

By the powerset, pair, and union axioms, $V_1 = V_0 \cup \powerset V_0$ exists and is a set.  $0 \in V_1$, as are all other sets of atoms.  $\{0\}$ is not
in $V_1$, but it is a subset of $V_1$ and is therefore in $V_2 = V_1 \cup \powerset V_1$, along with all other sets of sets of atoms.  We define `1' in this context to mean `$\{0\}$'. 
Similarly we define `2' to mean `$\{ 0, 1 \} = \{ 0, \{ 0 \} \} =
\{ \{\}, \{ \{\}\} \}$.  Leaving aside the rest of the universe for the moment, we continue in this vein to define an
infinite sequence $0, 1 = \{ 0 \}, 2  = \{ 0, 1 \}, \dots, n = \{ 0, 1, \dots, n-1 \}, \dots$, of sets such that each set in the
sequence is the set of elements in the sequence that precede it.  We call these sets
the \emph{natural numbers} or the \emph{finite ordinals}.  Using
`$+ 1$' to denote the successor operation we have
\[
n + 1 = n \cup \{ n \},
\]
for all finite ordinals $n$.
\begin{definition}
\label{ebu}
A class $X$ is \emph{transitive} \tiffdef
\[
(\forall x \in X)\, x
\subseteq X.
\]
\end{definition}
\begin{definition}
\label{ebo}
A class $X$ is \emph{well-ordered} by a binary relation $\prec$ iff
it is strictly linearly ordered by $\prec$ and every nonempty
subset of $X$ has a $\prec$-least member.  An \emph{ordinal} is a
transitive set that is well-ordered by $\in$.  We define $\Omega$
to be the class of ordinals.
\end{definition}
One can show that $\Omega$ is well-ordered by $\in$, so it is a
proper class, for if it were a set then both $\Omega$ and $\Omega' = \Omega \cup \{ \Omega\}$ would be sets well-ordered by $\in$, i.e., ordinals, and would therefore both be in $\Omega$, but $\{ \Omega, \Omega' \}$ is not strictly ordered by $\in$, as both $\Omega' \in \Omega$ and $\Omega \in \Omega'$, contradicting the fact that $\Omega$ is well-ordered, \emph{a fortiori} strictly ordered, by $\in$. 

Using the replacement axiom and the fact that $\Omega$ is
well-ordered by $\in$, one can show that (\ref{ebm})
defines a unique $V_\alpha$ for each ordinal $\alpha$.  We define
$V_\Omega$ to be the class of all elements $x$ such that $x \in
V_\alpha$ for some $\alpha \in \Omega$.  Clearly,
\[
x \in V_\Omega \implies x \subseteq V_\Omega.
\]
Conversely, given a set $x \subseteq V_\Omega$, we use the
replacement axiom to show that there exists $\alpha \in \Omega$
such that $x \subseteq V_\alpha$.  It follows that $x \in V_{\alpha
+ 1}$.  Hence, for any set $x$,
\begin{equation}
\label{ebp}
x \subseteq V_\Omega \implies x \in V_\Omega.
\end{equation}
\begin{definition}
\label{def rank set}
For $x \in V_\Omega$, we define $\rho(x)$, the \emph{rank of $x$},
to be the least $\alpha \in \Omega$ such that $x \subseteq
V_\alpha$.  (Note that $x \in V_{\alpha+1} \setminus V_\alpha$.)
\end{definition}
\begin{definition}
\label{ebt}
Suppose $E$ is a binary relation on a class $X$.
\begin{enumerate}
\item For any $Y \subseteq X$, $E\invimage Y \eqdef \{ x \in X \mid
(\exists y \in Y)\,x\,E\,y \}$.
\item $E$ is \emph{extensional} \tiffdef for all $x, x' \in X$, if
$E\invimage \{x\} = E\invimage \{x'\}$ then $x = x'$.
\item $E$ is \emph{well-founded} \tiffdef
\begin{enumerate}
\item for every for every $x \in X$, $E\invimage \{x\}$ is a set,
and
\item for every nonempty set $Y \subseteq X$ there exists $y \in Y$
such that $E\invimage \{y\} \cap Y = \emptyset$.
\end{enumerate}
\end{enumerate}
\end{definition}
The well-foundedness of a relation $E$ on a class $X$ allows for \emph{proof by induction} (specifically, $E$-induction), which consists of inferring that a formula $\phi$ holds for any $x \in X$ from the hypothesis that for any $x \in X$, $\phi$ holds for $x$ if $\phi$ holds for all $y \in X$ such that $yEx$.  To justify the method, we reason from the given hypothesis as follows.  Suppose $\phi$ fails for some $x \in X$.  Use the infinity and replacement axioms to show that there is a set $Y \subseteq X$ such that $x \in Y$ and $E\invimage Y \subseteq Y$.  Use the comprehension axiom and the well-foundedness of $E$ to infer that there exists $y \in Y$ such that $\lcneg \phi(y)$ and $(\forall y' \in X) (y'Ey \implies \phi(y'))$, which contradicts the hypothesis.

Well-foundedness also permits \emph{definition by recursion} (specifically, $E$-recursion), in which a function $f$ with domain $X$ is defined at $x \in X$ in terms of $f(y)$ for all $y \in X$ such that $yEx$.  Briefly, one shows by induction that for each $x \in X$, there exists a function $F$ such that $x \in \dom F$ and $E\invimage \dom F \subseteq \dom F$ and $F$ satisfies the recursion formula, whatever it may be.  We then show that any two such functions $F$ agree at $x$, and we define $f(x)$ to be this common value for each $x \in X$.  Then $f$ satisfies the recursion formula throughout its domain, which is $X$.

The special case of $X = V^\Omega$ and $E =\, \in$ is of particular importance, and proof by $\in$-induction and definition by $\in$-recursion are indispensable tools in set theory.

The \emph{\bf infinity} axiom states that there is a set that contains
every finite ordinal.\footnote{For the purpose of this axiom we may take `finite ordinal' to mean an ordinal $\alpha$ such that each nonzero $\beta \in \alpha$ has an immediate $\in$-precursor in $\alpha$.}  By the comprehension axiom there is therefore a
set that contains exactly the finite ordinals.  We call this set
`$\omega$'.  It then follows from the replacement axiom \emph{et al}, that $V_\omega =
\bigcup \{ V_n \mid n \in \omega \}$ exists and is a set.  We can continue way
beyond this point.  $\omega + 1, \omega + 2 = \omega + 1 + 1,
\dots, \omega \cdot 2 = \omega + \omega, \dots, \omega^2 = \omega \cdot \omega, \dots, \omega^\omega$, and similarly describable sets
barely begin to suggest the complexity of just the countable
members of this sequence.

The \emph{\bf foundation} or \emph{\bf regularity} axiom states that the
membership relation is well-founded---i.e., that any nonempty class
$X$ has a member $x$ that has no member in $X$.
\begin{prop}
\label{ebr}
The foundation axiom implies that no set is a member of itself,\footnote{If $x \in  x$, let $y = \{ z \subseteq x \mid x \in z \}$, which is a set by the powerset and comprehension axioms.  Note that
$x \in y$.  By the foundation axiom, for some $z \in y$, $z \cap y
= \emptyset$.  Since $z \in y$,  $x \in z$, and as we have just noted $x \in y$,
which is a contradiction.} and $V_\Omega = V$.  In fact, the foundation axiom is
equivalent to the latter statement.
\end{prop}
The foundation axiom brings the entire set-theoretical universe within the scope of proof by $\in$-induction and definition by $\in$-recursion.

The axioms we have just listed, viz., existence, extensionality,
comprehension, pair, powerset, union, replacement, infinity, and
foundation, constitute the theory $\tzfa$, the
\emph{Zermelo-Fraenkel} theory with atoms.  If atoms are excluded
we have $\tzf$.

The foundation axiom has a somewhat different character from the
other axioms, which mostly say that for a particular sort of property that
elements might have, for any property of that sort there is a set that consists of exactly the
elements with that property.  The other exception is the extensionality
axiom, whose purpose is more to say what we \emph{mean} by `class' than to say anything \emph{about} classes.  The
foundation axiom, on the other hand, may strike one as a gratuitous
limitation on the existence of sets---why shouldn't there be a set
not in $V_\Omega$?  One reason is that in applications of set
theory---when we are not concerned with sets \emph{in abstracto}
but rather with sets as they embody and pertain to concepts that
arise in other branches of mathematics and its applications, including physics---sets
outside $V_\Omega$ do not naturally arise.

An occurrence of a variable in an expression is \emph{bound} iff it
falls within the scope of a quantifier phrase that contains the
variable; otherwise it is \emph{free}.  Thus, for example, in
`$\forall x (x \in y \iff \forall y (y \in z \implies y \in x))$',
the occurrences of `$x$' are both bound, and the second and third
occurrences of `$y$' are bound.  The first occurrence of `$y$' is
free.  (Variable symbols in quantifier phrases are not considered
occurrences.)  If $\phi$ and $\tau$ are respectively a formula and
a term in the language of set theory, $\phi^\tau$ is defined as the
formula obtained from $\phi$ by restricting every bound variable in
$\phi$ to $\tau$.  In other words, if $u$ is a variable, the
quantifier phrases $\forall u$ and $\exists u$ are replaced by
$\forall u \in \tau$ and $\exists u \in \tau$.  (If necessary we
first effect a change of bound variables in $\phi$ to variables that do not occur in $\tau$.)  We call $\phi^\tau$ the \emph{relativization of $\phi$
to $\tau$}.

In $\tzfa$ without the foundation axiom one can prove
$\sigma^{V_\Omega}$ for any sentence $\sigma \in \tzfa$, including
the foundation axiom itself.  In a sense, therefore, $\tzfa$
without foundation incorporates $\tzfa$.  In particular, if $\tzfa$
without foundation is consistent then $\tzfa$ is
consistent.\footnote{Given a proof of a contradiction from $\tzfa$
we could construct a proof of a contradiction from $\tzfa -
\mbox{\bf foundation}$ by relativizing everything to $V_\Omega$.  In this
connection it should be noted that by G\"odel's incompleteness
theorem, Prop.~\ref{Con T}, if $\tzfa$ is consistent it is not capable of proving its
own consistency.}

The \emph{axiom of choice} states that if $x$ is a set of disjoint
nonempty sets then there exists a set $y$ such that for all $x' \in
x$ there exists a unique $y' \in y$ such that $y' \in x'$.  There
are a number of useful formulation of this axiom that are
equivalent (over $\tzfa$).  A particularly handy form is the
statement that every set can be well-ordered.  It is known that if
$\tzf$ is consistent then the axiom of choice can be neither
disproved nor proved in $\tzf$ (\emph{a fortiori} not in $\tzfa$);
in other words it is consistent with $\tzfa$ to assume either the
axiom of choice or its negation. 

\subsection{Models of set theory and generic extensions}

\begin{prop}
\label{ebv}
\disptitle{Theorem of $\tzfa$}
Suppose $E$ is an extensional well-founded relation on a set $X$.  Then there
exists a unique transitive set $Y$ and a unique isomorphism of
$(X,E)$ to $(Y,\in)$, i.e., a unique bijection $j : X \to Y$ such
that for all $x, x' \in X$, $x\,E\,x' \iff j(x) \in j(x')$.  We call $Y$ the \emph{transitive collapse} of $(X,E)$.
\end{prop}
\begin{pf} The set $Y$ and the isomorphism $j$ are constructed by
recursion relative to $E$.\qed\end{pf}

Recall that `$\structure M \models \theory T$' means that the structure
$\structure M$ satisfies every sentence of the theory $\theory T$. 
We are particularly interested in structures of the form
$\structure M = (M, E)$ that model $\tzfa$.  In the case that $E$ is
just the membership relation on $M$, we may omit to mention it.

Let $\con \theory T$ be the sentence `$\theory T$ is consistent'. 
Although we have indicated that it is a theorem of $\tzfa$, the following is a theorem of any theory that can interpret logic, including the weak theory of arithmetic alluded to earlier.  All we need for the present purpose, however, is that it is a theorem of $\tzfa$.
\begin{prop}
\label{eca}
\disptitle{Theorem of $\tzfa$}
Suppose $\tzfa \cup \{ \con \tzfa \}$ is consistent.  Then
\[
\tzfa \not\proves (\con \tzfa \implies \mbox{there is a transitive set
model of $\tzfa$}).
\]
\end{prop}
\begin{pf} Suppose $\tzfa \cup \{ \con \tzfa \}$ is consistent, and
suppose toward a contradiction that
\begin{equation}
\label{ecb}
\tzfa \proves (\con \tzfa \implies \mbox{there is a transitive set model
of $\tzfa$}),
\end{equation}
i.e., $\tzfa \cup \{ \con \tzfa \} \proves$ there is a transitive set
model of $\tzfa$.  We will describe a proof of $\con \big(\tzfa \cup
\{ \con \tzfa \} \big)$ from $\tzfa \cup \{ \con \tzfa \}$, the
existence of which contradicts the incompleteness theorem.

Arguing in $\tzfa \cup \{ \con \tzfa \}$ for the duration of this paragraph, we first show that there is
a transitive set model $(M, \in )$ of $\tzfa$, which we can do by hypothesis.  Let $N \subseteq
M$ be $V_\omega$ in the sense of $M$.  It is straightforward to
show that $N$ is the true $V_\omega$.  $\con \tzfa$ is a statement
about finitary objects, represented as members of $V_\omega$, so $\con \tzfa \implies (M, \in) \models \Con \tzfa$.  Hence, $(M, \in) \models \tzfa \cup \{ \con \tzfa \}$, from which it follows that $\tzfa \cup \{ \con \tzfa \}$ is consistent.

As stated above, the existence of the proof just described
contradicts the incompleteness theorem, so our assumption
(\ref{ecb}) is false.\qed\end{pf}

By the L\"owenheim-Skolem theorem---the proof of which is quite straightforward---for any set $M$, $(M, \in)$ has a
countable elementary substructure, i.e., a subset $N$ such that for
all formulas $\phi(x_0, \dots, x_{n-1})$ with the free variables
shown, and all $a_0, \dots, a_{n-1} \in N$,
\[
\big((N,\in) \models \phi[a_0, \dots, a_{n-1}]\big)
 \iff
\big((M,\in) \models \phi[a_0, \dots, a_{n-1}]\big).
\]
Thus, if $M$ is a transitive set model of $\tzfa$ then $(N,\in)$ is a countable well-founded model of
$\tzfa$, and its transitive collapse is a countable transitive model
of $\tzfa$.  Thus the hypothesis of the existence of a countable transitive model of $\tzfa$ is no stronger than the hypothesis of the existence of a transitive set model of $\tzfa$.

In the following discussion $M$ will be supposed to be a transitive class model of $\tzfa$.  We allow the possibility that $M = V$, the class of all sets.  As noted above, we cannot treat proper class structures as we do set structures \visavis the satisfaction relation.  In particular, if $M$ is a proper class we cannot actually even ``say'' that $(M, \in) \models \tzfa$; rather, for each axiom $\sigma$ of $\tzfa$ we have to say that $(M, \in) \models \sigma$, \ie, that $\sigma$ holds relative to $M$.  In general we cannot do this any more simply than by asserting $\sigma^M$, the relativization of $\sigma$ to $M$.  Since $\tzfa$ has axioms with arbitrarily large quantifier depth (by virtue of the schemas of replacement and comprehension, which are stated for all formulas) no single formula can say that $(M, \in) \models \tzfa$.  We treat `$M \models \tzfa$' as a schema when $M$ is a proper class.  Clearly $V \models \sigma$ is just $\sigma$ for any sentence $\sigma$.

Suppose $\mathbb P = (P, \le) \in M$ is a separative partial order
with a maximum element, 1.  A filter $F$ on $\mathbb P$ is $M$-generic iff $F$ meets every dense subset of $P$ that is in
$M$.  Leaving aside the question of existence for the moment, suppose $G$ is an $M$-generic filter on
$\mathbb P$.  We define a map $x \mapsto x^G$ from $M$ into $V$ by induction on
the rank of $x$ according to
\begin{equation}
\label{ecr}
x^G \eqdef
\begin{elscases}
 x& \mbox{if $x \in V^M_0$}\\
 \big\{ y^G\ \big|\ (\exists p \in G) [y,p] \in x \big\} &\mbox{otherwise}.
\end{elscases}
\end{equation}
Let
\begin{equation}
\label{ecs}
M[G] \eqdef \big\{ x^G\ \big|\ x \in M \big\}.
\end{equation}
Clearly $M[G]$ is transitive.

For $a \in M$ we define $\check a$ by induction on the rank of $a$
according to
\[
\check a \eqdef
\begin{elscases}
 a& \mbox{if $a \in V^M_0$}\\
 \{ [\check b, 1] \mid b \in a \} &\mbox{otherwise}.
\end{elscases}
\]
Since any filter on $\mathbb P$ contains 1 (the maximum element of
$\mathbb P$), it follows by induction on the rank of $a \in M$ that
for any filter $G$ on $\mathbb P$
\begin{equation}
\label{eq a check}
\check a^G = a.
\end{equation}
Thus $M \subseteq M[G]$.

Note that
\begin{equation}
\label{eq G check}
\{ [\check p, p] \mid p \in P \}^G = G
\end{equation}
for any filter $G$ on $\mathbb P$.  Hence $G \in M[G]$.

In general, $G \notin M$.  In particular, if $\mathbb P$ is atomless, which is the only interesting case, then for any filter $F \in M$, $P \setminus F$ is a dense subset of $P$ and is in $M$; as $F$ does not meet $P \setminus F$, $F$ is not $M$-generic.  We are assuming $G$ is $M$-generic, so $G \notin M$, and $M[G]$ properly includes $M$.

We are interested in the theory of $\structure M[G] = (M[G], \in, \Mtilde, \dots, \hat x, \dots)$, where
\begin{enumerate}
\item $\in^{\structure M[G]}$ is the membership relation restricted to $M[G]$,
\item $\Mtilde$ is a predicate symbol with $\Mtilde^{\structure M} = M$, and
\item for each $x \in M$, $\hat x$ is a constant symbol with $\hat x^{\structure M} = x^G$.
\end{enumerate}
Let $\lang L^{M, \mathbb P}$ be the language of $\structure M[G]$ in the usual sense except that we allow conjunction $\btbigwedge \mathcal E$ and disjunction $\btbigvee \mathcal E$ of arbitrary sets of formulas $\mathcal E \in M$.

$\lang L^{M, \mathbb P}$ differs from the language $\lang L^{\mathbb P}$ of Section~\ref{sec forcing language} in several ways:
\begin{enumerate}
\item $\lang L^{M, \mathbb P}$ has the binary relation symbol `$\in$' (in addition to `$=$', which is in all languages by our convention);
\item $\lang L^{M, \mathbb P}$ has the predicate symbol $\Mtilde$;
\item the constant symbols of $\lang L^{M, \mathbb P}$ are indexed by $M$, as opposed to $|\mathbb P|$; and
\item conjunction and disjunction are restricted to sets $\mathcal E$ of formulas such that $\mathcal E \in M$, whereas in $\lang L^{\mathbb P}$ they are unrestricted.
\end{enumerate}
For the sake of definiteness we provide a specific implementation of this grammar, which we have so far described only in the abstract.  The details of this implementation are unimportant as long as all syntactical relationships are simply definable in $(M,\mathbb P,\in)$.

We let
\[
\hat x \eqdef [0,x],
\]
for all $x \in M$.  For each $a \in M$ we let
\[
\tilde a \eqdef \hat{\check a} = [0, \check a].
\]
By virtue of (\ref{eq a check}), $\tilde a^{\structure M[G]} = a$ for all $a \in M$.  Note that $\tilde M$ does not fall under this definition since $M \notin M$, and our use of this notation is a mnemonic convenience based on the fact that `$\tilde M$' always denotes $M$.  We need not assign a specific denotation to `$\tilde M$' as it only occurs in expressions of the form `$\tilde M(\tau)$', which are defined below as elements of $M$.

We let
\[
\Gcheck \eqdef \{ [\check p,p] \mid p \in |\mathbb P| \}\widehat{\ }
 = \Big[ 0, \{ [\check p,p] \mid p \in P \}\Big].
\]
By virtue of (\ref{eq G check}), $\Gcheck^{\structure M[G]} = G$ for any filter $G$ on $\mathbb P$.  Note that $\Gcheck$ is in the list $\dots, \hat x, \dots$ of constant symbols for $\lang L^{M, \mathbb P}$, so we do not have to list it separately as we did in the description of $\lang L^{\mathbb P}$ in Section~\ref{sec forcing language}.

We define the variables of $\lang L^{M, \mathbb P}$ to be sets of the form $[1,n]$ for $n \in \omega$.  We refer to the constants and the variables collectively as the \emph{terms} of $\lang L^{M,\mathbb P}$.  The \emph{primitive formulas} are those that state the membership of a term in $M$ or a relationship of identity or membership between terms.  We represent the operations of forming the latter formulas by the boldface versions of the usual relation symbols.  Specifically, we let
\begin{eqnarray*}
\Mtilde(\tau) &\eqdef& \big[2, \tau \big]\\
\tau \opeq \tau' &\eqdef & \big[3, [\tau, \tau']\big]\\
\tau \opin \tau' &\eqdef & \big[4, [\tau, \tau'] \big].
\end{eqnarray*}
We will also write `$\tau \opin \Mtilde$' for `$\Mtilde(\tau)$'.  Similarly we use the boldface version of the usual logical connectives to indicate appropriate formula-forming operations.  For example, using `$\phi$' and `$\phi'$' to represent arbitrary formulas---which are, bear in mind, elements of $M$---we let
\begin{eqnarray*}
\opneg \phi &\eqdef &[5, \phi ]\\
\phi \opand \phi' &\eqdef & \big[6, [\phi, \phi'] \big]\\
\phi \opor \phi' &\eqdef & \big[7, [\phi, \phi'] \big]\\
\phi \opimplies \phi' &\eqdef & \big[8, [\phi, \phi'] \big]\\
\phi \opiff \phi' &\eqdef & \big[9, [\phi, \phi'] \big],
\end{eqnarray*}
where $\opneg$, $\opand$, $\opor$, $\opimplies$, and $\opiff$ correspond respectively to negation, conjunction, disjunction, implication, and equivalence.

Quantifier expressions may be handled similarly.  For example, using `$\phi$' again to represent an arbitrary formula, and `$u$' to represent an arbitrary variable, we let
\begin{eqnarray*}
\opforall u\, \phi &\eqdef & \big[10, [u, \phi] \big]\\
\opexists u\, \phi &\eqdef & \big[11, [u, \phi] \big].
\end{eqnarray*}
As a concrete example, take the existence axiom.  As an element of $\lang L^{M, \mathbb P}$, it is $\opexists u\, u \opeq u$, where $u$ is a variable, i.e., $u = [1,n]$ for some $n \in \omega$.  For definiteness, suppose $u = [1,0]$.  Then
\begin{equation}
\label{eq meta}
\opexists u\, u \opeq u
 =
[11, [[1,0], [2, [[1,0], [1,0]]]]]. 
\end{equation}
Note the difference between `$\opeq$' and `=' in (\ref{eq meta}).  The former denotes the $\lang L^{M,\mathbb P}$-formula-forming operation $\tau, \tau' \mapsto \big[2, [\tau, \tau']\big]$, while the latter asserts the identity of $\opexists u\, u \opeq u$ and $[11, [[1,0], [2, [[1,0], [1,0]]]]]$.  We refer to $\lang L^{M, \mathbb P}$ as the \emph{object language} in this context, as it is the object of our consideration.

$[11, [[1,0], [2, [[1,0], [1,0]]]]]$ is an element of the object language, which we see here is an $\lang L^{M, \mathbb P}$-formula.  Eq.~\ref{eq meta}, on the other hand, is an element of the \emph{metalanguage} for this discussion, the language we are using to talk about the object language---in other words, the language in which this paper is written.

The 30-character string `$[11, [[1,0], [2, [[1,0], [1,0]]]]]$' is a term of the metalanguage that denotes the set $[11, [[1,0], [2, [[1,0], [1,0]]]]]$, which is a $\lang L^{M, \mathbb P}$-formula.

We now define corresponding notions of forcing and boolean value for sentences of $\lang L^{M, \mathbb P}$ related by
\begin{equation}
\label{eq force bv}
\bv\sigma^{M, \mathbb P}
 =
\{ p \in P \mid p \forces^{M, \mathbb P} \sigma \}.
\end{equation} 
The valuation algebra is the regular algebra of $\mathbb P$ as constructed in $(M, \mathbb P, \in)$.  We usually may omit the distinguishing superscript on `$\forces$' and `$\bv\cdot$' without loss of clarity.

The definition is designed to give the following theorem.
\begin{prop}
\label{prop forcing truth}
Suppose $M$ is a transitive class model of $\tzfa$, $\mathbb P = (P, \le) \in M$ is a separative partial order, $\sigma$ is a sentence of $\lang L^{M, \mathbb P}$, and $G$ is an $M$-generic filter on $\mathbb P$.  Then $\structure M[G] \models \sigma$ iff $(\exists p \in G) p \forces \sigma$.
\end{prop}
The following corollary is easily derived using the definability of $\forces$ in $(M, \mathbb P, \in)$ and fact that $p \forces \sigma \iff p \in \bv\sigma$, $\bv\sigma$ being a regular subset of $\mathbb P$.
\begin{cor}
Suppose $M$ is a transitive model of $\tzfa$, $\mathbb P = (P, \le) \in M$ is a separative partial order, and for all $p \in P$, there exists an $M$-generic filter $G$ containing $p$.  Then for any sentence $\sigma$ of $\lang L^{M, \mathbb P}$, $p \forces \sigma$ iff for all $M$-generic $G$ containing $p$, $\structure M[G] \models \sigma$.
\end{cor}
{\sc Remark }It should be noted that the existence hypothesis in the corollary is not inconsistent with $M$ being a proper class.  For example
\[
(M[G], \in)  \models \mbox{$M$ is a proper class and $G$ is $M$-generic}.
\]

It is convenient to give the proof of the proposition concurrently with the definitions.  We will use $\forces$ and $\bv\cdot$ interchangeably according to (\ref{eq force bv}).  We initially define these without reference to $M$.  $\bv\cdot^{M, \mathbb P}$ and $\forces^{M, \mathbb P}$ will be defined by relativization to $M$, \ie, by restriction of quantified variables to $M$.
\begin{pf}
We define $\bv\sigma$ by recursion on the complexity of $\sigma$.  We begin with the primitive sentences $\hat x_0 \opin \hat x_1$ and $\hat x_0 \opeq \hat x_1$, which we well-order as follows.  Recall that $\Omega$ is the class of ordinals.  Let $\prec$ be the well-ordering of $\Omega \times \Omega \eqdef \{ [\alpha_0, \alpha_1] \mid \alpha_0, \alpha_1 \in \Omega \}$ as follows:
\begin{eqnarray*}
[\alpha_0, \alpha_1]
 \prec
[\beta_0, \beta_1]
&\iffdef&\max\{\alpha_0, \alpha_1\} < \max \{\beta_0, \beta_1\}\\
&&\mathor\Big(\max\{\alpha_0, \alpha_1\} = \max \{\beta_0, \beta_1\}\\
&&\phantom{\mathor\Big(} \mathand\big( \min\{\alpha_0, \alpha_1\} < \min \{\beta_0, \beta_1\}\\
&&\phantom{\mathor\Big( \mathand\big(}\mathor( \min\{\alpha_0, \alpha_1\} = \min \{\beta_0, \beta_1\}\\
&&\phantom{\mathor\Big( \mathand\big(\mathor(}\mathand \alpha_0 < \beta_0)\big)\Big).
\end{eqnarray*}
We define the \emph{rank} $r[x_0, x_1]$ of an ordered pair of sets to be the position of $[\rho(x_0), \rho(x_1)]$ in this ordering, where $\rho(\cdot)$ is the rank function of Def.~\ref{def rank set}; and we define the rank $r(\sigma)$ of a primitive sentence $\hat x_0 \opin \hat x_1$ or $\hat x_0 \opeq \hat x_1$ as $r[x_0, x_1]$.  We observe that $r[\cdot,\cdot]$ is monotone increasing in (the rank of) each argument separately, or, to put it in the form we will use, decreasing (the rank of) either argument in an ordered pair decreases the value of the rank function.  We also observe that an unordered pair $\{ \alpha_0, \alpha_1 \}$ gives rise either to one ordered pair, if $\alpha_0 = \alpha_1$, or two ordered pairs, if $\alpha_0 \neq \alpha_1$, such that $r[\alpha_0, \alpha_1]$ and $r[\alpha_1, \alpha_0]$ are consecutive ordinals.  It follows that decreasing (the rank of) either argument in an ordered pair as above and then reversing its arguments also necessarily decreases the value of the rank function.  This is all we need to know about $r[\cdot,\cdot]$ (in addition to its well-orderedness).

In the following discussion, `$p$', `$q$', `$r$', and related variables range over $P$, and $\bar p \eqdef \lceil p \rceil$ (to reduce notational clutter).  As we have assumed $\mathbb P$ is separative, $\bar p \in \regalg \mathbb P$.  We define
\begin{eqnarray}
\bv{\hat x_0 \opin \hat x_1}
 &=&
\bigvee_{[y,p] \in x_1} (\bar p \wedge \bv{\hat y \opeq \hat x_0})\label{def memb}\\
\bv{\hat x_0 \opeq \hat x_1}
 &=&
\left(\bigwedge_{[y,p] \in x_0} (\bar p \lcimplies \bv{\hat y \opin \hat x_1})\right)
 \wedge
\left(\bigwedge_{[y,p] \in x_1} (\bar p \lcimplies \bv{\hat y \opin \hat x_0})\right)\label{def equal},
\end{eqnarray}
where $\lcimplies$ is the boolean operation defined by (\ref{eq ba implies}).  Since $[y,p] \in x \implies \rho(y) < \rho(x)$, we see that (\ref{def memb}) and (\ref{def equal}) define $\bv\sigma$ for a given primitive sentence $\sigma$ in terms of $\bv{\sigma'}$ for sentences $\sigma'$ of lower rank.  $\bv\sigma$ is therefore uniquely defined for all primitive $\sigma$ by induction on the rank of $\sigma$.

Note that we do not have to specify the initial step of this recursion separately, since it is given by the general formula as
\begin{eqnarray*}
\bv{\hat 0 \opin \hat 0}
 =&
\bigvee_{[y,p] \in 0} (\bar p \wedge \bv{\hat y \opeq \hat 0}) &= 0\\
\bv{\hat 0 \opeq \hat 0}
 =&
\left(\bigwedge_{[y,p] \in 0} (\bar p \lcimplies \bv{\hat y \opin \hat 0})\right)
 \wedge
\left(\bigwedge_{[y,p] \in 0} (\bar p \lcimplies \bv{\hat y \opin \hat 0})\right) &= 1.
\end{eqnarray*}
Note also that using the identity $\bar p = \bv {\tilde p \opin \Gcheck}$ we can rewrite the right side of (\ref{def memb}) as $\bigvee_{[y,p] \in x_1} \bv{\tilde p \opin \Gcheck \opand \hat y \opeq \hat x_0}$, and the right side of (\ref{def equal}) can be similarly expressed.  Thus (\ref{def memb}) and (\ref{def equal}) permit the reduction of any primitive formula to an equivalent $P$-ary boolean expression applied to the basic formulas $\epsilon_p = \tilde p \opin \Gcheck$, \ie, an expression in $\lang L^{\mathbb P}$ in the sense of Section~\ref{sec forcing language}, with the same boolean value---which is to say, given the proposition, having the same truth value when interpreted in $M[G]$ for any $M$-generic $G$.  Note finally that since $M$ is assumed to be countable, this construction carried out ``in $M$'', \ie, relativized to $M$, identifies each primitive formula of $\lang L^{M, \mathbb P}$ with a Borel expression in $\lang L^{\mathbb P}$.

Turning now to the proof of the proposition by induction, we note that there is no need to treat the initial cases $\hat 0 \opin \hat 0$ and $\hat 0 \opeq \hat 0$ separately, as their proofs are covered by the general case (the induction hypothesis being vacuous).  Thus, we suppose $\sigma$ is $\hat x_0 \opin \hat x_1$ or $\hat x_0 \opeq \hat x_1$ and we suppose the proposition to be true for all primitive sentences with rank less than $r[x_0, x_1]$.  We claim that it is true for $\sigma$.  We note that for all filters $G$
\begin{eqnarray*}
\structure M[G] \models \hat x_0 \opin \hat x_1 &\iff& x_0^G \in x_1^G\\
\structure M[G] \models \hat x_0 \opeq \hat x_1 &\iff& x_0^G = x_1^G.
\end{eqnarray*}
Note also that
\[
((\exists p \in G) p \forces \sigma) \iff G \cap \bv\sigma \neq \emptyset.
\]
We therefore claim that for all $M$-generic $G$
\begin{eqnarray}
x_0^G \in x_1^G &\iff& G \cap \bv{\hat x_0 \opin \hat x_1} \neq \emptyset\label{eq in}\\
x_0^G = x_1^G &\iff& G \cap \bv{\hat x_0 \opeq \hat x_1} \neq \emptyset\label{eq eq}.
\end{eqnarray}
\begin{pf} [(\ref{eq in})] By the definition of $\cdot^G$, $x_0^G \in x_1^G$ iff there exists $[y,p] \in x_1$ such that $p \in G$ and $y^G = x_0^G$.  Since $[y,p] \in x_1 \implies \rho(y) < \rho(x_1)$, by the induction hypothesis the left side of (\ref{eq in}) is equivalent to
\begin{equation}
\label{eq in left}
(\exists [y,p] \in x_1)(p \in G \mathand G \cap \bv{\hat y \opeq \hat x_0} \neq \emptyset).
\end{equation}
By (\ref{def memb}) the right side of (\ref{eq in}) is equivalent to
\begin{equation}
\label{eq in right}
(\exists [y,p] \in x_1)(\exists p' \in G) (p' \le p \mathand p' \in \bv{\hat y \opeq \hat x_0}.
\end{equation}
Suppose (\ref{eq in left}) is true and that $y$ and $p$ witness this, \ie, $[y,p] \in x_1 \mathand p \in G \mathand G \cap \bv{\hat y \opeq \hat x_0} \neq \emptyset$.  Let $p'$ be in $G \cap \bv{\hat y \opeq \hat x_0}$, and let $p''$ be a common extension of $p$ and $p'$ in $G$, which must exist since $G$ is a filter.  Then $p'' \in \bar p$ and $p'' \in \bv{\hat y \opeq \hat x_0}$ so by  $p'' \in \bv{\hat x_0 \opin \hat x_1}$, so (\ref{eq in right}) is true.

Conversely, suppose $y$, $p$, and $p'$ witness (\ref{eq in right}).  Then $p \in G$ because $p \ge p' \in G$ (and $G$ is a filter), and $G \cap \bv{\hat x_0 \opin \hat x_1}$ contains $p'$, so (\ref{eq in left}) is true.\qed [(\ref{eq in})]
\end{pf}
Of note, the preceding argument does not require that $G$ be $M$-generic, only that it be a filter.  The proof of (\ref{eq eq}) uses the genericity hypothesis.
\begin{pf} [(\ref{eq eq})] We will prove (\ref{eq eq}) in its contrapositive form:
\begin{equation}
x_0^G \neq x_1^G \iff G \cap \bv{\hat x_0 \opeq \hat x_1} = \emptyset\label{eq eq contra}
\end{equation}
As before we use the definition of $\cdot^G$ and the induction hypothesis to write the left side of (\ref{eq eq contra}) as
\begin{equation}
\label{eq eq left}
\begin{array}{r}
(\exists [y,p] \in x_0)(p \in G \mathand G \cap \bv{\hat y \opin \hat x_1} = \emptyset)\\
 \mathor
(\exists [y,p] \in x_1)(p \in G \mathand G \cap \bv{\hat y \opin \hat x_0} = \emptyset)
\end{array}
\end{equation}
Suppose (\ref{eq eq left}) is true.  Without loss of generality, suppose $y$ and $p$ witness the first clause, so that $[y,p] \in x_1$, $p \in G$, and $G \cap \bv{\hat y \opin \hat x_1} = \emptyset$.  Then
\begin{equation}
\label{eq G not meet bar p}
G \cap (\bar p \lcimplies \bv{\hat y \opin \hat x_1}) = \emptyset.
\end{equation}
For suppose toward a contradiction that $p' \in G$ and $p' \in \bar p \lcimplies \bv{\hat y \opin \hat x_1} = \compl \bar p \vee \bv{\hat y \opin \hat x_1}$, \ie, $(\forall q \le p')(\exists r \le q) (r \in \compl \bar p \mathor r\in \bv{\hat y \opin \hat x_1}$.  Since $G$ is $M$-generic, there exists $r \in G$ such that $r \in \compl \bar p$ or $r\in \bv{\hat y \opin \hat x_1}$.  $r$ cannot be in $\compl \bar p$ because it is then incompatible with $p$, which contradicts $G$ being a filter.  Hence $r\in \bv{\hat y \opin \hat x_1}$, which contradicts our assumption that $y$ witnesses the first clause of (\ref{eq eq left}).  (\ref{eq G not meet bar p}) implies that $G$ does not meet the right side of (\ref{def equal}), so $G \cap \bv{\hat x_0 \opeq \hat x_1} = \emptyset$, as claimed.

Conversely, suppose $G \cap \bv{\hat x_0 \opeq \hat x_1} = \emptyset$, \ie, $G$ does not meet the right side of (\ref{def equal}).  Since $G$ is $M$-generic, it fails to meet $\bar p \lcimplies \bv{\hat y \opin \hat x_1}$ for some $[y,p] \in x_0$ or $\bar p \lcimplies \bv{\hat y \opin \hat x_0}$ for some $[y,p] \in x_1$.  Suppose without loss of generality that $[y,p] \in x_0$ and $G$ fails to meet $\bar p \lcimplies \bv{\hat y \opin \hat x_1} = \compl \bar p \vee \bv{\hat y \opin \hat x_1}$.  Since $G$ is $M$-generic and does not meet $\compl \bar p$, it does meet $\bar p$, so $p \in G$, and therefore $y^G \in x_0^G$.  But $G$ does not meet $\bv{\hat y \opin \hat x_1}$, so by the induction hypothesis $y^G \notin x_1^G$.  Therefore $x_0^g \neq x_1^G$, as claimed.\qed [(\ref{eq eq}] \end{pf}

This concludes the definition and proof for the sentences $\hat x_0 \opin \hat x_1$ and $\hat x_0 \opeq \hat x_1$.  There is one more type of primitive sentence, \viz, $\tilde M(\hat x)$, and we now let
\begin{equation}
\label{def tilde M}
\bv{\tilde M(\hat x)}
 \eqdef
\bigvee_{a \in M} \bv{\hat x \opeq \tilde a}.
\end{equation}
The proposition for this case says that for any $M$-generic $G$, $\structure M[G] \models \tilde M(\hat x)$ iff $G \cap \bigvee_{a \in M} \bv{\hat x \opeq \tilde a} \neq \emptyset$.  Since $G$ is $M$-generic, $G \cap \bigvee_{a \in M} \bv{\hat x \opeq \tilde a} \neq \emptyset$ iff $\exists a \in M) G \cap \bv{\hat x \opeq \tilde a} \neq \emptyset$.  We know by now that this is equivalent to $(\exists a \in M) \structure M[G] \models \hat x \opeq \tilde a$, \ie, for some $a \in M$, $x^G = \check a^G = a$, which is to say, $\structure M[G] \models \tilde M(\hat x)$.

We are finished with the primitive sentences.  Now we define $\bv\sigma$ for complex sentences $\sigma$ by recursion on the complexity of $\sigma$, i.e., on the number of logical connectives and quantifiers in $\sigma$, in the natural way:
\begin{equation}
\label{def conn quant}
\begin{array}{rcl}
\bv{\opneg \sigma}
&=&
\compl \bv\sigma\label{def log conn a}\\
\bv{\sigma_0 \opand \sigma_1}
&=&
\bv{\sigma_0} \wedge \bv{\sigma_1}\label{def log conn b}\\
\bv{\sigma_0 \opor \sigma_1}
&=&
\bv{\sigma_0} \vee \bv{\sigma_1}\label{def log conn c}\\
\bv{\sigma_0 \opimplies \sigma_1}
&=&
\bv{\sigma_0} \lcimplies \bv{\sigma_1}\label{def log conn d}\\
\bv{\sigma_0 \opiff \sigma_1}
&=&
\bv{\sigma_0} \lciff \bv{\sigma_1}\label{def log conn e}\\
\bv{\opforall u\, \phi(u)}
&=&
\bigwedge_{x \in M} \bv{\phi(\hat x)}\label{def quant a}\\
\bv{\opexists u\, \phi(u)}
&=&
\bigvee_{x \in M} \bv{\phi(\hat x)}\label{def quant b}.
\end{array}
\end{equation}
The inductive proof of the proposition for complex sentences is straightforward and is left as an exercise for the reader.\qed [Proposition~\ref{prop forcing truth}]\end{pf}

Note that (\ref{def memb}), (\ref{def equal}), (\ref{def tilde M}), and (\ref{def conn quant}) make no mention of filters on $\mathbb P$, and they make perfect sense without any assumption about the existence of $M$-generic filters.  We may regard $\lang L^{M, \mathbb P}$ as referring to a \emph{boolean-valued structure}---specifically a $\regalg{\mathbb P}$-valued structure, which we call $M^{\regalg{\mathbb P}}$ or simply $M^{\mathbb P}$.  $M$ may be any class model of $\tzfa$; in particular, $M$ may be $V$, the class of all sets, i.e., the set-theoretical universe.  We call $V^{\mathbb P}= V^{\regalg{\mathbb P}}$ the $\regalg{\mathbb P}$-valued universe.

The notion of an $\elsalg A$-valued structure, where $\elsalg A$ is a complete BA, is a straightforward generalization of the usual notion of a structure $\structure S$ if we regard the latter as a $\boldsymbol 2$-valued operational structure $\structure S^{\boldsymbol 2}$, where $\boldsymbol 2$ is the 2-element BA, whose only elements are 0 and 1, i.e. false and true.  It is sufficient for the present purpose to consider structures appropriate to a given purely relational language $\lang L$.  For each relation symbol $R$ of $\lang L$ we let $\hat R$ be a distinct operation symbol of the same rank.  These are the operation symbols of $\structure S^{\boldsymbol 2}$.  For each relation symbol $R$ we define
\[
\hat R^{\structure S^{\boldsymbol 2}}(x_0, \dots, x_{n-1})
 \eqdef
\begin{elscases}
 1 & \mbox{if $\langle x_0, \dots, x_{n-1} \rangle \in R^{\structure S}$}\\
 0 & \mbox{if $\langle x_0, \dots, x_{n-1} \rangle \notin R^{\structure S}$},
\end{elscases}
\]
where $n$ is the rank of $R$ and $\hat R$.  To obtain an $\elsalg A$-structure $\structure S$ for $\lang L$ we simply replace $\boldsymbol 2$ by $\elsalg A$.  $\structure S$ is defined by the interpretations of $\hat R^{\structure S}$, which are $|\elsalg A|$-valued functions with arguments in $|\structure S|$.  The satisfaction operation for $\structure S$, which we denote by `$\bv\cdot^{\elsalg A}$', gives a value in $|\elsalg A|$ for each formula $\phi(a_0, \dots, a_{n-1})$ of $\lang L$ with parameters $a_0, \dots, a_{n-1} \in |\structure S|$, and is defined in the natural way.

The case of present interest is that $\elsalg A = \regalg{\mathbb P}$ and $\system S$ is the class of terms $\hat x$ of the forcing language $\lang L^{M, \mathbb P}$.
\begin{definition}
We say that a formula $\phi$ of $\lang L^{M, \mathbb P}$ with $n$ free variables is \emph{valid at} $\langle x_0, \dots, x_{n-1} \rangle$ iff $\bv{\phi(\hat x_0, \dots, \hat x_{n-1})}^{\regalg{\mathbb P}} = 1$.  A sentence in the language of set theory (without the additional constants $\hat x$ of the forcing language) is \emph{$\regalg{\mathbb P}$-valid} iff it is valid when regarded as a sentence of $\lang L^{V, \mathbb P}$ (there are no free variables in $\sigma$ to particularize).
\end{definition}

We will not have need of the following two propositions, but they are of great interest in their own right and provide an additional perspective on the results of this paper.  The proof of the first proposition is rather pedestrian and is left as an exercise for the interested reader.
\begin{prop}
\label{prop val ded closed}
For any separative partial order (SPO) $\mathbb P$, the set of $\regalg{\mathbb P}$-validities is deductively closed.
\end{prop}
To prove this one of course needs a precise definition of the deducibility relation $\proves$; all (classical ones) are equivalent, and any will do.
\begin{prop}
\label{prop axioms forced}
If $\sigma$ is an axiom of $\tzfa$, then $\tzfa \cup \{\mbox{$\mathbb P$ is a SPO},\ \Set(\mathbb P)\} \proves \bv\sigma^{\regalg{\mathbb P}} = 1$.  Moreover, $\tzfca \proves \bv{\elsaxiom{AC}}^{\regalg{\mathbb P}} = 1$.
\end{prop}
The proof of this proposition is a bit more involved, albeit not particularly difficult, but it (the proof, that is, not the proposition) provides no insight of particular value for the purposes of this paper, so we will omit it.

As noted previously, (\ref{eq gen fil}),
\[
\bv{\mbox{$\Gcheck$ is a generic filter on $\check{\mathbb P}$}}^{\regalg{\mathbb P}} = 1.
\]
What makes these propositions so useful is that $\mathbb P$ may be chosen in such a way as to create many validities in addition to those just listed.  The following theorem shows how to use this capability to prove relative consistency results in set theory.
\begin{prop}
Suppose  $\tzfa$ is consistent, $\sigma$ is a sentence in the language of $\tzfa$ and $\tzfa \proves \exists_S \mathbb P\, (\mbox{\normalfont$\mathbb P$ is a SPO and }\bv\sigma^{\regalg{\mathbb P}} = 1)$.  Then $\tzfa \cup \{ \sigma \}$ is consistent.  The same holds as well for any of the theories $\tzf$, $\tzfc$, and $\tzfca$.
\end{prop}
\begin{pf} Suppose toward a contradiction that $\tzfa \cup \{ \sigma \}$ is inconsistent, and suppose that $\pi$ is a proof of a contradiction, say $\theta \opand \opneg \theta$, from $\tzfa \cup \{ \sigma \}$.  Like all proofs, $\pi$ is finite and it makes use of only finitely many axioms of $\tzfa$.\footnote{Although we have given the axioms of $\tzfa$ as a finite list, several members of the list are schemas that have infinitely many instances.}

We construct a proof in $\tzfa$ as follows.  We begin by giving a proof of
\[
\exists_S \mathbb P\, (\mbox{\normalfont$\mathbb P$ is a SPO and }\bv\sigma^{\regalg{\mathbb P}} = 1).
\]
We then say `let $\mathbb P$ be such a SPO', and we show that each of the axioms of $\tzfa$ used in $\pi$ is a $\regalg{\mathbb P}$-validity, which we know we can do by Prop.~\ref{prop axioms forced}.  We now use Prop.~\ref{prop val ded closed} and the existence of $\pi$ to infer that $\theta \opand \opneg \theta$ is a $\regalg{\mathbb P}$-validity, i.e., $\bv{\theta \opand \opneg \theta}^{\regalg{\mathbb P}} = 1$.  But $\bv{\theta \opand \opneg \theta}^{\regalg{\mathbb P}} = \bv\theta^{\regalg{\mathbb P}} \wedge \compl \bv\theta^{\regalg{\mathbb P}} = 0$.  Hence, $\emptyset = 0^{\regalg{\mathbb P}} = 1^{\regalg{\mathbb P}} = |\mathbb P|$.  But a partial order is by definition nonempty, so this is a contradiction.

The existence of this proof (of a contradiction from $\tzfa$) contradicts our hypothesis that $\tzfa$ is consistent; and this contradiction establishes the untenability of our supposition that $\tzfa \cup \{ \sigma \}$ is inconsistent.  So $\tzfa \cup \{ \sigma \}$ is consistent, as claimed.\qed\end{pf}

\subsection{The ontology of hidden variables}
\label{sec ontol hid var}

We now have the means in hand to address definitively the issue of the existence of generic filters, which as we have seen is the crux of the hidden-variables program for reductive propositional systems:
\begin{enumerate}
\item One approach is to posit a countable transitive model $M$ of some suitable fragment of $\tzfa$---which may be taken to be finite, in which case the existence of $M$ is provable in $\tzfa$.  $M$ may be taken large enough to comprehend all structures of interest to us and all sets with respect to which we want pseudoclassical states to be generic.  As $M$ is countable, $M$-generic filters provably exist.
\item A second approach is simply to suppose that $M$-generic filters exist, even though $M$ may not be
countable.  The consistency of this supposition is proved as
outlined above.  As we have previously noted, of course, an $M$-generic filter cannot be in $M$ except in trivial cases, so if $M$-generic filters exist then $M$ is not the entire universe $V$ of sets.
\item The third approach is to suppose that $M =
V$ and to treat generic filters as having only a sort of
\emph{potential} existence.  The statement `$p \forces \sigma$',
where $\sigma$ is a sentence of $\lang L^{M, \mathbb P}$ may now be
interpreted informally as saying `if the universe were larger than it is and there were a generic (i.e., $V$-generic) filter $G$ with $p \in G$, then $\sigma$ would be true in $V[G]$'.  As we have seen
above, this is an interesting notion---and the set of sentences
with $\regalg{\mathbb P}$-value 1 is an interesting theory---even if generic
filters do not properly exist.  One conceptual advantage of this approach is that we do not
have to qualify the term `generic'; a filter is generic iff it
meets \emph{all} dense sets, not just those in some \emph{ad hoc}
collection.
\end{enumerate}  

We may tie the question of existence of generic filters in the mathematical sense (which has to do with the existence axioms of set theory) more closely to the question of physical existence of pseudoclassical states if, instead of thinking of the set-theoretical world as something distinct from the physical world, we regard the physical and mathematical worlds as one.  This
is not a bizarre conceit; we are quite accustomed, for example, to
treating certain sets of (physical) atoms, with certain relations, as
molecules; certain sets of molecules as, say, biological organisms;
certain sets of organisms as species, etc.

As noted above, it is perfectly possible to develop a theory of
sets that allows for the existence of objects that are not sets,
that we call for mathematical purposes \emph{urelements} or
\emph{atoms}.  This is the theory $\tzfa$ that we have been using above.  We may therefore define $V$ to be the
set-theoretical universe erected on a base consisting of purely
physical entities as urelements.  It is even conceivable that the
physical world is entirely set-theoretical, or---to put it perhaps
a bit more palatably---that the physical world is pure structure.

It is hard to imagine that anyone one conversant with historical
trends in physics could be confident that this is not so, but even if we adopted this point of view we would still have a need for a sense of physical existence distinct from mathematical existence, the former implying the latter but not \emph{vice versa}.  Thus, if pseudoclassical states in a given context then the corresponding generic filters must exist, but not \emph{vice versa}.  As with all physical theories, in practice the important question is that of utility:  Do we get a more useful theory by positing the existence of pseudoclassical states?  The answer to this question is almost surely `no', even assuming that a proper quantum theory can be constructed on this basis.  (See Bohm\cite{Bohm:1952a,Bohm:1952b} for an early attempt at this.)

Returning now to the three approaches to the existence of generic filters listed above, we see that the first approach may be interpreted as saying that pseudoclassical states for countably many posable questions---\viz, those in some countable model $M$ of $\tzfa$---could physically exist, since the corresponding generic filters mathematically exist.  The hiddenness of these states lies in the fact that their complete description is more complicated than any posable question---is not in fact in $M$.

The second approach suggests that we regard $M$ as containing the world of quantum mechanics in the conventional sense, including all ordinary quantum states and everything we can do with them, while pseudoclassical states reside in the $V \setminus M$.  The hiddenness of these states is due to the fact that they are not in the ``ordinary'' world.

The third approach suggests that we think of physical pseudoclassical states as having the same sort
of potential as opposed to actual existence as the $V$-generic filters that are their mathematical counterparts.  To summarize the main result of this paper in this context, let $\elsalg A = \regalg{\mathbb P^{\elsalg R}}$, where $\elsalg R$ is a reductive propositional system consisting of observables of a physical system $\system S$.  Let $V^{\elsalg A}$ be the $\elsalg A$-valued universe erected over the physical world.  Ordinary statements in the theory of $V$ with the additional unary predicate symbol `$\Gcheck$' may be interpreted in $V^{\elsalg A}$ and have a value in $\elsalg A$.  The set $\theory T^{\elsalg A}$ of statements with value 1 is the \emph{theory of $V^{\elsalg A}$}.  $\theory T^{\elsalg A}$ contains `$\Gcheck$ is a $V$-generic filter on $\mathbb P^{\elsalg R}$', and it contains `$f^{\tilde{\boldsymbol P}}_\Gcheck \in \bv{\tilde T}^{\hat\iota}$' for every $\omega$-sequence $\boldsymbol P$ of commuting propositions in $\elsalg R$ and every boolean expression $T$ such that $\mu^{\boldsymbol P}_\psi \big(\bv T^\iota\big) = 1$ for all nonzero $\psi \in \vecsp V$ (i.e., every test of randomness), where $\iota$ is the standard interpretation $\sigma \mapsto I_\sigma$ in the sense of $V$ and $\hat\iota$ is the same thing in the sense of $V^{\elsalg A}$.  In other words, $\Gcheck$ may be regarded as (representing) the pseudoclassical state of $\system S$ \visavis $\elsalg R$, and $f^{\tilde{\boldsymbol P}}_\Gcheck$---the sequence of values of $\langle P_n \mid n \in \omega \rangle$ for $\Gcheck$---looks for all the world like a sequence generated according to the stochastic process (i.e., probability measure) $\mu^{\boldsymbol P}_\psi$ for any nonzero $\psi$.  

This approach to the issue of hidden variables seems the most natural.  It makes no assumption about the existence of generic filters, either by imposing a restriction---such as countability---on the set of ``posable questions'', or by supposing that the (actual) set-theoretical universe is a generic extension of some subuniverse that contains all such questions.  Of course, we sacrifice the notion of a pseudoclassical state as a definite entity, but upon reflection this appears as not so much a defect of the interpretation as an insight into the natural status of hidden variables in the quantum theory.

A final word to dispel any doubt that may linger on the subject.  Boolean-valued logic (other than ``classical'' $\boldsymbol 2$-valued logic) has appeared in this paper only as an object of study, not as a mode of discourse.  The theorems are all $\boldsymbol2$-valued, in fact $\{ 1 \}$-valued. 

\appendix*
\section{Logical and set-theoretic conventions}

The essential relations of set theory are those of identity, denoted by `$=$' or `is', and membership, denoted by `$\in$', `is in', or `is a member of'.  A \emph{class} is any entity whose identity is determined entirely by its members---in other words, an object $A$ is a class iff for any $B$, $A=B \iff (\forall x)(x \in A \iff x \in B)$.  A \emph{set} is a class that is a member of some class.  A class is \emph{proper} iff it is not a set.  The necessity of distinguishing sets as special classes is most simply illustrated by \emph{Russell's paradox}, which inquires whether the class $A$ of all classes that are not members of themselves is a member of itself.  The resolution lies in noting that $A$ is ill defined, as it cannot contain \emph{all} classes that are not members of themselves, but only those that are sets.  If we so modify the definition then we may conclude that $A$ is not a member of itself and is therefore a proper class.  (In standard set theory, no set is a member of itself, so $A = V$, the class of all sets.)

`Set theory' is used generically to refer to any mathematical theory of membership.  In a \emph{pure class theory} we assume that all objects are classes; in a \emph{pure set theory} we assume that all objects are sets.  If we wish we may admit objects that are not classes.  In the context of set theory we call such objects \emph{atoms} or \emph{urelements}.  By definition they have no members.  By convention, they are permitted to be members of classes.  \big(An entity that neither has nor is a member (of anything) has no role in a theory of membership.\big)  They are regarded for set-theoretic purposes as primitive and devoid of intrinsic structure.  All their properties are attributable to their membership in various classes.  An \emph{element} is either a set or an urelement; thus, it is something that can be a member of a class.

Set theory with atoms subsumes ordinary mathematical practice and, indeed, ordinary scientific practice.  In chemistry, for example, we may regard (chemical) atoms as atoms (i.e., urelements).  Chemical bonds may also be regarded as urelements.  A molecule is a set of atoms and bonds related in a certain way, and so forth.  Physical quantities may be regarded as primitive entities with certain relations and operations; for example, there are physically definable operations of addition for temperature, energy, and many other physical quantities.  The elaborate mathematical formalism of modern physics consists of statements interpretable in a set-theoretical superstructure erected on top of the physical world, as it were.

`$\{ \cdots \}$' is used to denote a class with its members indicated in some fashion.  For example, $\{x,y\}$ is the (unordered) pair of $x$ and $y$ and \mbox{$\{ x \mid \phi(x) \}$} is the class of all elements $x$ that satisfy the formula $\phi$.  $[x,y] \eqdef \{ x, \{ x, y\} \}$\footnote{  `$\eqdef$' is used to mean `is defined to be', and `$\iffdef$' is used to mean `is defined to be true iff', i.e., roughly, `is defined to mean'.  `$\mapsto$' means `maps to' and provides a handy way of naming functions.  For example, `$x,y \mapsto x^y$' names the exponentiation operation.  A free variable may also be represented by a dot, so that, for example, `$|\cdot|$' has the same meaning as `$x \mapsto |x|$'.} is the \emph{ordered pair} of $x$ and $y$.  Note that $x$ and $y$ must be elements in order that $[x,y]$ exist, and if $x,y,x',y'$ are elements, then $[x,y] = [x',y'] \iff (x=x' \mathand y=y')$.  A \emph{binary relation} is a class of ordered pairs.  The
\emph{domain} $\dom R$ and \emph{image} $\im R$ of a binary
relation $R$ are
\label{ecd}
\begin{eqnarray}
\dom R
&\eqdef
\{ x \mid \exists y\, [x,y] \in R \}\\
\im R
&\eqdef
\{ y \mid \exists x\, [x,y] \in R \}.
\end{eqnarray}
If $R$ is a relation and $X$ is a class, $R\image X \eqdef \{ y \mid (\exists x \in X) [x,y] \in R \}$ is the (forward) image of $X$ under $R$, and $R\invimage X \eqdef \{ x \mid (\exists y \in X) [x,y] \in R \}$ is the inverse image of $X$ under $R$.

A \emph{function} is a binary relation $F$ with the property that
\begin{equation}
\label{ece}
(\forall x \in \dom F)(\exists! y) [x,y] \in F,
\end{equation}
where `$\exists!$' means `there exists a unique'.  For $x \in \dom
F$, $F(x)$ is that element $y$ such that $[x,y] \in F$.  If $F$ is a function and $X$ is a class, $F\image X = \{ F(x) \mid x \in X \}$ and $F\invimage X = \{ x \in \dom F \mid F(x) \in X \}$.

We will also find it convenient to name functions by expressions of the form
\begin{equation}
\label{ecf}
\langle \tau(x) \mid \phi(x) \rangle
 \eqdef
\{ [x, \tau(x)] \mid \phi(x) \}.
\end{equation}
where $\phi(x)$ is a formula, i.e., it is true or false depending
on $x$, and $\tau(x)$ is a term i.e., it denotes a element that is
uniquely determined by $x$ for all $x$ such that
$\phi(x)$.  $\phi$ and $\tau$ may have free variables other than $x$.  

$\emptyset$ is the empty set, i.e., the (unique) set with no
members.  We also use `0' to denote $\emptyset$.  1 is defined as
$\{ 0 \}$, 2 is $\{ 0, 1\}$, 3 is $\{ 0,1,2\}$, etc.  In other
words, the natural numbers are defined in such a way that each
natural number $n$ is the set of numbers that precede it (of which
there are $n$).  These are also referred to as the \emph{finite
ordinals}.  $\omega \eqdef \{ 0, 1, \dots \}$ is the set of all
finite ordinals.  For $n \in \omega$, a \emph{sequence of length $n$} or \emph{$n$-sequence} or \emph{$n$-tuple} is a function with domain
$n$.  Note that if $\sigma$ is an $n$-sequence then $\sigma = \langle \sigma(m) \mid m \in n \rangle$.  We may also denote such a sequence $\sigma$ by indicating its elements as an explicit list between angle brackets.  Thus $\sigma = \langle \sigma(0), \sigma(1), \dots, \sigma(n-1) \rangle$.  Similarly, an \emph{$\omega$-sequence}---also called an \emph{infinite sequence}---is a function with domain $\omega$, and we may indicate such a sequence $\sigma$ by `$\langle \sigma(0), \sigma(1), \dots \rangle$'.  \emph{Concatenation} of finite sequences is defined and indicated as follows:  $\sigma\concat\sigma' = \langle \sigma(0), \dots, \sigma(n-1), \sigma'(0), \dots, \sigma'(n'-1) \rangle$, where $\sigma$ and $\sigma'$ have length $n$ and $n'$, respectively.

Note that $\langle
x,y \rangle \neq [x,y]$.  The former is
\[
\Big\{ \big\{ 0, \{ 0, x \} \big\}, \big\{ 1, \{ 1, y \} \big\}
\Big\},
\]
while the latter is
\[
\big\{ x, \{ x, y \} \big\}.
\]

(\ref{ecf}) stresses the view of a function as an indexed
family of elements.  We will also have occasion for a single expression
that denotes an indexed family of classes in general (whether sets
or proper classes).  For this purpose we make the following
definition.  If $R$ is a binary relation and $x$ is a element,
\begin{equation}
\label{ech}
R\famind x \eqdef \{ y \mid [x,y] \in R \}.
\end{equation}
Note that $R\famind x$ is defined for all $x$, even if $x \notin
\dom R$, in which case $R\famind x = \emptyset$.  If $C$ is a class
and $R$ is a binary relation then we may consider $R$ as a
\emph{$C$-indexed family} (of classes) if we agree that we will use
only members of $C$ as indices.  We represent this
\emph{$C$-indexed family} by
\begin{equation}
\label{eci}
( R\famind x \mid x \in C ).
\end{equation}
Analogously to elements, we define a finite sequence of classes to be
an $n$-indexed family for a finite ordinal $n$.  $(), (a), (a,b)$,
etc., are respectively 0-, 1-, 2-sequences, and so on.  Note that
(\ref{eci}) allows us to efficiently index families that
contain $\emptyset$ and families that do not.  We use the notation `$(\cdots)$' for an $n$-indexed family, where $n$ is a natural number and $\cdots$ is a list of $n$ classes, defining it to be the smallest class that serves the purpose.  Thus, for example,
\begin{eqnarray*}
(X, Y, Z)
 =
\{ [0,a] \mid a \in X \} \cup \{ [1,a] \mid a \in Y \}\\ \cup \{ [2,a] \mid a \in Z \}.
\end{eqnarray*}
It is perhaps a useful exercise to compare the three ways we have of forming an ``ordered pair'':
\begin{eqnarray*}
[x,y]
&=
\{ x, \{ x, y \} \}\\
\langle x, y \rangle
&=
\{ [0,x], [1,y] \}
 =
\{ \{ 0, \{ 0, x \} \}, \{ 1, \{ 1, y \} \} \}\\
&=
\{ \{ \{\,\}, \{ \{\,\}, x \} \}, \{ \{ \{\, \}\}, \{ \{ \{\, \}\}, y \} \} \}\\
(x,y)
&=
\{ [0,a] \mid a \in x \} \cup \{ [1,a] \mid a \in y \}.
\end{eqnarray*}

A general \emph{relation of rank $n$} may be regarded as a class of $n$-tuples.  \big(Note that this imposes a distinction between a binary relation (a class of ordered pairs) and a relation of rank 2 (a class of 2-tuples)---a distinction that we would not wish to be required to maintain scrupulously.\big)  For example, the order relation $<$ on the real number field $\mathbb R$ is the set of $2$-tuples $\langle x, y \rangle$ with $x,y \in \mathbb R$ and $x < y$.  A general operation of rank $n$ is a relation $F$ of rank $n+1$ with the property that for all $x_1, \dots, x_n$ there exists at most one $x$ such that $\langle x, x_1, \dots, x_n \rangle \in F$, and we define $F(x_1, \dots, x_n)$ to be this $x$ if it exists. 

A \emph{structure} is, most simply, a class of individuals together with some relations and operations on this class.  For example, the field of real numbers may be regarded as a structure $\mathbb R = (R, +, \cdot)$, where $R$ is the set of real numbers, and $+$ and $\cdot$ are respectively the operations of addition and multiplication.  Note that we use the `$(\cdots)$' method of indicating a sequence of classes, and we list the class of individuals first.  We refer to this class also as the \emph{universe} or \emph{domain} of the structure.  Given a structure $\structure S$, we define $|\structure S|$ to be its domain.  Thus, in general, $\structure S = ( |\structure S|, \dots)$.

If $A$ is a set and $B$ is a class we define $\preset{A}B$ (read
``$B$ pre $A$'') to be the class of all functions from $A$ to $B$. 
In particular, for $n \in \omega$, $\preset{n}B$ is the set of sequences of
members of $B$ of length $n$, and $\preset{\omega}B$ is the set of all
infinite sequences from $B$.  We define $\preset{<\omega}B$ to be $\bigcup_{n \in \omega} \preset nB$.
\begin{definition}
\label{def PO}
A \emph{partial order (PO)} $\mathbb P$ is a structure $\mathbb P = (|\mathbb P|, \le)$, where $|\mathbb P|$ is a nonempty class and $\le$ is a binary relation such that for all $p,q,r \in |\mathbb P|$,
\begin{enumerate}
\item  $p \le p$,
\item $(p \le q\ \&\ q \le p) \implies p = q$, and
\item $(p \le q\ \&\ q \le r) \implies p \le r$.
\end{enumerate}
\end{definition}
Note that we have distinguished the PO $\mathbb P$, which is a structure, from the class $|\mathbb P|$ of its individuals, which is in general an arbitrary set.  $\mathbb P$ is $|\mathbb P|$ together with the relation $\le$.  This distinction is logically significant and is necessary in any general discussion of structure.  We often emphasize the relationship between a structure $\structure S$ and its domain of individuals by writing `$|\structure S|$' for the latter.  In some instances, as in the case of a collection of sets viewed as a partial order with $\le = \subseteq$ for example, the relations and operations of a structure are definable from the individuals, so specifying the domain of individuals is sufficient to determine the structure.  In such cases we frequently conflate a structure with is domain of individuals, typically using the structure name without the flanking $|$'s for the domain of individuals.

We say \emph{$p$ extends $q$} iff $p \le  q$.  Given $X \subseteq |\mathbb P|$, we let
\begin{eqnarray*}
\lceil X \rceil &\eqdef \{ p \in |\mathbb P| \mid (\exists q \in X)\, p \le q \}\\
\lfloor X \rfloor &\eqdef \{ p \in |\mathbb P| \mid (\exists q \in X)\, p \ge q \},
\end{eqnarray*}
and we let $\lceil p \rceil$ and $\lfloor p \rfloor$ be respectively $\lceil \{ p \} \rceil$ and $\lfloor \{ p \} \rfloor$ for $p \in |\mathbb P|$.  $ p, q \in |\mathbb P|$ are \emph{compatible} iff they have a common extension.  $p|q$ iff $p$ is incompatible with $q$.  An \emph{antichain} in $\mathbb P$ is a set of pairwise incompatible members of $|\mathbb P|$.  $\mathbb P$ has the \emph{countable chain condition} iff every antichain in $\mathbb P$ is countable.

A subset $D$ of $|\mathbb P|$ is \emph{directed} iff $(\forall p, q \in D)(\exists r \in D)(r \le p \mathand r \le q)$.  A \emph{filter} is a directed set $F$ that is closed upward, i.e., any member of $|\mathbb P|$ with an extension in $F$ is in $F$, or, equivalently, $\lfloor F \rfloor = F$.  We will usually refer to directed sets as \emph{prefilters}.  Given a prefilter $D$, $\lfloor D \rfloor$ is clearly the smallest filter that extends $D$.\footnote{In set-theoretical practice there is usually no reason to deal with prefilters that are not filters.  Note, however, that if one PO $\mathbb P$ is properly included in another PO $\mathbb P'$, and $D$ is a prefilter in $\mathbb P$, then $\lfloor D \rfloor$ in $\mathbb P'$ may not be the same as in $\mathbb P$.  For this technical reason, we find it convenient to work explicitly with prefilters.}  Any $p \in |\mathbb P|$ imposes a condition on a directed set $D$, viz., that $p$ be in $D$ and we often refer to the members of $|\mathbb P|$ in this context as \emph{conditions}.

A prefilter is \emph{maximal} iff it is not properly included in any prefilter.  Clearly, a maximal prefilter is a filter.
\begin{prop}
\label{prop max filter}
Any prefilter can be extended to a maximal filter.
\end{prop}
\begin{pf} This follows from the axiom of choice in the form of Zorn's lemma and the observation that if $\mathcal F$ is a set of prefilters linearly ordered by $\subseteq$ then $\bigcup \mathcal F$ is a prefilter.\qed\end{pf}
\begin{definition}
\label{def bool alg}
A boolean algebra is a structure $\elsalg A = ( |\elsalg A|, \compl, \vee, \wedge, 1, 0 )$ with the following properties:
\begin{enumerate}
\item $0 \neq 1$.
\item For all $P, Q, R \in |\elsalg A|$
\begin{eqnarray*}
P \vee P = P, \quad P \wedge P = P,\\
P \vee Q = Q \vee P, \quad P \wedge Q = Q \wedge P,\\
P \vee ( Q \vee R ) = (P \vee Q) \vee R,\\
P \wedge (Q \wedge R) = (P \wedge Q) \wedge R,\\(P \vee Q) \wedge R = (P \wedge R) \vee (Q \wedge R),\\(P \wedge Q) \vee R = (P \vee R) \wedge (Q \vee R).
\end{eqnarray*}
\item For each $P \in |\elsalg A|$, $\compl P$ is the unique element of $|\elsalg A|$ such that
\[
P \vee \compl P = 1 \quad\mbox{and}\quad P \wedge \compl P = 0.
\]
\item For all $P, Q \in |\elsalg A|$
\[
\compl(P \vee Q) = \compl P \wedge \compl Q
 \quad\mbox{and}\quad \compl(P \wedge Q) = \compl P \vee \compl Q.
\]
\end{enumerate}
\end{definition}

We write `$P-Q$' for `$P \wedge (\compl Q)$'.

If $\elsalg A$ is a BA we let $\elsalg A^-$ be the partial order defined by
\begin{enumerate}
\item $|\elsalg A^-| = |\elsalg A| \setminus \{0\}$;
\item $\forall P, Q \in |\elsalg A^-|\, (P \leq Q \iff P = P \wedge Q)$.
\end{enumerate}
We define \emph{compatibility}, \emph{filter}, \emph{antichain}, etc., for $\elsalg A$ in terms of the corresponding notions for $|\elsalg A^-|$ as a PO.  In terms of $\leq$, the meet and join of
$P$ and $Q$ are respectively the greatest lower bound and the least
upper bound of $\{ P, Q\}$.  The following is easily proved.
\begin{prop}
Suppose $F$ is a filter on a BA $\elsalg A$.  Then the following are equivalent:
\begin{enumerate}
\item $F$ is maximal.
\item $\forall P \in |\elsalg A^-|\, (P \in F \mathor (\exists Q \in F) P|Q)$.
\item $\forall P \in |\elsalg A^-|\, (P \in F \mathor \compl P \in F)$.
\end{enumerate}
\end{prop}
A maximal filter in a BA is also called an \emph{ultrafilter}.

A boolean algebra $\elsalg A$ is
\emph{complete} iff if every subset of $\elsalg A$ has a greatest
lower bound, which we call the \emph{meet} of the set.  Every set
in a complete boolean algebra also has a least upper bound, which
we call the \emph{join} of the set.

A boolean algebra is \emph{countably complete} iff every countable
set of elements has a meet (equivalently, a join) in the
algebra.
\begin{prop}
\label{dun}
If a BA $\elsalg A$ is countably complete and satisfies the
countable chain condition then $\elsalg A$ is complete.  
\end{prop}
\begin{pf} The proof depends on the axiom of choice.  On this
assumption any set can be put in one-one correspondence with a
von Neumann cardinal, i.e., an ordinal $\kappa$ such that there does not exist a function from any ordinal $\lambda < \kappa$ onto $\kappa$.  We proceed by induction.  Suppose joins
exist for all sets of cardinality less than $\kappa$, where
$\kappa$ is a cardinal, and suppose $X = \{ x_\alpha \mid \alpha
\in \kappa \}$ is a subset of $\elsalg A$.  Let $y_\alpha =
\bigvee \{ x_\beta \mid \beta \in \alpha \}$ for each $\alpha \in
\kappa$.  The elements $x_\alpha - y_\alpha$, $\alpha \in
\kappa$, are pairwise incompatible, so by the countable chain
condition, only countably many of them are nonzero, so the join
of the corresponding $y_\alpha$s exists.  This is clearly the
least upper bound of $X$.\qed\end{pf}


\end{document}